\documentclass[preprint,showpacs,aps,preprintnumbers,amsmath,amssymb,nofootinbib]{revtex4}
\usepackage{amsmath}
\usepackage{epsfig}
\usepackage{dcolumn}
\usepackage{bm}

\begin{document}

\preprint{version 4}
\bigskip

\title{Theory of the two-proton radioactivity in the continuum shell model}

\author{J. Rotureau$^{1-4}$}
\author{J. Oko{\l}owicz$^{1,5}$}
\author{M. P{\l}oszajczak$^{1}$}

\affiliation{%
$^1$Grand Acc\'{e}l\'{e}rateur National d'Ions Lourds (GANIL),CEA/DSM -- 
CNRS/IN2P3, BP 55027, F-14076 Caen Cedex 05, France\\
$^2$Department of Physics and Astronomy, University of Tennessee,
Knoxville, TN 37996 \\
$^3$Physics Division, Oak Ridge National Laboratory, P.O.~Box 2008,
Oak Ridge, TN 37831 \\
$^4$Joint Institute for Heavy-Ion Research, Oak Ridge, TN 37831 \\
$^5$Institute of Nuclear Physics, Radzikowskiego 152, PL-31342 Krak\'ow, Poland
}

\date{\today}

\begin{abstract}
We develop the microscopic description of the two-nucleon radioactivity in the framework of the Shell Model Embedded in the Continuum. This approach is applied for the description of spontaneous two-proton radioactivity in $^{45}$Fe,  $^{48}$Ni and $^{54}$Zn.
\end{abstract}


\maketitle

\section{Introduction}
Most of states of a nucleus are embedded in the continuum of decay channels due to which they get a finite lifetime. That means: the discrete states of a nucleus shade off into resonance states with complex energies $(E_i-\frac{i}{2}\Gamma_i)$. $E_i$ give the positions in energy of the resonance states while the widths $\Gamma_i$ are characteristic of their lifetimes. The $E_i$ may be different from the energies of the discrete states, and the widths $\Gamma_i$ may be large corresponding to a short lifetime. Nevertheless, there is a well defined relation between the discrete states characterizing the closed system, and the resonance states appearing in the open system. The main difference in the theoretical description of quantum systems without and with coupling to an environment of the decay channels is that the function space of the system is supposed to be complete in the first case while this is not so in the second case. Accordingly, the Hamilton operator is Hermitian in the first case, and the eigenvalues are discrete. The resonance states, however, characterize a subsystem described by a non-Hermitian Hamilton operator with complex eigenvalues. The function space containing 'everything' consists, in the second case, of system plus environment.

The mathematical formulation of this problem goes back to Feshbach \cite{feshbach} who introduced the two subspaces of the Hilbert space
containing the (i) discrete and (ii) real-energy scattering states, respectively. This has given foundation of the Continuum Shell Model (CSM) \cite{mahaux,philpott,rotter}. Number of particles in
the scattering states provides in CSM a natural hierarchy of approximations which, phenomenologically,  is associated with a successive opening of more and more complicated decay channels at higher excitation energies. The technical difficulties in the practical implementation of this strategy are such that all past numerical applications of the CSM have been restricted to the description of processes involving one-nucleon decay channels only. Few attempts to treat the multiparticle continuum \cite{eppel,wendler} proposed numerical schemes which have never been adopted in realistic multiconfiguration mixing calculations.
  
Feshbach succeeded in formulating a unified description of nuclear reactions with both direct processes (the short-time scale) and compound nucleus processes (the long-time scale). A unified description of nuclear structure and nuclear reaction aspects is much more complicated and became possible only at the turn of the last century \cite{karim,Benb} in the framework of the so-called Shell Model Embedded in the Continuum (SMEC) (see Ref. \cite{opr} for recent review) which is parented to the CSM. 
The SMEC has been applied for the description of spectra and reactions involving one particle in the scattering continuum, like the $(p,p')$ reaction, the proton/neutron radiative capture reactions  \cite{karim,Benb,karim1}, the Coulomb dissociation reaction \cite{shyam}, or the first-forbidden $\beta$-decay \cite{karim2}. Further applications of the SMEC with one-particle continuum involved the study of the continuum effects in the binding systematics of neutron-rich nuclei in $sd$ shell \cite{masses}, and the statistical aspects of the continuum coupling for unbound states in $^{24}$Mg \cite{drozdz}.

 First multiconfigurational shell model (SM) approach with no restriction on the number of particles in the continuum has been proposed recently \cite{Mic02,betan} in the complete Berggren single-particle (s.p.) basis \cite{berggren} which consists of Gamow (or resonant) states and the complex non-resonant continuum of scattering states. The s.p. Berggren basis is generated by a finite-depth potential, and the many-body states are obtained as the linear combination of Slater determinants spanned by resonant and non-resonant s.p. basis states. GSM is a tool {\em par excellence} for nuclear structure studies which includes all continuum effects and correlations between nucleons simultaneously. However, an absence of the projection on the asymptotic decay channels does not allow at present for the application of GSM to the description of nuclear reactions.

In this work, we shall formulate the SMEC with the two-particle scattering continuum and present the theory of the two-proton (2p) decay.
The 2p radioactivity, which has been observed in 2002 by Pf\"utzner {\em et al.} \cite{pfutzner} and Giovinazzo {\em et al.} \cite{giovinazzo}, is a new spontaneaous decay mode in Nature which may appear in even-$Z$ nuclei beyond 2p drip-line. Various theoretical models have been proposed to describe this new form of the radioactivity \cite{grigorenko,fedorov,Barb,rop1}.

The profound relation between the appearance of the 2p radioactivity in even-$Z$ systems and the short-ranged pairing correlations gives a hope that in future studies of the proton-proton spatial-momentum correlations in the asymptotic state of 2p decay, one will learn about the basic features of the pairing field, such as its radial dependence or the multipole structure. A possible influence of the nucleon-nucleon (NN) correlations in the decaying even-$Z$ nucleus on the asymptotic state of two protons, can be extracted from the data only in the unified theoretical framework which contains all ingredients which are necessary for a description of the initial state of $A$-nucleon system, the final state of $A-2$-nucleon system and the asymptotic state of two emitted protons. 
In the SMEC formalism with the two-particle continuum, one can describe: (i) the asymptotic physical  two-particle states, (ii) the configuration mixing in the wave functions of the decaying system with $A$ nucleons and the daughter systems with A-1 and $A-2$ nucleons, (iii) the coupling of discrete states in the parent system $A$ to the decay channels $[(A-1)\otimes (1)]$ and $[(A-2)\otimes (2)]$, as well as (iv) the complete antisymmetrization of the wave-functions in both parent $(A)$ and daughter $(A-1)~ {\rm and} ~(A-2)$ systems. Hence, the SMEC allows to extract features of the NN correlations not only from the width of the decaying state but also from the correlations both between the emitted protons and between the emitted proton(s) and the daughter nuclei $(A-1)$, $(A-2)$. Presently available experimental data allows to address only the first of those two theoretical challenges.  

In this work, we shall apply the SMEC formalism for the description of the half-lifes for the 2p decays from the ground state (g.s.) of $^{45}$Fe, $^{48}$Ni and $^{54}$Zn. This formalism has been applied before for the study of the 2p decay from the excited state $1_2^-$ in $^{18}$Ne \cite{rop1}. Here, we shall compare a diproton decay (a direct 2p decay) with a sequential 2p emission through either the continuum states correlated by weakly bound states of A-1 nucleus, or the resonance(s) of the nucleus A-1. The residual interaction between discrete states and the scattering continuum states is the Wigner-Bartlett zero-range interaction what restricts our description of the diproton decay with three-body asymptotics to the scenario, introduced first in the $R$-matrix studies by Barker \cite{barker1}, consisting of the emission of $(2p)$ cluster and incorporating the final-state interaction between two protons in terms of the $s$-wave phase shift. 

The paper is organized as follows. The essential elements of the SMEC formalism with one particle in the scattering continuum are discussed in chapter II. The extension of the SMEC formalism to include both one- and two-particle continuum states will be discussed in chapter III. Chapter IV is devoted to the detailed presentation of the theory of the 2p emission. We shall discuss different limiting situations corresponding to (i) the direct 2p decay with three-body asymptotics (cf sect. IV.C), (ii) the direct (2p) cluster emission (cf sect. IV.B), and (iii) the sequential 2p emission through either the correlated continuum in the absence of intermediate resonance(s) or through the resonance(s) in the intermediate A-1 nucleus (cf sect. IV.A). All details which are not necessary to follow the main ideas of the extended SMEC formalism and the 2p emission theory are put in the appendices. Application of the SMEC formalism to the description of the 2p decay from the g.s. of $^{45}$Fe, $^{48}$Ni and $^{54}$Zn is discussed in chapter V. The main conclusions are given in chapter VI. 


\section{SMEC with one particle in the scattering continuum}
The SMEC describes the nucleus as an open quantum system (OQS) \cite{opr}. The total function space consists of two sets: the space of $L^2$-functions used in the standard SM and the function space of the scattering states into which the SM states are embedded. Discrete SM states of the $A$-nucleon system can decay only when their coupling to the scattering wave functions is taken into account. 

The two sets of wave functions are defined by solving the Schr\"odinger equation:
\begin{eqnarray}
\label{SMeq1}
H_{SM}|\Phi_i\rangle=E_i^{(SM)}|\Phi_i\rangle 
\end{eqnarray}
for the discrete SM states of the closed quantum system (CQS), and the Schr\"odinger equation:
\begin{eqnarray}
\label{SMeq2}
\sum_{c'}(E-H_{cc'})\xi_E^{c'(+)}=0
\end{eqnarray}
for the scattering states of the environment. Here, $H_{SM}$ is the standard SM Hamiltonian and $H_{cc}=H_0+V_{cc}$ is the standard Hamiltonian used in the coupled-channel (CC) calculations. The channels are determined by the motion of unbound particle relative to the residual nucleus with A-1 bound particles in a certain state $|\Phi_j^{A-1}\rangle$. 
The states $\{\Phi_j^{A-1}\}$ of a daughter nucleus are discrete SM states, $\xi_E^{c(+)}$ are scattering states projected on the channel $c$.
The channel numbers $c$ are defined by the quantum numbers of the states $j$ of the daughter nucleus and those of the unbound particle which are coupled to the total angular momentum and parity. The states of the A-1 system are assumed to be stable. 

By means of the two functions sets: ${\cal Q}\equiv \{|\Phi_i^{A}\rangle\}$, ${\cal P}\equiv \{|\xi_E\rangle\}$, one can define the projection operators:
\begin{eqnarray}
{\hat Q}&=&\sum_{i=1}^N|\Phi_i^A\rangle\langle\Phi_i^A| \nonumber \\
{\hat P}&=&\int_{0}^\infty dE|\xi_E\rangle\langle\xi_E|
\end{eqnarray}
with
\begin{eqnarray}
{\hat Q}|\xi_E\rangle&=&0 \nonumber \\
{\hat P}|\Phi_i^A\rangle&=&0
\end{eqnarray}
and the projected Hamiltonians: ${\hat Q}H{\hat Q}\equiv H_{QQ}$ and ${\hat P}H{\hat P}\equiv H_{PP}$. We identify $H_{SM}$ with $H_{QQ}$ the CQS Hamiltonian and $H_{cc}$ with $H_{PP}$. The Schr\"odinger equation in the total function space is
\begin{eqnarray}
(H-E)|\Psi_E\rangle=0
\end{eqnarray}
Assuming ${\cal Q}+{\cal P}=I_d$, one can determine a third wave function $|\omega_i^{(+)}\rangle$ which is the continuation of the SM
 wave function $|\Phi_i^A\rangle$ into the continuum ${\cal P}$. Function $|\omega_i^{(+)}\rangle$ is obtained by solving the CC equations
with the source term {\bf{$|w_i\rangle=H_{PQ}|\Phi_i\rangle$}}:
\begin{eqnarray}
\label{SMeq5}
|\omega_i^{(+)}(E)\rangle=G_P^{(+)}(E)|w_i\rangle
\end{eqnarray} 
where
\begin{eqnarray}
\label{extra_00}
G_P^{(+)}(E)={\hat P}(E-H_{PP})^{-1}{\hat P}
\end{eqnarray}
is the Green's function for the motion of a single nucleon in the ${\cal P}$ subspace, $E$ is the total energy of the nucleus $A$ and $H_{PQ}\equiv {\hat P}H{\hat Q}$. Using the three function sets: $\{|\Phi_i^A\rangle\}$, $\{|\xi_E\rangle\}$ and $\{|\omega_i^{(+)}\rangle\}$, one constructs the solution 
$|\Psi_E\rangle={\hat Q}|\Psi_E\rangle+{\hat P}|\Psi_E\rangle$ in the total function space with:
\begin{eqnarray}
{\hat Q}|\Psi_E\rangle &=&(E-{\cal H}_{QQ}(E))^{-1} H_{QP} |\xi_E\rangle \nonumber \\
{\hat P}|\Psi_E\rangle&=&|\xi_E\rangle+G_P^{(+)}(E)H_{PQ} {\hat Q}|\Psi_E\rangle
\end{eqnarray}
One obtains:
\begin{eqnarray}
|\Psi_E\rangle=|\xi_E\rangle+\sum_{i,k}(|\Phi_i^A\rangle+|\omega_i^{(+)}(E)\rangle)\langle\Phi_i^A|
(E-{\cal H}_{QQ}(E))^{-1}|\Phi_k^A\rangle\langle\Phi_k^A|H_{QP}|\xi_E\rangle
\end{eqnarray}
In the above equations, ${\cal H}_{QQ}(E)$ stands for the energy dependent effective Hamiltonian:
\begin{eqnarray}
\label{SMeq3}
{\cal H}_{QQ}(E)=H_{QQ}+H_{QP}G_P^{(+)}(E)H_{PQ}
\end{eqnarray}
 in the function space of discrete states which takes into account the modification of the CQS Hamiltonian $H_{QQ}$ by the coupling to the scattering states. ${\cal H}_{QQ}$ is therefore the OQS Hamiltonian in ${\cal Q}$ subspace. 
 
 The energies and widths of the resonance states follow from the solutions of the fixed-point equations:
 \begin{eqnarray}
 \label{fixed_point}
 E_i&=&{\tilde E}_i(E=E_i) \nonumber \\
 \Gamma_i&=&{\tilde \Gamma}_i(E=E_i)
 \end{eqnarray}
where functions ${\tilde E}_i(E)$ and ${\tilde \Gamma}_i(E)$ follow from the eigenvalues of the OQS energy-dependent Hamiltonian in ${\cal Q}$ subspace. The identification of $E_i$ and ${\Gamma}_i$ in (\ref{fixed_point}) with standard spectroscopic observables is justified by an adequate definition of the two subspaces ${\cal Q}$ and ${\cal P}$.

\section{SMEC with two particles in the scattering continuum}

\subsection{Effective Hamiltonian in ${\cal Q}$  \label{df}}
Let us denote by ${\cal  T}$ a subspace of the Hilbert space with the two-particles in the continuum and by ${\hat T}$ the corresponding projection operator. We assume:
\begin{eqnarray}
{\cal Q}+{\cal P}+{\cal T}=I_{d}
\end{eqnarray}
which allows to formulate the completness relation in the total function space. Consequently, one can decompose the Hamiltonian $H$ into the parts acting in different subspaces and the coupling terms between those different subspaces:
\begin{eqnarray}
H&=&({\hat Q}+{\hat P}+{\hat T})H({\hat Q}+{\hat P}+{\hat T}) \nonumber  \\  &=&H_{QQ}+H_{QP}+H_{QT}+H_{PQ}+H_{PP}+H_{PT}+H_{TQ}+H_{TP}+H_{TT}
\end{eqnarray}
One can show (cf appendix \ref{H_eff}) that the effective Hamiltonian ${\cal H}_{QQ}(E)$, 
which takes into account the modification  of the CQS Hamiltonian $H_{QQ}$ by the couplings to the ${\cal P, T}$ subspaces, can be written in the form:
\begin{eqnarray}
{\cal H}_{QQ}(E)&=&H_{QQ}+H_{QP}G_{P}^{+}(E)H_{PQ} \nonumber \\
&+&[H_{QT}+H_{QP}G_{P}^{+}(E)H_{PT}]\tilde{G}_{T}^{+}(E)[H_{TQ}+H_{TP}G_{P}^{+}(E)H_{PQ}] 
\label{H_eff_T}
\end{eqnarray}
 which separates terms due to the coupling with one- and two-particle continuum of scattering states. In the above equation, 
$\tilde{G}_{T}^{+}(E)$ is the Green's function in ${\cal T}$ modified by the coupling with ${\cal P}$:
\begin{eqnarray}
\tilde{G}_{T}^{+}(E)=\lim_{\epsilon \to 0}[E+i\epsilon-H_{TT}-H_{TP}G_{P}^{+}(E)H_{PT}]^{-1} 
\label{H_eff_P}
\end{eqnarray}
Similarly as in the standard SMEC with one particle in the scattering continuum, we define the source term:
 \begin{eqnarray}
 \label{intro_01}
|w_i\rangle=[H_{TQ}+H_{TP}G_{P}^{(+)}(E)H_{PQ}]|\Phi_i\rangle  \label{source_T} 
\end{eqnarray}
and the continuation $|\omega^{+}_i\rangle$ of the SM wave function $|\Phi_i\rangle$ into the continuum 
${\cal T}$:
\begin{eqnarray}
\label{intro_02}
|\omega^{+}_i\rangle=\tilde{G}_{T}^{(+)}(E)[H_{TQ}+H_{TP}G_{P}^{(+)}(E)H_{PQ}]|\Phi_i\rangle 
=\tilde{G}_{T}^{(+)}(E)|w_i\rangle \label{omega_T}
\end{eqnarray}
Matrix elements  of $[H_{QT}+H_{QP}G_{P}^{(+)}(E)H_{PT}]\tilde{G}_{T}^{(+)}(E)[H_{TQ}+H_{TP}G_{P}^{(+)}(E)H_{PQ}]$
in ${\cal H}_{QQ}(E)$, which correspond to the coupling with ${\cal T}$, can be expressed as 
the overlap between the source term  (\ref{intro_01}) and the function $\omega^{+}_j$ (eq. (\ref{intro_02})).

One may notice that  ${\cal H}_{QQ}(E)$ can be also written in the form:
\begin{eqnarray}
{\cal H}_{QQ}(E)&=&H_{QQ}+H_{QT}G_{T}^{(+)}(E)H_{TQ} \nonumber \\ 
&+&[H_{QP}+H_{QT}G_{T}^{(+)}(E)H_{TP}]\tilde{G}_{P}^{(+)}(E)[H_{PQ}+H_{PT}G_{T}^{(+)}(E)H_{TQ}] 
\label{H_eff_T2}
\end{eqnarray}
which separates explicitely the direct coupling term between ${\cal Q}$ and ${\cal T}$ subspaces. Here
 $\tilde{G}_{P}^{(+)}(E)$ stands for the Green's function in ${\cal P}$ modified by the couplings with ${\cal T}$:
\begin{eqnarray}
\tilde{G}_{P}^{(+)}(E)=\lim_{\epsilon \to 0}[E+i\epsilon-H_{PP}-H_{PT}G_{T}^{(+)}(E)H_{TP}]^{-1} 
\end{eqnarray}
The expression (\ref{H_eff_T2}) is obtained by a permutation of projection operators ${\hat P}$ and ${\hat T}$ in eq. (\ref{H_eff_T}).

The coupling between ${\cal Q}$, ${\cal P}$ and ${\cal T}$ subspaces yields also effective Hamiltonians ${\cal H}_{PP}(E)$ and ${\cal H}_{TT}(E)$ in ${\cal P}$ and ${\cal T}$ subspaces, respectively.
For example, the unperturbed Hamiltonian $H_{PP}$ becomes:
\begin{eqnarray}
{\cal H}_{PP}(E)&=&H_{PP}+H_{PQ}G_{Q}(E)H_{QP} \nonumber \\ 
&+&[H_{PT}+H_{PQ}G_{Q}(E)H_{QT}]\tilde{G}_{T}^{(+)}(E)[H_{TP}+H_{TQ}G_{Q}(E)H_{QP}] 
\end{eqnarray}
in the OQS formalism. Similarly, $H_{TT}$ becomes:
\begin{eqnarray}
{\cal H}_{TT}(E)&=&H_{TT}+H_{TQ}G_{Q}(E)H_{QT} \nonumber \\ 
&+&[H_{TP}+H_{TQ}G_{Q}(E)H_{QP}]\tilde{G}_{P}^{(+)}(E)[H_{PT}+H_{PQ}G_{Q}(E)H_{QT}] 
\end{eqnarray}
In the following, we shall be interested in applying the SMEC formalism for a particular problem of the 2p emission, {\em i.e.} the decay from ${\cal Q}$ to ${\cal T}$, for which a relevant operator is ${\cal H}_{QQ}(E)$.


\section{Description of 2p emission in SMEC}

The effective Hamiltonian ${\cal H}_{QQ}(E)$ (cf (\ref{H_eff_T}) and (\ref{H_eff_T2})) takes into account couplings between ${\cal Q}$ subspace and the subspaces with one-  (${\cal P}$-subspace) and two-nucleons (${\cal T}$-subspace) in the scattering continuum. In this expression, all possible emissions of two protons as well as one proton are implicitely included. In real 2p decays, certain emission scenarios may be less probable, so it is interesting to consider limiting cases of the general emission process and isolate appropriate terms describing them. 

If nuclei A and A-1 respect the diproton emission condition considered by Goldansky \cite{gold60}:
\begin{eqnarray}
E_{A-1}-\frac{1}{2}\Gamma_{A-1} > E_{A}+\frac{1}{2}\Gamma_{A}  
\label{cond}
\end{eqnarray}	
then one can suppose that effects of the coupling involving  ${\cal P}$ subspace, {\it i.e.} the terms $H_{QP}$, $H_{TP}\dots$  in (\ref{H_eff_T}) and (\ref{H_eff_T2}), are less important. In this case,  
${\cal H}_{QQ}(E)$ can be approximated by:
\begin{eqnarray}
{\cal H}_{QQ}^{(dir)}(E)=H_{QQ}+H_{QT}G^{(+)}_{T}(E)H_{TQ}
 \label{H_eff_direct}
\end{eqnarray}
${\cal H}_{QQ}^{(dir)}(E)$ is the effective Hamiltonian in ${\cal Q}$ subspace describing the {\em direct} emission from ${\cal Q}$ to ${\cal T}$. 
If the condition (\ref{cond}) is not satisfied, then 
an intermediate system A-1 plays an important role in the 2p emission. Neglecting the direct couplings between ${\cal Q}$ and ${\cal T}$ subspaces in the expression (\ref{H_eff_T}), one obtains:
 \begin{eqnarray}
{\cal H}_{QQ}^{(seq)}(E)=H_{QQ}+H_{QP}\tilde{G}_{P}^{(+)}(E)H_{PQ} 
\label{H_eff_seq2}   
 \end{eqnarray}

This operator describes
emission of two protons through the resonance of an intermediate nucleus A-1 and in this case, the emission of the first particle implies automatically the emission of a second particle. 

The mechanism of sequential 2p emission process may also occur via the continuum states correlated by either weakly bound states of a nucleus A-1, or by weakly unbound states in nucleus A-1. Such a physical situation has been studied recently in the 2p decay of the $1_2^-$ state in $^{18}$Ne \cite{campo,rop1}.
The effective Hamiltonian describing this situation is:
 \begin{eqnarray}
{\cal H}_{QQ}^{(seq)}(E)=H_{QQ}+H_{QP}G_{P}^{(+)}(E)H_{PQ} 
+[H_{QP}G_{P}^{(+)}(E)H_{PT}]\tilde{G}_{T}^{(+)}(E)[H_{TP}G_{P}^{(+)}(E)H_{PQ}] \nonumber \\
\label{H_eff_seq1} 
\end{eqnarray}
As before, this expression has been  derived from  (\ref{H_eff_T2}) neglecting the direct couplings between   ${\cal Q}$ and ${\cal T}$ subspaces.  The third term in (\ref{H_eff_seq1}) describes a sequential 2p decay whereas the second term is responsible for the 1p decay.
In the following, we shall calculate the width for 2p emission assuming either a sequential emission, {\it i.e.} two succesive and independent proton emissions, or a direct emission of (2p)  cluster. The intereference between these two limiting processes will be neglected in the present studies. 
%
%


\subsection{Sequential 2p emission \label{sub_seq}} 
The sequential 2p emission may occur either through the resonance of an intermediate nucleus A-1 or through the correlated continuum of nucleus A-1. In the following, we shall be interested in the latter case. 
Matrix elements corresponding to the direct ${\cal Q}$-${\cal T}$ coupling are usually much smaller than those of the ${\cal Q}$-${\cal P}$ coupling due to the smaller Coulomb barrier in the latter case. For that reason, we rewrite the effective Hamiltonian (\ref{H_eff_T}) (cf eq. (\ref{H_eff_seq1}) as:
\begin{eqnarray}
{\cal H}_{QQ}^{(seq)}(E)=H_{QQ}+H_{QP}G_{P}^{(+)}(E)H_{PQ}
+H_{QP}\tilde{G}_PH_{PT}G_{T}^{(+)}(E)H_{TP}G_{P}^{(+)}(E)H_{PQ} 
\label{3_express}
\end{eqnarray}
{\em i.e.} we neglect $H_{QT}$ couplings (the direct 2p emission) but keep $H_{QP}$ couplings (the 1p emission). Hence, the interference between the 1p emission and the sequential 2p emission decay can be investigated both for close (virtual ${\cal Q}$-${\cal P}$ excitations) and open (a 'true' 1p decay) 1p emission channels. 

 One can see in (\ref{3_express}) the source terms for both the 1p emission and the sequential 2p emission. Diagonalizing ${\cal H}_{QQ}^{(seq)}(E)$, one obtains energies  of states in the nucleus A  as well as their widths associated with the emission of one and two protons. 
 
 In principle, one cannot separate the partial widths for each of those decay modes. However, since the couplings corresponding to the 2p emission are  in most cases significantly smaller than those associated with the 1p emission, we shall first diagonalize ${\cal H}_{QQ}^{(seq)}(E)$  in the SM basis $\{|\Phi^{A}\rangle\}$  neglecting the 2p emission. This provides new basis states $\{{|\tilde \Phi}^{A}\rangle\}$ which are linear combinations of SM states in ${\cal Q}$.
Using these new {\em mixed} SM {\em states}, we calculate the 2p emission width for a sequential decay, {\it i.e.} we calculate the matrix element:
\begin{eqnarray}
\delta(E)=\langle {\tilde \Phi}^{A}_{i}|H_{QP}\tilde{G}_P
H_{PT}G_{T}^{(+)}(E)H_{TP}G_{P}^{(+)}(E) H_{PQ}|{\tilde \Phi}^{A}_{i}\rangle 
\label{seq_me}
\end{eqnarray}
describing the sequential 2p emission of the decaying state $|{\tilde \Phi}^{A}_{i}\rangle$. One should notice, that mixing of SM states $\{|\Phi_i^{A}\rangle\}$ due to the sequential 2p emission can be neglected because the dominant term is the 1p emission.

In the following, we shall assume that the subsequent proton emissions are independent, {\it i.e.} we neglect correlations between the two protons in the continuum and describe the interaction of the first emitted proton with other A-1 nucleons by a mean-field ${\hat p}h^{(seq)}{\hat p}$, where ${\hat p}$ denotes the projection operator on the one-particle continuum states. For ${\hat p}h^{(seq)}{\hat p}$, we take a diagonal potential which enters in the CC equations. This implies (cf  appendix \ref{seq_formalism}):
\begin{itemize}
\item 
$H_{PP}\rightarrow H_{Q'Q'}+{\hat p}h^{(seq)}{\hat p}$ : $H_{PP}$ is divided into $H_{Q'Q'}$
and ${\hat p}h^{(seq)}{\hat p}$. $H_{Q'Q'}$  acts in the ${\cal Q'}$ subspace containing the (quasi-)bound states of a nucleus A-1.
\item 
$H_{TT}\rightarrow H_{P'P'}+{\hat p}h^{(seq)}{\hat p}$ : $H_{TT}$ is divided into $H_{P'P'}$ and
${\hat p}h^{(seq)}{\hat p}$. $H_{P'P'}$ acts in the subspace ${\cal P'}$ which contains the states of
 A-1 nucleons out of which A-2 are (quasi-)bound and one proton occupies a continuum state.
\item 
$H_{PT}\rightarrow H_{Q'P'}\bigotimes I_d(A)$ : {\it i.e.} matrix elements of the residual two-body interaction involving the first emitted proton are neglected.
\end{itemize}
With these assumptions, the matrix element $\delta(E)$ takes a form (cf appendix \ref{seq_formalism}):
\begin{eqnarray}
\delta(E)&=&\langle {\tilde \Phi}^{A}_{i}|H_{QP}\frac{1}{E^{+}-{\hat p}h_{0}^{(seq)}{\hat p}-H_{Q'Q'}-H_{Q'P'}(E^{+}-{\hat p}h_{0}^{(seq)}{\hat p}-H_{P'P'})^{-1}H_{P'Q'}} H_{Q'P'} \nonumber \\ 
\nonumber  \\
&\times& \frac{1}{E^{+}-{\hat p}h_{0}^{(seq)}{\hat p}-H_{P'P'}}  H_{P'Q'} 
\frac{1}{E^{+}-{\hat p}h_{0}^{(seq)}{\hat p}-H_{Q'Q'}}H_{PQ}|{\tilde \Phi}^{A}_{i}\rangle  
\label{seq_me2}
\end{eqnarray}
and the width of a state $|{\tilde \Phi}^{A}_{i}\rangle$ is given by:
\begin{eqnarray}
\label{szer}
\Gamma(E)=-2 {\cal I}m(\delta(E))
\end{eqnarray}
The width of a physical resonance state follows then from the solution of the fixed-point equations (\ref{fixed_point}) on condition that all subspaces of the Hilbert space involved in the description of the resonance decay are defined adequately.

In the case of a sequential decay through a resonance in the intermediate  nucleus A-1, the effective Hamiltonian is given by the expression (\ref{H_eff_seq2}). With the same assumptions as used in the derivation of eq. (\ref{seq_me2}), one obtains (cf  appendix \ref{seq_formalism}):
\begin{eqnarray}
\delta(E)&=&\langle {\tilde \Phi}^{A}_{i}|H_{QP}\frac{1}{E^{+}-{\hat p}h_{0}^{(seq)}{\hat p}-H_{Q'Q'}-H_{Q'P'}(E^{+}-{\hat p}h_{0}^{(seq)}{\hat p}-H_{P'P'})^{-1}H_{P'Q'}} H_{PQ}|{\tilde \Phi}^{A}_{i}\rangle \nonumber \\  
\label{seq_me3}
\end{eqnarray}


\subsection{(2p)  cluster emission \label{cluster_theo}}

In this section, we shall consider the emission of two correlated protons in a form of the (2p)  cluster. The effective Hamiltonian for this process is given  in eq. (\ref{H_eff_direct}). Moreover, if  one includes couplings to the one-nucleon decay channels which are responsible for 
the {\em external mixing} of SM states in ${\cal Q}$, then ${\cal H}_{QQ}^{(dir)}$ becomes:
\begin{eqnarray}
\label{eqham}
{\cal H}_{QQ}^{(dir)}(E)=H_{QQ}+H_{QP}G_{P}^{(+)}(E)H_{QP}+H_{QT}G_{T}^{(+)}(E)H_{TQ}
\end{eqnarray}
Formally, this expression is derived from eq. (\ref{H_eff_T}) (or eq. (\ref{H_eff_T2})), neglecting couplings between ${\cal P}$ and  ${\cal T}$ subspaces. 

In the following, we assume the two-step scenario for the 2p decay \cite{Barb}. In the first step, two protons are emitted in a form of a (2p)  cluster. In the second step, the cluster desintegrates outside of the nuclear potential  of nucleus A-2 due to the final state interaction \cite{Wat,Mig}. The final state pp-interaction is taken into account by the density $\rho(U)$ of
 proton states corresponding to the intrinsic energy $U$ in the proton-proton system \cite{Bara,Barb}. The calculation of the density   $\rho(U)$ is based on the $s$-wave phase shift in $pp$ collision.

Matrix element of the effective Hamiltonian describing the (2p)  cluster emission is:
\begin{eqnarray}
\delta(E)=\langle{\tilde \phi}^{(int)}_{i}|H_{QT}G_{T}^{(+)}(E)H_{TQ}|{\tilde \phi}^{(int)}_{i}\rangle  ~ \ ,
\label{cluster_mat_elt}
\end{eqnarray}
where $|{\tilde \phi}^{(int)}_{i}\rangle$ is the intrinsic state corresponding to the SM state $|{\tilde\Phi}_{i}\rangle$ mixed by either by coupling to one-proton decay channels or diproton decay channes (cf sect. \ref{sect_dipr_Fe}). Working with intrinsic states $\{{\tilde \phi}^{(int)}_{i}\}$, allows to take into account the recoil correction for the daughter system A-2. 

Let us consider the completness relation:
\begin{eqnarray}
\int_{0}^{+\infty} dR R^{2} \int_{0}^{+\infty} dr r^{2} \sum_{c=[t^{(int)},(l_{x},l_{y})^{L},(L,S)^{J_{2p}}]}|c,r,R\rangle \langle c,r,R|=I_{d} 
\label{compl}
\end{eqnarray}
where $c$ is a channel characterized by  the intrinsic state $t^{(int)}$ of a daughter nucleus A-2, the intrinsic angular momentum of the subsystem formed by two protons $l_{x}$, the
spin of two protons $S$, and the relative angular momentum $l_{y}$ between two protons and a daughter nucleus A-2. $l_{x}$  and $l_{y}$ are coupled to $L$, $L$ and $S$ are coupled to
 $J_{2p}$ and, finally, the total angular momentum of nucleus A-2
 and $J_{2p}$ are coupled to $J$. $r$ is the intrinsic radial variable of the cluster and
 $R$ is the distance between the center of mass of the cluster and the daughter nucleus A-2.
 Intrinsic state $0s$ of a (2p)  cluster is described in the harmonic oscillator basis. Since the intrinsic angular momentum of the (2p)  cluster is $l_{x}=0$, therefore its spin is $S=0$ due to the antisymmetry of the wave function. 

Sharing of the total energy between the intrinsic energy of cluster and the energy associated with the center of mass motion of cluster is taken into account phenomenologically by the proton states density $\rho(U)$, {\it i.e.} we suppose that the emission of two protons in $0s$ intrinsic state with the intrinsic energy $U$ is distributed according to the density $\rho(U)$. In this case, the problem of 2p emission with the three-body asymptotics reduces to a problem with the 
two-body asymptotics because the degrees of freedom corresponding to the intrinsic movement of protons are described phenomenologically by the proton states density $\rho(U)$. The corresponding
completness relation (\ref{compl} ) takes a form:
\begin{eqnarray}
\int_{0}^{+\infty} dR R^{2} \int_{0}^{+\infty} dU \sum_{c(U)=[t^{(int)},0s(U),L,S=0]} \rho(U) |c(U),R\rangle \langle c(U),R|=I_{d}
\end{eqnarray}
Hence, the matrix element (\ref{cluster_mat_elt}) can be written as:
\begin{eqnarray}
&&\int_{0}^{+\infty}dR R^{2}
\int_{0}^{+\infty} dU \rho(U) \langle {\tilde \phi}^{(int)}_{i}|H_{QT}  \sum_{c(U)}|c(U),R\rangle \langle c(U),R|G_{T}^{(+)}(E)H_{TQ}|{\tilde \phi}^{(int)}_{i}\rangle \nonumber \\
=&&\int_{0}^{+\infty}\int_{0}^{+\infty}dU dR R^{2} w_{i,c}^{*}(U,R) \rho(U) \omega^{(+)}_{i,c}(U,R) 
\label{mat_elt_cluster} 
\end{eqnarray}
where $w_{i,c}(U,r)$ is the projection of the source term $|w_i\rangle=H_{TQ}|{\tilde \phi}^{(int)}_{i}\rangle$
on the channel $c(U)$:
\begin{eqnarray}
w_{i,c}(U,R)=R\langle c(U),R|w\rangle=R \langle c(U),R|H_{TQ}|{\tilde \phi}^{(int)}_{i}\rangle
\end{eqnarray}
The calculation of the source term is given in the appendix \ref{source_annexe}. The projected source $w_{i,c}(U,r)$ does not depend explicitely on  $U$ because this dependence in the two-step emission scenario follows from the  cluster emission process. In the following, the projected cluster source will be denoted by $w_{i,c}(r)$. 

The coupling term $H_{TQ}$ is given by the two-body residual interaction:
\begin{eqnarray}
H_{TQ}={\hat T}\left(\sum_{i}h(i)+\sum_{i<j} V^{(res)}(ij)\right){\hat Q} 
={\hat T}\left(\sum_{i<j} V^{(res)}(ij)\right){\hat Q} 
\label{hqt} 
\end{eqnarray}
The Coulomb interaction is included as the average Coulomb field in $h$ and does not enter in  $H_{TQ}$.

The function $\omega^{(+)}_{i,c}(U,R)$ in (\ref{mat_elt_cluster}) is the projection on channel
 $c(U)$  of the continuation $|\omega_i^{(+)\rangle}=G_{T}^{(+)}(E)H_{TQ}|{\tilde \phi}^{(int)}_{i}\rangle$ of an intrinsic state  $|{\tilde \phi}^{(int)}_{i}\rangle$  in ${\cal T}$ subspace:
\begin{eqnarray}
\omega^{(+)}_{i,c}(U,R)=R\langle c(U),R| G_{T}^{(+)}(E)H_{TQ}|{\tilde \phi}^{(int)}_{i}\rangle \label{omega_eq}
\end{eqnarray}

For the Hamiltonian  $H_{TT}$, we suppose that the total system ${\rm [A-2]}\otimes{\rm [(2p)]}$ can be considered as a two-body system in the mean-field  $U_{0}$, {\it i.e.} we are considering
 $H_{TT}$ in a following form:
\begin{eqnarray}
H^{(cl)}_{TT}= {\hat T}^{(cl)} \left [ {\tilde H}^{(A-2)}+ {\tilde H}^{(cl)}+ \frac{P^{2}}{2\mu} + U_{0} \right ] {\hat T}^{(cl)}
\label{H_tt_cluster}
\end{eqnarray}
where ${\tilde H}^{(A-2)}$ is the intrinsic Hamiltonian of the daughter nucleus  A-2, and  ${\tilde H}^{(cl)}$ is the intrinsic Hamiltonian of the (2p)  cluster. The cluster is described as a particle of mass $M=2M_{p}$ ($M_{p}$ denotes a proton mass) and charge $Z=2$. $P^{2}/2\mu$ is the intrinsic kinetic energy of the system ${\rm [A-2]}\otimes{\rm [(2p)]}$, and $\mu$  is the reduced mass of the system.  ${\hat T}^{(cl)}$ is the projection operator on the subspace of cluster states in the continuum of the potential $P^{2}/2\mu + U_{0}$. Calculation of the projected function $\omega^{(+)}_{i,c}(U,R)$ corresponds to solving an inhomogeneous 
Schr\"{o}dinger equation:
\begin{eqnarray}
(E-H^{(cl)}_{TT})|\omega_i^{(+)}\rangle
=|w_i\rangle  
\label{eq_omega_prev}
\end{eqnarray}
Projecting eq. (\ref{eq_omega_prev}) on a channel $c(U)$, one obtains:
 \begin{eqnarray}
 \left [E-({\tilde E}^{(A-2)}+U)-{\hat T}^{(cl)}\left(\frac{P^{2}}{2\mu} + U_{0}\right){\hat T}^{(cl)}\right ] 
\omega^{(+)}_{i,c}(U,R)=w_{i,c}(U,R) 
\label{eq_omega}
\end{eqnarray}
In this representation, ${\tilde E}^{(A-2)}$ is the intrinsic energy of a daughter nucleus and $U$ is the intrinsic energy of a (2p)  cluster which is distributed according to the density $\rho(U)$. 
%
%


\subsection{Direct 2p emission with three-body asymptotics} \label{para_direct}

In this section, we shall describe direct 2p emission without a simplifying two-step scenario in which the first step consists of the emission of the (2p)  cluster which, subsequently, decays due to the final-state interaction. Matrix elements of the effective Hamiltonian ${\cal H}_{QQ}^{(dir)}(E)$ (cf (\ref{H_eff_direct}))  in the SM basis $\{|\Phi_i^{A}\rangle\}$ are:
\begin{eqnarray}
\label{extra_01}
\langle\Phi_i^{A}| {\cal H}_{QQ}^{(dir)}(E)|\Phi_j^{A}\rangle=E_i^{(SM)}\delta_{ij} 
+\langle w_i|\omega^{(+)}_j\rangle
\end{eqnarray}
where $|w_i\rangle=H_{TQ}|\Phi_i^{A}\rangle$ is the source term and $|\omega^{(+)}_j\rangle=G_{T}^{(+)}(E)|w_j\rangle $  is the continuation of the state $|\Phi_j^{A}\rangle$ 
 in the ${\cal T}$ subspace.

\begin{figure}[h]
\begin{center}
\mbox{\epsfig{file=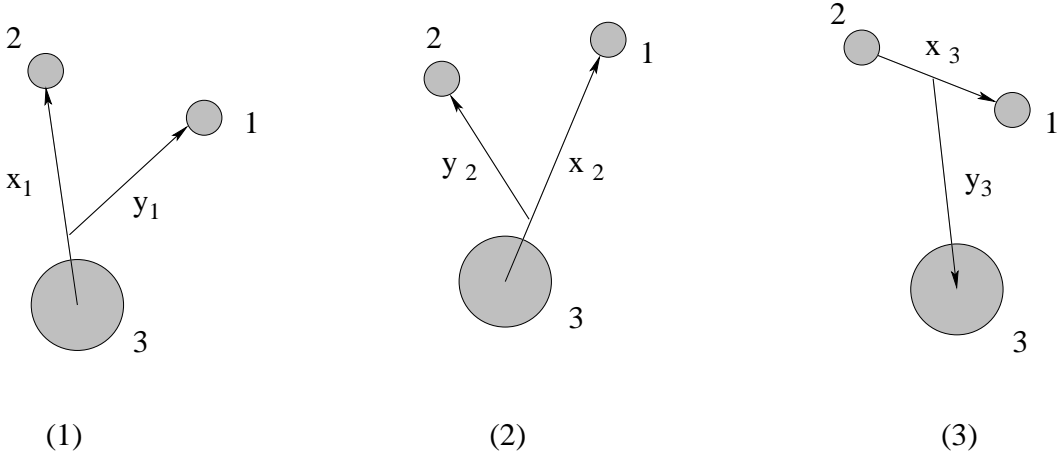,height=6cm,width=14cm}}
\end{center}
\vspace{-1cm}
\caption{Different sets of Jacobi coordinates for a three-body system.}
\label{jaz_fig}
\end{figure}

 Calculation of the energy correction to the SM eigenvalue in (\ref{extra_01}) requires a formulation for the three-body asymptotic in Jacobi coordinates (see Fig. \ref{jaz_fig}):
\begin{eqnarray*}
&&{\bf x}_k=\sqrt{\mu_{ij}}{~}{\bf r}_{ij}  {~~~~~~~~~~~~} \rm{with} \it {~~~} \mu_{ij}=\frac{A_{i}A_{j}}{A_{i}+A_{j}} \\ \\
&&{\bf y}_{k}=\sqrt{\mu_{(ij)k}}{~}{\bf r}_{(ij)k}  {~~~~~~~} \rm{with} \it  {~~~} \mu_{(ij)k}=
\frac{(A_{i}+A_{j})A_{k}}{A_{i}+A_{j}+A_{k}}
\end{eqnarray*}
where $A_{i}=m_{i}/m$, $m_{i}$ is the mass of a particle  $i$ and $m$ is the nucleon mass.
The source term  $|w_i\rangle$ and the function  $|\omega^{(+)}_j\rangle$ are calculated using the expansion in basis of hyperspherical harmonics ${\cal Y}_{KL}^{l_{x},l_{y}}$ \cite{Del}. 
The convenient variables in this basis are:
\begin{itemize}
\item the hyperradius $\rho$ defined by :~~~~$\rho=\sqrt{x_k^{2}+y_k^{2}}$ 
\item the hyperangle $\alpha_k$ defined by :~~~~$\arctan \left(x_k/y_k\right)$ 
\item the ensemble of angles $(\theta_{x_k},\phi_{x_k},\theta_{y_k},\phi_{y_k})$ 
associated with the direction ${\bf{x}}_k$ and ${\bf{y}}_k$.
\end{itemize}

The hyperradius  $\rho$ is the same in all three ensembles of the Jacobi coordinates shown in Fig. \ref{jaz_fig} but the hyperangles  $\alpha_k$ differ in systems  {\bf Y} (the coordinate systems (1) and (2) in Fig. \ref{jaz_fig}) and {\bf T} (the coordinate system  (3) in Fig. \ref{jaz_fig}). 
The hyperspherical harmonics  ${\cal Y}_{KL}^{l_{x_k},l_{y_k}}(\Omega^{k}_{5})$ are defined as:
\begin{eqnarray}
{\cal Y}_{KL}^{l_{x_k},l_{y_k}}(\Omega^{k}_{5})=\Psi_{K}^{l_{x_k},l_{y_k}}(\alpha)\left[
Y_{l_{x_k}}(\hat{x_k})\otimes Y_{l_{y_k}}(\hat{y_k})  \right]^{L} ~ \ ,
 \label{def_hypq}
\end{eqnarray}
where $\Omega^{k}_{5}\equiv (\alpha_k,\hat{x_k},\hat{y_k})$, and functions
$\Psi_{K}^{l_{x_k},l_{y_k}}(\alpha)$  are defined using integer order Jacobi polynomials: \begin{eqnarray}
\label{extra_02}
\Psi_{K}^{l_{x_k},l_{y_k}}(\alpha)=N_{K}^{l_{x_k},l_{y_k}}(\sin \alpha)^{l_{x_k}}(\cos \alpha)^{l_{y_k}}
P_{n}^{l_{x_k}+1/2,l_{y_k}+1/2}(\cos(2\alpha))
\end{eqnarray}
$N_{K}^{l_{x_{k}},l_{y_{k}}}$ is the normalization constant, and  $n=(K-l_{x_k}-l_{y_k})/2$.  
$Y_{l_{x_k}}(\hat{x_k})$ and $Y_{l_{y_k}}(\hat{y_k})$ in (\ref{def_hypq})
are the spherical harmonics associated with $\hat{x}$ and $\hat{y}$, and $K$ is the hypermoment.
The source term  $|w_i\rangle$ and the function  $|\omega^{(+)}_j\rangle$ are calculated in
the coordinate system {\bf T}. To simplify notation, vectors  (${\bf x}_3$, ${\bf y}_3$) in the definition of system {\bf T} will be denoted by (${\bf x}$,${\bf y}$). 

A decay channel $c$ in the ${\cal T}$ subspace is defined as:
\begin{eqnarray}
c=(t,K,(l_x,l_y)L,S;J_{2p};J) \label{r_canal}
\end{eqnarray}
where $t$ is a state of a daughter nucleus  A-2, and K is the hypermoment associated with the hyperspherical harmonic function ${\cal Y}_{KL}^{l_{x},l_{y}}$. The angular momenta
 $l_x$ and $l_y$, which are associated with the directions  {\bf{x}} and {\bf{y}}, are coupled to
 $L$. $S$ is the total spin of two protons. $L$ and $S$ are coupled to  $J_{2p}$, and the total angular momentum $I_t$ of a daughter nucleus in a state $t$ is coupled with  $J_{2p}$ to $J$. 
The antisymmetrization of the wave functions for two protons in the continuum is taken into account by choosing an even value for $l_x+S$. Details of the calculation of the source term are given in the appendix \ref{annexe_source_dir}.

The function $\omega^{(+)}_j$, which is a continuation of the SM state $|\Phi_j^{A}\rangle$  in ${\cal T}$, is a solution of the inhomogeneous equation:
\begin{eqnarray}
(E-H_{TT})|\omega^{(+)}_j\rangle=|w_j\rangle 
\label{o_q}
\end{eqnarray}
with:
\begin{eqnarray}
\label{extra_03}
H_{TT}={\hat T} \left[ {\tilde H}^{(A-2)} +\hat{{\cal K}} + v_{0}(A) + v_{0}(A-1) 
+ \sum_{i\leq j}^{j\geq A-1}V^{(res)}(i,j)+V^{(C)}(A-1,A)\right]{\hat T} 
\end{eqnarray}
where ${\tilde H}^{(A-2)}$ is the intrinsic Hamiltonian of the A-2 daughter nucleus, $\hat{{\cal K}}$ is the intrinsic kinetic energy of a three-body system:  the daughter nucleus  and the two protons in the continuum.
 $v_{0}(A-1)$ and $v_{0}(A)$ are the one-body potentials acting on the two protons, denoted by an index A-1 and $A$, respectively.   $V^{(res)}(i,j)$  for $j\geq A-1$  corresponds to the residual interaction between nucleons of the daughter nucleus and the protons in the continuum. $V^{(res)}(A-1,A)+V^{(C)}(A-1,A)$ is the sum of the residual and Coulomb interactions between the two emitted protons. In the basis of hyperspherical harmonics, eq. (\ref{o_q}) takes a form of the CC equations (cf appendix \ref{Annexe_direct}):
\begin{eqnarray}
\sum_{c'} H_{cc'}(\rho)\omega^{(+)}_{j,c'}(\rho)=w_{j,c}(\rho)
\label{m_cc}
\end{eqnarray}
where $H_{cc'}(\rho)$ is the coupling potential between channels  $c$ and $c'$. 

The infinite range of the Coulomb interaction in $H_{cc'}$ does not allow to decouple CC equations  at infinity. Consequently, an asymptotic behavior of $\omega^{(+)}_{j,c'}(\rho)$ cannot be defined without an approximation. One way to proceed is to neglect the off-diagonal potentials  $H_{cc'}(\rho)$ for $\rho > \rho_0$ and to define an effective Sommerfeld parameter from the diagonal potentials $H_{cc}(\rho)$ \cite{Grig}. In this approximation, eqs. (\ref{m_cc}) for $\rho > \rho_0$ become a system of decoupled two-body Coulomb equations.

If the residual interaction  $V^{(res)}$ is a contact force:
$V^{(res)}({\bf r}_1-{\bf r}_2)={\bar V}_0 \delta({\bf r}_1-{\bf r}_2)$
, then the contribution to $H_{cc}(\rho)$ due to the two-body interaction between two protons in the continuum:
\begin{eqnarray}
&&\langle K,l_x,l_y,L,S,J_{2p}, \rho|V^{(res)}|K,l_x,l_y,L,S,J_{2p}, \rho \rangle 
 \nonumber \\   
&&\propto  \frac{1}{\rho^{3}} {\bar V}_0 \int d\alpha \cos^{2}(\alpha) \sin^{2}(\alpha) 
\Psi_{K}^{l_{x},l_{y}}(\alpha)\Psi_{K}^{l_{x},l_{y}}(\alpha) 
\frac{\delta(\cos(\alpha)/\sqrt{\mu_x})}{(\cos(\alpha)/\sqrt{\mu_x})^{2}}  
\nonumber \\
&& \propto -\frac{1}{\rho^{3}}
\end{eqnarray}
has an ultraviolet divergence for $\rho\simeq 0$.  In this case, solutions of CC equations oscillate with a frequency which tends to infinity as $\rho \rightarrow 0$. In general, attractive potentials 
$V(\rho)\sim \rho^{-\tau}$ with $\tau >2$ have an infinite number of bound states \cite{Case,Efi,Niel} and have to be regularized. Unfortunately, the standard cutoff procedure for the  singular potential at  
$\rho < \rho_0$ cannot be applied since the solution of CC equations would then depend in a random way on the value of the cutoff radius  $\rho_0$. Hence, the {\em finite-range} $V^{(res)}$  is obligatory in solving the most general problem of the direct 2p emission. In this case, the CC coupling potentials
$H_{cc'}(\rho)$ contain a non-local term in  $\rho$ 
(cf appendix \ref{Annexe_direct}) and, consequently, CC equations  (\ref{m_cc}) become the integro-differential equations. Numerical solution of those equations for the problem of the 2p decay will be addressed in a future paper.


\section{Discussion of the results}
\label{disc_res}

Nuclear decays with three fragments in the final state are very exotic processes. The 2p radioactivity is an example of such a process which can occur for even-Z nuclei beyond the proton drip line: if the
sequential decay is energetically forbidden, a simultaneous 2p decay becomes the only possible decay branch. The diproton decay may also be observed in a situation where a 
1p decay is allowed, as found in the SMEC study \cite{rop1} of the decay of $1_2^-$ state at 6.15 MeV in $^{18}$Ne \cite{campo}. However, in this case the diproton decay is strongly influenced by an interplay between the external mixing (through the ${\cal Q}-{\cal P}$ coupling) and the internal mixing (inside of the ${\cal Q}$ subspace) in SM wave functions \cite{rop1}, which invalidates an idealized picture of 
an independent decay mode associated with the pairing field. For that reason, it is important to search experimentally for the g.s. 2p decay in those nuclei beyond the 2p drip-line which are stable with respect to the 1p emission. One should stress that the external mixing of SM wave functions is effective also in nuclei with closed 1p decay channels, as has been pointed in the studies of the binding energy systematics in $sd$-shell nuclei \cite{masses}. Hence, the many-body states close to the 1p emission threshold can be modified strongly by the residual coupling between ${\cal Q}$ and ${\cal P}$ subspaces.
Below, we shall study this aspect of a diproton decay for $^{45}$Fe and $^{48}$Ni. 

In this chapter, we shall describe spontaneous diproton decays from the g.s. of $^{45}$Fe, $^{48}$Ni and $^{54}$Zn, which have been observed recently.
The valence space used to describe these nuclei consists of $1s0d0f1p$ shells for $^{45}$Fe and $0f1p$ shells for $^{48}$Ni and $^{54}$Zn.
As a residual interaction between different subspaces we use the Wigner-Bartlett contact force:
\begin{eqnarray}
\label{force_con}
V^{(res)}={\bar V}_{0}[\alpha + \beta P^{\sigma}]\delta(\bf{r_{2}}- \bf{r_{1}})
\end{eqnarray}
where $\alpha+\beta=1$, ${\bar V}_0$ is the strength parameter and $P^{\sigma}$ is the spin-exchange operator. In the following, we shall take $\alpha=0.73$ \cite{Benb} and 
${\bar V}_0=-900$ MeV$\cdot$fm$^{3}$.


\subsection{The decay of $^{45}$Fe}
\label{Fe45}

The 2p radioactivity of $^{45}$Fe has been reported recently \cite{pfutzner,giovinazzo,dossat} on the basis of accumulated experimental evidence which can be consistently explained assuming an important 2p decay branch. The reported decay energy \cite{dossat} is $Q_{2p}=1.154(16)$ MeV. The half-life fit of the decay-time spectrum yields \cite{dossat} $T_{1/2}=1.6^{+0.5}_{-0.3}$ ms, a somewhat lower value than reported previously \cite{pfutzner,giovinazzo} ($T_{1/2}=4.7^{+3.4}_{-1.4}$ ms). The 2p decay competes with the $\beta$-decay and the estimated 2p branching ratio is 0.57(10) \cite{dossat}. The sequential decay through the intermediate g.s. of $^{44}$Mn or the correlated continuum above this state was estimated as less probable \cite{giovinazzo} based on the model predictions for the $Q_{1p}$ which range from -24 keV to +10 keV  \cite{Brown91,Ormand96,Cole96}.

\subsubsection{Calculation of the source term and the function $\omega^{(+)}$ for the diproton decay}

The source term for the direct 2p decay is expanded in the harmonic oscillator basis (cf appendix \ref{annexe_source_dir}). The  g.s. decay of $^{45}\rm{Fe}$ occurs in the channel: 
$c=(t^{(int)},0s,L=0,S=0,J^{\pi}=3/2^{+})$,
where $t^{int}$ is the (intrinsic) $J^{\pi}=3/2^+$ g.s. of $^{43}\rm{Cr}$, and $L$ is the relative angular momentum between the (2p)  cluster and $^{43}\rm{Cr}$. The internal 
state of the cluster is $0s$ in the harmonic oscillator basis and spins of two protons are coupled to the total spin $S=0$. The radial wave functions which appear in the source term calculation are generated using the Woods-Saxon potential with the Blomqvist-Wahlborn parametrization \cite{Blomquist},  the diffuseness parameter $a=0.67$ fm, and the radius: $R_{0}=1.27 (A-1)^{1/3}$.
These states are the one-body resonances which are regularized using a cut-off procedure \cite{opr}. The cut-off radius is fixed at $R_{cut}=8~\rm{fm}$ for $1p_{1/2}~,~1p_{3/2}$ s.p. resonances and $R_{cut}=7~\rm{fm}$ for  $0f_{5/2}~,~0f_{7/2}$ s.p. resonances.
Those two values of $R_{cut}$ correspond approximately to the top of the potential barrier for those states. The diffuseness of the cut-off function is $a_{cut}=1$ fm in both cases \cite{rop1}.

\vskip 1.5cm
\begin{figure}[h]
\begin{center}
\mbox{\epsfig{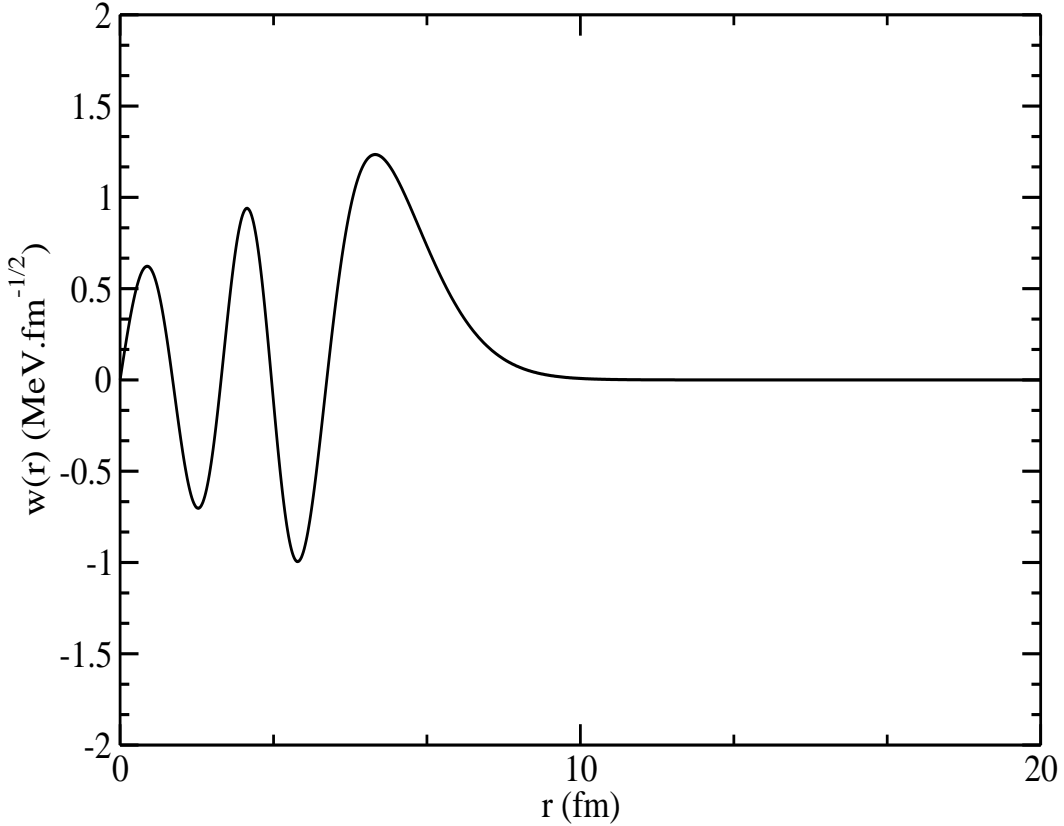}}
\end{center}
\caption{Real part of the diproton source function for the g.s. decay of $^{45}\rm{Fe}$.}
\label{source_Fe45_fig}
\end{figure}
In the calculation of function $\omega^{(+)}$, we assume that an interaction between the (2p)  cluster and the daughter nucleus $^{43}\rm{Cr}$ is described by the average potential $U_{0}$ which is a sum of the central Wood-Saxon potential and the Coulomb potential. The parameters of $U_{0}$ are deduced from  the deuteron scattering data \cite{daeh}. The depth of the Woods-Saxon potential in $U_0$ is adjusted to obtain the $s$-wave resonance  for a particle of mass $2m_p$ and charge $Z=2$ at the available energy for the 2p decay. 

The real part of the diproton source function corresponding to the diproton decay of $^{45}$Fe (see Table \ref{tablo_fe})
is shown in Fig \ref{source_Fe45_fig}. The imaginary part of the source function, generated by an external mixing due to the coupling to 1p decay channels, is very small in comparison with the real part. This is due to weak external mixing of different $J^{\pi}=3/2^{+}$ SM states in $^{45}\rm{Fe}$. The external mixing is included by considering coupling to the channels $(J^{\pi},l_j)^{3/2^+}$ with $J^{\pi}=2_1^-$ (ground state) and $J^{\pi}=3_1^-$ (first excited state) states of $^{44}$Mn (A-1 system) and $l_j=p_{1/2},p_{3/2},f_{5/2},f_{7/2}$ waves for the proton in the continuum. 

 
\subsubsection{Diproton decay of $^{45}$Fe\label{sect_dipr_Fe}}

In the calculations for $^{45}$Fe, we take IOKIN interaction ($1s0d0f1p$ shells) \cite{IOKIN}  in $H_{QQ}$. This force contains the USD Hamiltonian for the $sd$ shell \cite{wildenthal} and the KB$^{'}$ interaction for the $pf$ shell \cite{strasbourg}. The cross-shell interaction is the $G$-matrix \cite{lks}. Configurations with the excitations from $1s0d$ to $0f1p$ shells are excluded. The results for diproton decay half-lifes are summarized in Table \ref{tablo_fe}.
\vskip 0.5cm
\begin{table} [here]
\begin{center}
\begin{tabular}{|c|c|c|c|c|}
\hline 
$Q_{2p}$ (MeV)~~& $~~T_{1/2}$ (ms)~~&~~$T_{1/2}^{Q-T}$ (ms)~~&$~~T_{1/2}$ (ms)~~ $Q_{1p}=-0.1$ $\rm{MeV}$ &  $~~T_{1/2}$ (ms)~~$Q_{1p}=0.05$ $\rm{MeV}$  \\ \hline 
1.138 &  21.46 & 21.42 & 19.80  & 19.77  \\
1.154 &  13.33 & 13.31 & 12.30  & 12.28\\
1.170 &  8.37 & 8.35 & 7.72   & 7.71\\ \hline 
\end{tabular}
\end{center}
\caption[T4]{Half-lifes for the diproton decay of the g.s. of $^{45}$Fe for different values of decay energies. IOKIN effective SM interaction is used in ${\cal Q}$ subspace. The strength of the residual interaction (\ref{force_con}) is ${\bar V}_0=-900$ MeV$\cdot$fm$^{3}$. The second column corresponds to the approximation of a direct 2p decay without an external mixing. In the third column, external mixing of SM wave functions generated by the ${\cal Q}$-${\cal T}$ coupling is taken into account in the calculation of the diproton decay. In next two columns, we show results including external mixing generated by the ${\cal Q}$-${\cal P}$ coupling for $Q_{1p}=-0.1$ $\rm{MeV}$ (virtual $Q$-$P$ excitations) and $Q_{1p}=0.05$ $\rm{MeV}$ (open 1p decay channel).}
\label{tablo_fe}	
\end{table}

The calculated half-lifes are somewhat longer than found by Dossat et al. \cite{dossat}. The external ${\cal Q}$-${\cal P}$ mixing of SM wave functions, reduces the diproton half-lifes  by $\sim10\%$ for chosen values of $Q_{1p}$. Interestingly, this reduction is almost the same if the one-proton threshold in $^{45}$Fe is at 100 keV or if the 1p decay channel is opened with an available decay energy of 50 keV. 
One should stress, that this effect depends strongly on the value of $Q_{1p}$ (see Ref. \cite{rop1} for the discussion of the external mixing in the decay of $1_2^-$ state at 6.15 MeV in $^{18}$Ne for which $Q_{1p}=2.228$ MeV) and, hence, its experimental determination is mandatory for a full understanding of the 2p decay from the g.s. of $^{45}$Fe. External mixing of SM wave functions generated by the ${\cal Q}$-${\cal T}$ coupling gives a negligible correction to the diproton decay half-lifes (cf Table \ref{tablo_fe}) and can be neglected.


\subsubsection{Sequential 2p emission from the ground state of $^{45}$Fe}
Half-lifes for the diproton decay of $^{45}$Fe calculated in SMEC using IOKIN interaction are longer than reported by Dossat et al. \cite{dossat}. One can inquire whether the sequential 2p emission could play a significant role in the g.s. decay of $^{45}$Fe, thereby reducing the discrepancy with the data. 
 
We shall consider the sequential 2p emission through the continuum states correlated by the g.s. ($J^{\pi}=2^-$) and the first excited state ($J^{\pi}=3^-$) of $^{44}\rm{Mn}$. Excitation energy of the $J^{\pi}=3_1^-$ state with respect to the g.s. depends on the value of $Q_{1p}$. For
$Q_{1p}=1.154$ MeV  in $^{44}$Mn, it is $E^*=1.455$ MeV. In this case, both $J^{\pi}=2_1^-$ and $J^{\pi}=3_1^-$ states are the resonances decaying by 1p emission. 

The theoretical scheme for the sequential 2p emission has been described before. In a description of the first proton emission from the g.s. ($J^{\pi}=3/2_1^+$) of $^{45}$Fe 
we use the $'^{45}{\rm Fe}'$-reference potential. This potential 
is generated using the Woods-Saxon potential with Blomqvist-Walhborn parametrization \cite{Blomquist}, the diffuseness $a=0.67$ fm and the radius $R_0=1.27(A-1)^{1/3}$ fm. 
The depth $\hat{V_0}$ of the central part and the strength $\hat{V}_{ls}$ of the spin-orbit part are:
$\hat{V_0}=-46.50$ MeV and $\hat{V}_{ls}=-8.24$ MeV, respectively.
This reference Woods-Saxon potential is used to calculate the wave functions of s.p. states which are not affected by the continuum coupling.  For other s.p. states, we have to take into account the correction given by the diagonal potential $V_{cc}$ generated by the residual interaction in the CC equations \cite{opr}. 

To describe emission of the first proton 
to the $2^-$ and $3^-$ continuum of $^{44}$Mn, we consider $p_{1/2}$, $p_{3/2}$ and $f_{5/2}$, $f_{7/2}$ waves for the emitted proton. Let us consider the 1p decay channels: $(2^-,l_j)^{3/2^+}$
with two waves $p_{3/2}$ and $f_{7/2}$, {\it i.e.}
$c_{0}=(2^{-},p_{3/2})^{3/2^{+}}$,  $c_{1}=(2^{-},f_{7/2})^{3/2^{+}}$. 
The diagonal potentials $H_{cc}$ ($c=c_0,c_1$) appearing in the CC equations are renormalized  
 using a self-consistent procedure described in \cite{opr}. We adjust the depth ${\hat V}_{cc}$ of the central part of the Woods-Saxon potential in $H_{cc}$ for $c=c_0,c_1$ in order to reproduce energies of $1p_{3/2}$ and $0f_{7/2}$ s.p. states given by the $'^{45}{\rm Fe}'$-reference potential. 
 For the remaining channels $c_{2}=(2^{-},p_{1/2})^{3/2^{+}}$ and  $c_{3}=(2^{-},f_{5/2})^{3/2^{+}}$ involving $p_{1/2}$ and $f_{5/2}$ waves, we take the same depths of the Woods -Saxon potential in $H_{cc}$ as obtained before for $p_{3/2}$ and $f_{7/2}$ waves, respectively, {\it i.e.} ${\hat V}_{00}={\hat V}_{22}$ and ${\hat V}_{11}={\hat V}_{33}$. 
\begin{table} [here] 
\begin{center}
\begin{tabular}{|c|c|c|c|}
\hline 
$Q_{2p}$ (MeV)~~& $~~T_{1/2}$ (ms)~ $Q_{1p}=0.05$ $ \rm{MeV}$ ~&~~$T_{1/2}$ (ms)~ $Q_{1p}=0.0$ $ \rm{MeV}$  ~&~~$T_{1/2}$ (ms)~ $Q_{1p}=-0.1$ $ \rm{MeV}$ \\ \hline 
1.138 &  171.2 & 199.6 & 258.6  \\
1.154 &  109.9 & 127.8 & 164.9  \\
1.170 &  71.4  & 82.9  & 106.6  \\ \hline 
\end{tabular}
\end{center}
\caption[T4]{Half-lifes for the  sequential decay of the g.s. $J^{\pi}=3/2_1^+$ of $^{45}\rm{Fe}$ for different values of the available energy for the 2p decay and different $Q_{1p}$-values. External mixing of $3/2^+$ SM wave functions generated by the ${\cal Q}$-${\cal P}$ coupling is included. The strength of the residual interaction (\ref{force_con}) is ${\bar V}_0=-900$ MeV$\cdot$fm$^{3}$.} 
\label{tablo_seq_mixing}	
\end{table}
\begin{table} [here]
\begin{center}
\begin{tabular}{|c|c|c|c|}
\hline 
$Q_{2p}$ (MeV)~~& $~~T_{1/2}$ (ms)~ $Q_{1p}=0.05$ $\rm{MeV}$ ~&~~$T_{1/2}$ (ms)~ $Q_{1p}=0.0$ $ \rm{MeV}$  ~&~~$T_{1/2}$ (ms)~ $Q_{1p}=-0.1$ $ \rm{MeV}$ \\ \hline 
1.138 &  235.0  & 277.7 & 368.9 \\
1.154 &  151.1  & 178.1 & 235.6 \\
1.170 &  98.4   & 115.7 & 152.5 \\ \hline 
\end{tabular}
\end{center}
\caption[T4]{Half-lifes for the sequential decay of the g.s. $J^{\pi}=3/2_1^+$ of $^{45}\rm{Fe}$.
External mixing generated by the ${\cal Q}$-${\cal P}$ coupling is neglected. For other details, see the caption of Table \ref{tablo_seq_mixing}.} 
\label{tablo_seq_nomixing}	
\end{table}
For the channels $(3^-,l_j)^{3/2^+}$ 
, we take for each wave $l_j$ the same value of the depth of the Woods-Saxon potential in $H_{cc}$ as obtained previously in the corresponding channel $(2^-,l_j)^{3/2^+}$. 

The $'^{44}{\rm Mn}'$-reference potential used in the description of the emission of the second proton from $^{44}\rm{Mn}$ has the same diffuseness as the $'^{45}{\rm Fe}'$-reference potential and the radius $R_{0}=1.27 (A-2)^{1/3}$. The strength of the central part and the spin orbit part in 
$^{44}{\rm Mn}'$-reference potential are: $\hat{V_0}=-47.16$ MeV and 
$\hat{V}_{ls}=-8.37$ MeV, respectively. The diagonal potentials $H_{cc}$ are renormalized using the same procedure as described  for the emission of the first proton. The wave functions which are not renormalized by the continuum coupling are given by the $'^{44}{\rm Mn}'$-reference potential. 
The one-body operator ${\hat p}h^{(seq)}{\hat p}$, which appears in the expression for the width (cf eq. (\ref{seq_me3})), is identified with the $H_{cc}$ obtained for $'^{45}{\rm Fe}'$-reference potential.

\begin{figure}[h]
\begin{center}
\mbox{\epsfig{file=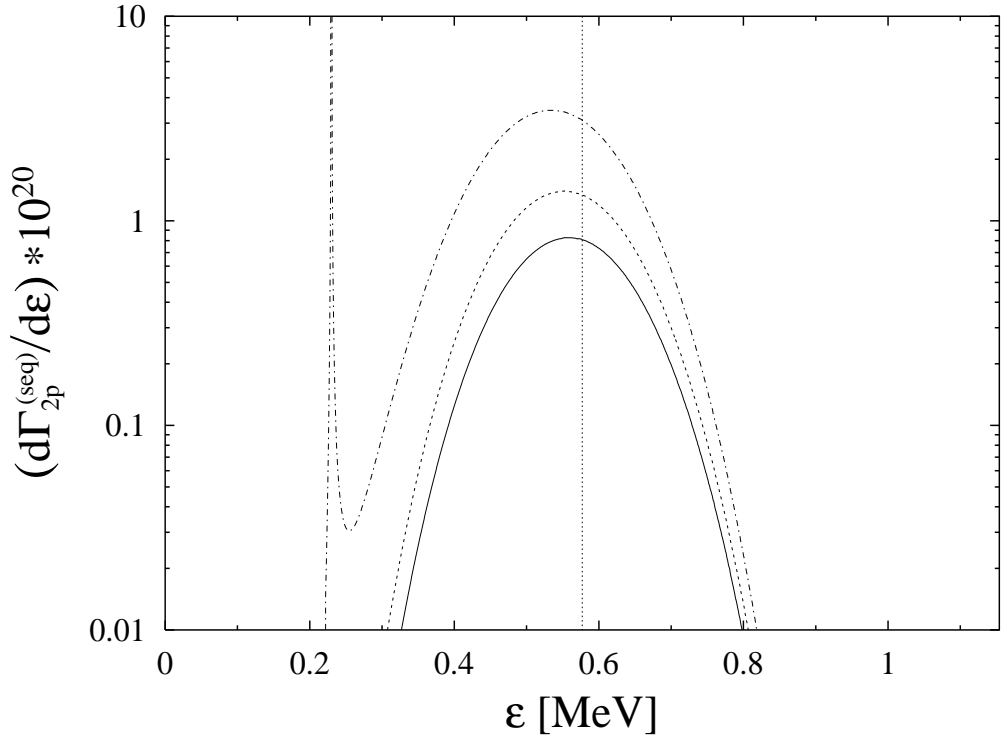,height=14cm,width=10cm,angle=-90}}
\end{center}
\caption{Energy distribution of the first emitted proton in the sequential 2p decay of the g.s. 
$J^{\pi}=3/2_1^+$ in $^{45}\rm{Fe}$.  The calculations have been performed for different energies of the g.s. $J^{\pi}=2^-$ of $^{44}{\rm Mn}$: $Q_{1p} = -0.1$ MeV (solid line), $0.05$ MeV (dashed line) and $0.23$ MeV (dashed-dotted line). 
The total energy for 2p decay is $Q_{2p}=1.154$ MeV and the vertical line denotes $Q_{1p}=Q_{2p}/2$. The calculations include effect of an external ${\cal Q}$-${\cal P}$ mixing.}
\label{seq_dep_eps_fig}
\end{figure}

Half-lifes for the sequential decay, as shown in Tables \ref{tablo_seq_mixing} and \ref{tablo_seq_nomixing}, are about one order of magnitude longer than half-lifes for the diproton decay
and, therefore,  cannot explain the discrepancy between the reported value \cite{dossat} and the SMEC results obtained with IOKIN interaction. The calculations are performed for $Q_{1p}=-0.1$ MeV (g.s. of $^{45}$Fe is stable with respect to the 1p emission), $Q_{1p}=0$ MeV (g.s. of $^{45}$Fe is at the 1p emission threshold) and $Q_{1p}=0.05$ MeV  (1p emission channel is opened in the g.s. of $^{45}$Fe).  
In all cases, the 2p emission channel is opened. 
For $Q_{1p}=-0.1$ MeV and $Q_{1p}=0$ MeV, the sequential 2p decay goes exclusively through the  
'ghost' of the g.s. in $^{44}$Mn (the continuum states correlated by the proximity of the 
$J^{\pi}=2_1^-$ g.s. of $^{44}$Mn) (cf Fig \ref{seq_dep_eps_fig}). 
The energy centroid of the 'ghost', {\em i.e.} the most probable energy of the first proton in the sequential 2p decays shown in Tables \ref{tablo_seq_mixing} and \ref{tablo_seq_nomixing}, is $\varepsilon_{1p}^{(ghost)}\leq 0.55$ MeV  (cf Fig \ref{seq_dep_eps_fig}). 
This centroid moves slightly with $Q_{1p}$ following the dependence of the correlated continuum on the position of $2_1^-$ state in $^{44}$Mn. The full width at half maximum  of the ghost is $\Gamma^{(ghost)}=0.197$ MeV, 0.199 MeV and 0.208 MeV for $Q_{1p}=-0.1$ MeV, 0.05 MeV and 0.23 MeV, respectively. For $Q_{1p}=0.05$ MeV, {\em i.e.} when the g.s. of $^{44}$Mn is inside of the available energy interval $[0,Q_{2p}]$ for the sequential 2p decay, the intermediate resonance ($J^{\pi}=2_1^-$ g.s. in $^{44}$Mn) contribution is totally screened by the Coulomb barrier and the 2p decay goes through 'ghost' far away from the resonance region $Q_{1p}\pm\Gamma_{2_1^-}/2$. For $Q_{1p}=0.23$ MeV (the dashed-dotted curve in Fig. \ref{seq_dep_eps_fig}), a fraction of the sequential 2p decay goes through the g.s. of $^{44}$Mn decreasing significantly half-life with respect to the extrapolation of the branch $T_{1/2}(Q_{1p})$ at $Q_{1p}<0.2$ MeV.

The energy distribution of the first proton in the sequential decay of the g.s. in $^{45}$Fe is shown in 
Fig.  \ref{seq_dep_eps_fig} for several values of $Q_{1p}$. One can clearly see the 'ghost' of the g.s. in $^{44}$Mn at $\varepsilon \sim 0.55$ MeV, which plays an important role in  the sequential 2p decay even if the energy of the g.s. in $^{44}$Mn is inside of the interval $[0,Q_{2p}]$ (cf the dashed and dashed-dotted curves in Fig. \ref{seq_dep_eps_fig} for  $Q_{1p}=0.05$ MeV and 0.23 MeV). In other words, the sequential 2p decay for $0<Q_{1p}<0.2$ MeV is predominantly related to the strength of the 'ghost' at $\varepsilon \simeq \varepsilon_{1p}^{(ghost)}\pm \Gamma^{(ghost)}/2$ and {\em not} to the strength of the 1p resonance at $Q_{1p}$. Therefore, the sequential 2p decay width in the interval $0<Q_{1p}<0.2$ MeV in the considered example of $^{45}$Fe g.s. decay does not reduce to the product of the width for the first step ($\Gamma_{1p}$) and the branching ratio for the second step, as one would obtain in the semiclassical analysis \cite{fedorov}. In the interval 
0.2 MeV$<Q_{1p}<0.275$ MeV, the sequential 2p decay the transition through the $J^{\pi}=2_1^-$ g.s. in $^{44}$Mn is progressively enhanced and becomes dominant for $Q>0.3$ MeV.
\begin{figure}[h]
\begin{center}
\mbox{\epsfig{file=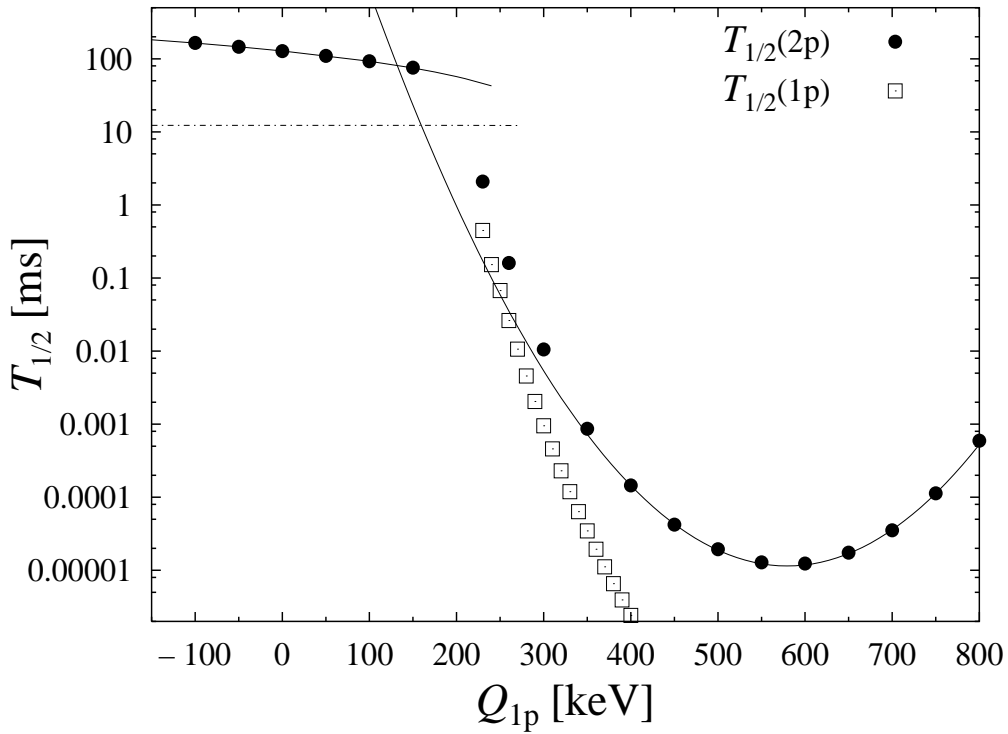,height=14cm,width=10cm,angle=-90}}
\end{center}
\caption{The half-life for the sequential 2p decay from the g.s. $J^{\pi}=3/2^+$ in $^{45}\rm{Fe}$ is shown as a function of $Q_{1p}$ (circle) along with the half-life for 1p decay (squares). The total energy for 2p decay is $Q_{2p}=1.154$ MeV. The calculations include effect of an external ${\cal Q}$-${\cal P}$ mixing. The dashed-dotted line shows the half-life for the diproton decay.}
\label{seq_dep_Q1_fig}
\end{figure}

The ratio between sequential and diproton half-lifes decreases going from negative values of $Q_{1p}$ to positive ones. External mixing of SM wave functions reduces the half-lifes for the sequential decay by about 30\% (cf Tables  \ref{tablo_seq_mixing} and \ref{tablo_seq_nomixing}). Couplings to the decay channels $(3^-,l_j)^{3/2^+}$ associated with the excited state of $^{44}$Mn are relatively important and reduce the sequential 2p emission half-life by $\sim30$\% for $Q_{1p}=0.05$ MeV and $\sim38$\% for $Q_{1p}=-0.1$ MeV. The above conclusions depend however strongly on the value of $Q_{1p}$ in $^{45}$Fe (cf Fig. \ref{seq_dep_Q1_fig}). 

According to our model, the sequential 2p emission and the diproton emission yield  comparable half-lifes already for $Q_{1p}\simeq 0.2$ MeV. 
For small positive values of $Q_{1p}$ ($0<Q_{1p}\leq 0.2$ MeV) as well as in the case of closed 1p decay channel(s) ($Q_{1p} < 0$), the sequential 2p decay half-life changes linearly with $Q_{1p}$ 
($T_{1/2}\sim Q_{1p}$). In this limit,  $\Gamma_{2p}^{(seq)}\gg \Gamma_{1p}$ and the 2p decay goes via the correlated continuum  (the 'ghost' of the g.s.) in the  available decay energy window $[0,Q_{2p}]$. For $Q_{1p}> 0.3$ MeV, one enters in the regime:  $\Gamma_{1p}\gg \Gamma_{2p}^{(seq)}$, where the role of $2_1^-$ resonance in the intermediate nucleus $^{44}$Mn is dominant in the $\Gamma_{2p}^{(seq)}$. In this regime, the dependence of $T_{1/2}$ on $Q_{1p}$ is Gaussian: 
$T_{1/2}\sim\exp -(Q_{1p}-Q_{1p}^{(0)})^2$.
The half-life of the sequential emission has a minimum close to $Q_{1p}^{(0)}\sim Q_{2p}/2$ and begins to grow again for larger $Q_{1p}$-values due to a smaller available energy for the second proton. In the minimum, $\Gamma_{1p}$ is about 3 orders of magnitude larger than $\Gamma_{2p}^{(seq)}$.
 Even though the value $Q_{1p}\simeq 0.2$ MeV for which a sequential 2p decay becomes comparable with a diproton decay lies outside of the range of $Q_{1p}$-values predicted in various calculations  \cite{Brown91,Ormand96,Cole96}, nevertheless it is close enough to raise doubts about the precise mechanism of the 2p decay from the g.s. of $^{45}$Fe which only devoted measurement of masses in this region can remove. 

 One should stress that the sequential 2p decay appears {\em always} whenever $Q_{2p}>0$, {\em independently} of the sign of $Q_{1p}$. In other words, the sequential 2p decay competes with the diproton mode  even for closed 1p emission channels. The separation of 2p decay modes into the diproton decay and the sequential 2p emission becomes unjustified if the two modes yield comparable partial decay widths, {\em i.e.} $\Gamma_{2p}^{(dir)}\simeq\Gamma_{2p}^{(seq )}$. In this limit, a distinction between sequential in time and instanteneous (a direct 2p emission) 2p emission processes looses its meaning. Consequently, the finite-range of  residual interactions cannot be neglected and the true three-body asymptotics  of decaying channels cannot be reduced to any sequence of two-body decays. 


\subsection{The decay of $^{48}$Ni}
\label{Ni48}

Recently, a tentative evidence for the 2p radioactivity in $^{48}$Ni has been reported by Dossat et al. \cite{dossat}. In this experiment, one event of 2p decay with the decay energy 
$Q_{2p}=1.35(2)$ MeV and a partial decay half-life $T_{1/2}=8.4^{+12.8}_{-7.0}$ ms has been identified \cite{dossat}. Below, we shall present the analysis of the situation in this nucleus from the SMEC perspective. 
 
 
\subsubsection{Diproton decay of $^{48}$Ni}
The g.s. decay of $^{48}\rm{Ni}$ occurs in the channel : 
$c=(t^{(int)},0s,L=0,S=0,J^{\pi}=0^{+})$,
where $t^{(int)}$ is the $J^{\pi}=0^+$ g.s. of the daughter nucleus $^{46}\rm{Fe}$, and  $L$ is the relative angular momentum between the (2p)  cluster and the daughter nucleus. 
Calculation of the source term and the function $\omega^{(+)}$ is similar as described above for $^{45}$Fe (see also  appendix \ref{annexe_source_dir}). In particular, radial wave functions are generated analogously as for $^{45}$Fe and the regularization procedure of the one-body resonances is identical as above.
\begin{table} [here]
\begin{center}
\begin{tabular}{|c|c|c|c|}
\hline 
$Q_{2p}$ (MeV)~~& $~~T_{1/2}$ (ms)~ (IOKIN)\cite{wildenthal,strasbourg,lks}~~&~~$T_{1/2}$ (ms)~ (KB3)\cite{strasbourg}~~&~~$T_{1/2}$ (ms)~ (GXPF1)\cite{honma}~~ \\ \hline 
1.33 &  10.3  & 11.4 & 12.3 \\
1.35 &  6.2 & 6.9 & 7.4   \\
1.37 &  3.75  & 4.2 & 4.5   \\ \hline 
\end{tabular}
\end{center}
\caption[T4]{Half-lifes for the diproton decay of the g.s. $J^{\pi}=0_1^+$ of $^{48}$Ni for different values of decay energies and different SM effective interactions. External mixing of SM wave functions is neglected. The strength of the residual interaction (\ref{force_con}) is ${\bar V}_0=-900$ MeV$\cdot$fm$^{3}$.} 
\label{tablo_ni}	
\end{table}

\begin{table} [here]
\begin{center}
\begin{tabular}{|c|c|c|c|c|c|}
\hline 
$Q_{2p}$ (MeV)~~& $~~T_{1/2}$ (ms)~~&~~$T_{1/2}^{Q-T}$ (ms)~~&$~~T_{1/2}$ (ms)~~$Q_{1p}=-100$ $\rm{keV}$~~&~~$T_{1/2}$ (ms)$~~Q_{1p}=50$ $\rm{keV}$ \\ \hline 
1.33 &  10.27 & 10.16 & 10.09  & 10.08  \\
1.35 &  6.18 & 6.11 & 6.07 & 6.06   \\
1.37 &  3.76 & 3.72 & 3.69  & 3.69   \\ \hline 
\end{tabular}
\end{center}
\caption[T4]{The same as in Table \ref{tablo_fe} but for the diproton decay of the g.s. $J^{\pi}=0_1^+$ of $^{48}$Ni.}
\label{tablo_ni_mix}	
\end{table}
For the effective interaction in $H_{QQ}$ we take IOKIN interaction in $psdpf$ shells, as well as KB3 \cite{strasbourg} and GXPF1 \cite{honma} interactions in $pf$ shells. The calculated half-lives (cf Tables \ref{tablo_ni} and \ref{tablo_ni_mix}) are compatible with the experimental value \cite{dossat} for all effective SM interactions used in the SMEC calculations.   
The effect of external ${\cal Q}-{\cal P}$ mixing on the calculated half-lifes (cf Table \ref{tablo_ni_mix} for IOKIN effective interaction) is negligible for chosen values of $Q_{1p}$.  External mixing generated by the ${\cal Q}-{\cal T}$ coupling is relatively more important than in $^{45}$Fe.


\subsubsection{Sequential 2p emission from the ground state of $^{48}$Ni}
Half-lifes for the sequential 2p decay are shown in Tables \ref{tablo_seq_ni_mixing} and \ref{tablo_seq_ni_nomixing} for IOKIN interaction.
 We consider the emission through the continuum states correlated by the g.s. ($J^{\pi}=3/2^-$) and the first excited state ($J^{\pi}=7/2^-$) of $^{47}$Co. The residual continuum coupling with ${\bar V}_0=-900$ MeV$\cdot$fm$^{3}$ reverses the position of $3/2_1^-$ and $7/2_1^-$ SM states. This flip, which appears for $-900$ MeV$\cdot$fm$^{3}$ $<{\bar V}_0<-700$ MeV$\cdot$fm$^{3}$, is a sensitive test of the strength of the residual coupling to the continuum states at the beginning of the $fp$ shell.
\begin{table} [here]
\begin{center}
\begin{tabular}{|c|c|c|c|}
\hline 
$Q_{2p}$ (MeV)~~& $~~T_{1/2}$ (ms)~ $Q_{1p}=0.05$ $\rm{MeV}$ ~&~~$T_{1/2}$ (ms)~ $Q_{1p}=0.0$ $ \rm{MeV}$  ~&~~$T_{1/2}$ (ms)~ $Q_{1p}=-0.1$ $ \rm{MeV}$ \\ \hline 
1.33 &  25.7  & 30.05 & 39.7 \\
1.35 &  16.55 & 19.3 & 25.4   \\
1.37 &  10.8  & 12.5 & 16.45  \\ \hline 
\end{tabular}
\end{center}
\caption[T4]{Half-lifes for the sequential decay of the g.s. $J^{\pi}=0_1^+$ of $^{48}\rm{Ni}$ for different values of the available energy for the 2p decay and different $Q_{1p}$-values. External mixing of $J^{\pi}=0^+$ SM wave functions via the ${\cal Q}$-${\cal P}$ coupling is taken into account.
The strength of the residual interaction (\ref{force_con}) is ${\bar V}_0=-900$ MeV$\cdot$fm$^{3}$ and the SM interaction is IOKIN \cite{wildenthal,strasbourg,lks}.}
\label{tablo_seq_ni_mixing}	
\end{table}

External mixing reduces the half-life by $\sim 15\%$, {\em i.e.} is less important than found in $^{45}$Fe. However, the sequential half-life is only a factor $\sim 2.5-4$ times longer than the diproton half-life. This ratio decreases going from negative to positive values of $Q_{1p}$. The sequential and diproton half-lifes become comparable for $Q_{1p}\sim 0.15$ MeV, in which case the sequential and diproton decay modes cannot be considered as independent ones.
\begin{table} [here]
\begin{center}
\begin{tabular}{|c|c|c|c|}
\hline 
$Q_{2p}$ (MeV)~~& $~~T_{1/2}$ (ms)~ $Q_{1p}=0.05$ $\rm{MeV}$ ~&~~$T_{1/2}$ (ms)~ $Q_{1p}=0.0$ $ \rm{MeV}$  ~&~~$T_{1/2}$ (ms)~ $Q_{1p}=-0.1$ $ \rm{MeV}$ \\ \hline 
1.33 &  29.4  & 34.4 & 45.4 \\
1.35 &  18.9 & 22.1 & 29.0   \\
1.37 &  12.3  & 14.3 & 18.8  \\ \hline 
\end{tabular}
\end{center}
\caption[T4]{Half-lifes for the sequential decay of the g.s. $J^{\pi}=0_1^+$ of $^{48}\rm{Ni}$. External mixing is neglected. For other details, see the caption of Table \ref{tablo_seq_ni_mixing}.}
\label{tablo_seq_ni_nomixing}	
\end{table}

Coupling to the decay channels $(7/2^-,l_j)^{0^+}$ associated with the excited state of $^{47}$Co is between $6\%$ and $9\%$ for all studied cases of the sequential 2p decay of $^{48}$Ni - again significantly less than found in $^{45}$Fe.
 

\subsection{The decay of $^{54}$Zn}
\label{Zn54}
 
First observation of $^{54}$Zn and its decay by 2p emission has been reported recently by Blank {\em et al.} \cite{blankzn} on the basis of both an experimental evidence and consistency arguments.  The reported decay energy is $Q_{2p}=1.48(2)$ MeV. The experimental partial half-life for the 2p emission is $T_{1/2}=3.7^{+2.2}_{-1.0}$ ms.  The estimated 2p branching ratio is $0.87^{+0.1}_{-0.17}$ \cite{blankzn}. 

The g.s. decay of $^{54}\rm{Zn}$ occurs in the channel : $c=(t^{(int)},0s,L=0,S=0,J^{\pi}=0^{+})$,
where $t^{(int)}$ is the $J^{\pi}=0^+$ g.s. of $^{52}\rm{Ni}$, and  $L$ is the relative angular momentum between the (2p)  cluster and $^{52}\rm{Ni}$. The calculation of the source term and the function $\omega^{(+)}$ is similar as described above for $^{45}$Fe.
 
\begin{table} [here]
\begin{center}
\begin{tabular}{|c|c|c|c|}
\hline 
$Q_{2p}$ (MeV)~&~~$T_{1/2}$ (ms)~ (t=3)~~&~~$T_{1/2}$ (ms)~ (t=4)~~&~~$~~T_{1/2}$ (ms)~ (t=5)~~ \\ \hline 
1.46  & 32.7 & 28.4  & 27.27    \\
1.48 &  20.6 & 17.5 &  16.96  \\
1.50 & 12.7 & 10.9 &   10.65  \\ \hline 
\end{tabular}
\end{center}
\caption[T4]{Half-lifes for the diproton decay of the g.s. $J^{\pi}=0_1^+$ of $^{54}$Zn for different values of 2p decay energies and different limitations of the ${\cal Q}$ subspace calculations with the KB3 interaction \cite{strasbourg}. The strength of the residual interaction (\ref{force_con}) is ${\bar V}_0=-900$ MeV$\cdot$fm$^{3}$.} 
\label{tablo_zn_kb3}	
\end{table}
As an effective interaction in the ${\cal Q}$ subspace, we take KB3 \cite{strasbourg} and GXFP1 \cite{honma} interactions. Results shown in Tables \ref{tablo_zn_kb3} (KB3 interaction) and \ref{tablo_zn_gxfp1} (GXFP1 interaction) does not include the external mixing, which is very small in this nucleus. Calculated half-lifes for the diproton decay are longer than reported by Blank {\em et al.} \cite{blankzn}. As before, the sequential 2p decay
cannot be excluded {\em a priori}. To achieve a better understanding of the 2p decay pattern of $^{54}$Zn, one has to know however the $Q_{1p}$ values in $^{54}$Zn and $^{53}$Cu. 
\begin{table} [h]
\begin{center}
\begin{tabular}{|c|c|c|c|}
\hline 
$Q_{2p}$ (MeV)~~& ~~$T_{1/2}$ (ms)~ (t=3)~~&~~$T_{1/2}$ (ms)~ (t=4)~~&~~$~~T_{1/2}$ (ms)~ (t=5)~~ \\ \hline 
1.46 & 26.6 & 23.0 &  22.18    \\
1.48 &   16.3 & 14.5 &  13.8   \\
1.50 &  10.3 & 9.1 & 8.67  \\ \hline 
\end{tabular}
\end{center}
\caption[T4]{The same as in Table \ref{tablo_zn_kb3} but for the GXPF1 interaction \cite{honma}. }
\label{tablo_zn_gxfp1}	
\end{table}

Results in Tables \ref{tablo_zn_kb3} and \ref{tablo_zn_gxfp1} show the convergence of calculated half-lifes with the truncation order $t$ which denotes the maximum number of nucleons which are allowed to be excited from the $f_{7/2}$ orbit to $p_{3/2}$, $f_{5/2}$ and $p_{1/2}$ orbits, relative to the lowest filling approximation \cite{honma}. We can see that the convergence is nearly attained
 for $t=5$. 

\section{Conclusions}
All studies of the OQS have shown that the coupling of the system to the environment of its decay channels may change the properties of the system \cite{opr}. 
These changes cannot be neglected when the coupling matrix elements  between system and environment  are of the same order  of magnitude as the level distance or larger. In this case, the changes can be described neither by perturbation theory nor by introducing statistical assumptions for the level distribution. The non-linear effects become important which cause a redistribution of the spectroscopic properties of the system. In general, the magnitude of the coupling between system and environment depends explicitly on the location of various emission thresholds and on the structure of poles of the scattering matrix ($S$-matrix). The latter feature is absent in the standard SM. For low-$l$ orbits ($l=0,1$), the coupling to environment becomes singular at the particle emission threshold if the the corresponding $l$-pole of the $S$-matrix is at the threshold \cite{inpc}. Such couplings may induce the non-perturbative rearrangement of many-body wave functions which contain those strongly coupled orbits. For high-$l$ orbits $(l\geq 2$), the coupling is non-singular and can be mocked to certain extent by introducing the dependence of the monopole terms in the effective SM interaction on the number of particles in valence orbits. Thus, in the same interval of excitation energies, even at low level densities, one may find coexisting many-body states with largely different susceptibility to the coupling to the environment. 

The resonance phenomena are described well by two ingredients. The first ingredient is the effective Hamiltonian ${\cal H}_{QQ}$ that contains all the basic structure information involved in the CQS Hamiltonian $H_{QQ}$. The ${\cal H}_{QQ}$ contains also the coupling matrix elements between discrete and continuous states and its (complex) eigenvalues determine both the positions of the resonance states and their widths. The second ingredient is the unitarity of the $S$-matrix which causes a non-trivial energy dependence of the coupling matrix elements between resonance states and continuum. 

The SMEC, which has all those features, has been extended in this work to describe the decay of the CQS due to the coupling to the environment of decay channels with one- and two-nucleons in the scattering continuum. This new theory has been applied here for the study of the 2p decay from the g.s. of $^{45}$Fe, $^{48}$Ni and $^{54}$Zn. The results are very sensitive to the $Q$-values which for these nuclei are either not known experimentally with a sufficient precision $(Q_{2p}$-values), or are unknown as in the case of $Q_{1p}$-values. Keeping this warning in mind, one finds that calculated values of  the diproton width in these nuclei agree rather well with the experimental predictions, in particular for even-even $^{48}$Ni and $^{54}$Zn nuclei. Different effective SM interactions (cf Tables \ref{tablo_ni}, \ref{tablo_zn_kb3} and \ref{tablo_zn_gxfp1}) give a similar results to within $\sim 10-20\%$. External mixing, generated by the ${\cal Q}$-${\cal P}$ coupling, modifies the diproton decay width by $<10\%$ in $^{45}$Fe and $\sim 2\%$ in $^{48}$Ni, {\em i.e.} the diproton mode seems to be rather close to a 'pure' mode predicted by Goldansky \cite{gold60}.
This is in contrast to  conclusions from the analysis of the diproton decay of $1_2^-$ state in $^{18}$Ne in Ref. \cite{rop1}. One should mention, however, that the separate treatment of sequential and diproton modes becomes invalid when $\Gamma_{2p}^{(seq)}\simeq\Gamma_{2p}^{(dir)}$. In this case, the 2p decay has to be described using a full Hamiltonian (\ref{H_eff_T}) with the three-body asymptotic.
In the studied case of $^{45}$Fe and $^{48}$Ni, this happens for
$Q_{1p}\simeq$ 0.2 MeV. The ${\cal Q}$-${\cal T}$ coupling provides a negligible contribution to the external mixing even in the region of $Q_{1p}$ values where $\Gamma_{1p}\ll\Gamma_{2p}^{(dir)}$.

 Somewhat worse agreement between calculated and reported diproton widths
 is found in $^{45}$Fe. Neither external mixing nor the sequential 2p emission process explain this difference if the values of $Q_{1p}$ are taken close to the limits suggested in Refs. \cite{Brown91,Ormand96,Cole96}. One should stress again that this conclusion depends strongly on the assumed values of both $Q_{2p}$ and $Q_{1p}$. For $Q_{1p}>0.22$ MeV, the half-life for a sequential process becomes shorter than the half-life of a pure diproton emission.

Recently, $R$-matrix approach has been applied for a description of the diproton decay in $^{45}$Fe \cite{Barb}, $^{48}$Ni \cite{dossat} and $^{54}$Zn \cite{blankzn}. In this model, external mixing is neglected and the emission process is described by a simple $R$-matrix ansatz. Smaller importance of the external mixing in those nuclei leads to a good agreement with SMEC results. For GXFP1 effective interaction, one finds: 
$T_{1/2}^{(SMEC)}=7.4_{-2.9}^{+4.9}$ ms and $T_{1/2}^{(R-matrix)}=8.4_{-7.0}^{+12.8}$ ms in $^{48}$Ni, and $T_{1/2}^{(SMEC)}=13.8_{-5.1}^{+8.4}$ ms and $T_{1/2}^{(R-matrix)}=10_{-4}^{+7}$ ms in $^{54}$Zn.

In our studies, we have employed a contact force with the the spin-exchange for the residual interaction between ${\cal Q}$, ${\cal P}$ and ${\cal T}$ subspaces. As a consequence, the three-body final state in the direct 2p decay could not be calculated (cf sect. \ref{para_direct}) and we used the two-step scenario \cite{barker1} to describe this decay. Application of the finite-range residual interaction allows in future to describe this most general case using the theoretical formalism which has been presented in this work.

\vspace{1.5cm}
{\bf Acknowledgements}\\
We wish to thank Alex Brown for useful communication.

\appendix{}

\section{Effective Hamiltonian in ${\cal Q}$ subspace}\label{H_eff}

Solution $|\Psi\rangle$ of the Schr\"{o}dinger equation in 
${\cal Q}+{\cal P}+{\cal T}=I_d$, has components in each of the considered subspaces, {\em i.e.}
$|\Psi\rangle=|\Psi_{Q}\rangle+|\Psi_{P}\rangle+|\Psi_{T}\rangle$.
Hence, the  Schr\"{o}dinger equation can be written as:
\begin{eqnarray}
(E-H)\left[|\Psi_{Q}\rangle+|\Psi_{P}\rangle+|\Psi_{T}\rangle\right]=0 \label{schro}
\end{eqnarray}
Projecting (\ref{schro}) on ${\cal Q}$, ${\cal P}$ and ${\cal T}$, one obtains:
\begin{eqnarray}
&&(E-H_{QQ})|\Psi_{Q}\rangle=H_{QP}|\Psi_{P}\rangle+H_{QT}|\Psi_{T}\rangle \label{eq_Q} \\
&&(E-H_{PP})|\Psi_{P}\rangle=H_{PQ}|\Psi_{Q}\rangle+H_{PT}|\Psi_{T}\rangle \label{eq_P} \\
&&(E-H_{TT})|\Psi_{T}\rangle=H_{TQ}|\Psi_{Q}\rangle+H_{TP}|\Psi_{P}\rangle \label{eq_T}
\end{eqnarray}
The component $|\Psi_{P}\rangle$ can be found from (\ref{eq_P}):
\begin{eqnarray}
|\Psi_{P}\rangle=G_{P}^{+}(E)\left[H_{PQ}|\Psi_{Q}\rangle+H_{PT}|\Psi_{T}\rangle \right] \label{eq_P2}
\end{eqnarray}
where  $G_{P}^{+}(E)$ is the Green's function in ${\cal P}$ (cf  eq. (\ref{extra_00})). 
Hence, $|\Psi_{T}\rangle$ can be expressed in the form  (cf (\ref{eq_T}), (\ref{eq_P2})):
\begin{eqnarray}
|\Psi_{T}\rangle=\tilde{G}_{T}^{+}(E)\left[ H_{TQ}|\Psi_{Q}\rangle + H_{TP}G_{P}^{+}(E)H_{PQ}|\Psi_{Q}\rangle \right] \label{eq_T2}
\end{eqnarray}
where $\tilde{G}_{T}^{+}(E)$  is the Green's function in  ${\cal T}$, modified by the coupling with  
${\cal P}$ (cf eq. (\ref{H_eff_P})).
Using eqs. (\ref{eq_Q}), (\ref{eq_P2}) and (\ref{eq_T2}), one can demonstrate that
$|\Psi_{Q}\rangle$ is a  solution of the equation:
\begin{eqnarray}
(E-H_{QQ})|\Psi_{Q}\rangle&=&H_{QT}\tilde{G}_{T}^{+}(E)\left [H_{TQ}+H_{TP}G_{P}^{+}(E)H_{PQ} \right ]|\Psi_{Q}\rangle \nonumber 
 \nonumber \\
&+&H_{QP}G_{P}^{+}(E)\left[H_{PQ}+ H_{PT}\tilde{G}_{T}^{+}(E)\left [H_{TQ}+H_{TP}G_{P}^{+}(E)
H_{PQ}\right] \right]|\Psi_{Q}\rangle \nonumber
\end{eqnarray}
The above equation can be rewritten in a form:
\begin{eqnarray}
(E-{\cal H}_{QQ}(E))|\Psi_{Q}\rangle=0
\end{eqnarray}
where ${\cal H}_{QQ}(E)$ is the energy-dependent  effective Hamiltonian in ${\cal Q}$:
\begin{eqnarray}
{\cal H}_{QQ}(E)&=&H_{QQ}+H_{QP}G_{P}^{+}(E)H_{PQ} \nonumber \\ 
&+&\left[H_{QT}+H_{QP}G_{P}^{+}(E)H_{PT}\right]
\tilde{G}_{T}^{+}(E) \left[H_{TQ}+H_{TP}G_{P}^{+}(E)H_{PQ}\right] 
\end{eqnarray}

\section{Matrix elements for the sequential 2p emission}
 \label{seq_formalism}

A part of the effective Hamiltonian ${\cal H}_{QQ}^{(seq)}$ in ${\cal Q}$ which describes the sequential 2p emission is (cf sect. (\ref{sub_seq})):
\begin{eqnarray}
H_{QP}\frac{1}{E^{+}-H_{PP}-H_{PT}G_{T}^{+}(E)H_{TP}}
H_{PT}G_{T}^{+}(E)H_{TP}G_{P}^{+}(E) H_{PQ} 
\label{hami_eff}
\end{eqnarray}
We assume that the first emitted proton interacts with the remaining A-1 nucleons through a mean-field ${\hat p}h^{(seq)}{\hat p}$. This implies:
\begin{eqnarray}
&&H_{PT}={\hat P} \left [ \sum_{i\leq j=1}^{A} V^{(res)}(i,j) \right] {\hat T} \Longrightarrow {\hat Q}'\left[ \sum_{i\leq j=1}^{A-1} V^{(res)}(i,j) \right ] {\hat P}'\otimes I_{d}(A)=H_{Q'P'}\otimes I_{d}(A)\nonumber \\ \label{assum1}  \\ \nonumber
%
%
&&H_{TP}={\hat T} \left[ \sum_{i\leq j=1}^{A} V^{(res)}(i,j) \right] {\hat P} \Longrightarrow {\hat P}'\left[ \sum_{i\leq j=1}^{A-1} V^{(res)}(i,j) \right]{\hat Q}' \otimes I_{d}(A)=H_{P'Q'}\otimes I_{d}(A) \nonumber \\ \label{assum2}
\end{eqnarray}
where ${\hat Q}'$ and ${\hat P}'$ are projection operators on the subspaces
 ${\cal Q}'$ and ${\cal P}'$  of (A-1)-nucleon states. $I_{d}(A)$ is the identity operator which acts on the first emitted proton. In ${\cal Q}'$, all   A-1 nucleons are in (quasi-)bound orbits. In ${\cal P'}$, A-2 nucleons are in (quasi-)bound states and one proton occupies a continuum state. Hence, after the first proton emission, all states of the system are separated in the two subspaces, denoted 
 ${\cal P'}$$\otimes$p and ${\cal Q'}$$\otimes$p, where   p is the space of s.p. states in the continuum. The particle occupying the state in p interacts with particles in  
  ${\cal Q'}$ and ${\cal P'}$ by the one-body potential ${\hat p}h^{(seq)}{\hat p}$. Hence, we have:  
\begin{eqnarray}
H_{PP}&=&{\hat P} \left[ \sum_{i=1}^{A} h(i)+\sum_{i\leq j=1}^{A} V^{(res)}(i,j) \right] {\hat P}
 \nonumber \\ \nonumber \\
&=&{\hat Q}'\left[ \sum_{i=1}^{A-1} h(i)+\sum_{i\leq j=1}^{A-1} V^{(res)}(i,j) \right] {\hat Q}' +{\hat p}h^{(seq)}{\hat p}  = H_{Q'Q'}+{\hat p}h^{(seq)}{\hat p}  \label{assum3}
\end{eqnarray}
where ${\hat p}$ is the projection operator on the space  p. Similarly, one obtains:
\begin{eqnarray}
H_{TT}= H_{P'P'}+{\hat p}h^{(seq)}{\hat p} \label{assum4}
\end{eqnarray}
Hence, the expression  (\ref{hami_eff}) can be written as:
\begin{eqnarray}	
&&H_{QP}\frac{1}{E^{+}-{\hat p}h^{(seq)}{\hat p}-H_{Q'Q'}-H_{Q'P'}
\left[{E^{+}-{\hat p}h^{(seq)}{\hat p}-H_{P'P'}}\right]^{-1}H_{P'Q'}}\nonumber  \\ 
&\times&H_{Q'P'}\frac{1}{E^{+}-{\hat p}h^{(seq)}{\hat p}-H_{P'P'}}H_{P'Q'} 
\frac{1}{E^{+}-{\hat p}h^{(seq)}{\hat p}-H_{Q'Q'}}H_{PQ} \nonumber   \\ 
&=&H_{QP}\frac{1}{E^{+}-M}H_{Q'P'}\frac{1}{E^{+}-{\hat p}h^{(seq)}{\hat p}-H_{P'P'}}H_{P'Q'}\frac{1}{E^{+}-{\hat p}h^{(seq)}{\hat p}-H_{Q'Q'}}H_{PQ} ~~\label{g1}
\end{eqnarray}
$M$ in the above equation stands for:
\begin{eqnarray}
M=H_{Q'Q'}+{\hat p}h^{(seq)}{\hat p}+H_{Q'P'}\frac{1}{E^{+}-{\hat p}h^{(seq)}{\hat p}-H_{P'P'}}H_{P'Q'}
\end{eqnarray}

Let us now calculate the matrix element of the operator in ($\ref{g1}$) for the state
 $|{\tilde \Phi}^{A}_{i}\rangle$.
Let us define the completness relation in  ${\cal P}$:
\begin{eqnarray}	
\sum_{t,l,j}\int_{0}^{+\infty} de~|t,e,l,j;J\rangle \langle t,e,l,j;J|=I_{d} \label{C6}
\end{eqnarray}
where $|t\rangle$ is an eigenvector of  $H_{Q'Q'}$ corresponding to an eigenvalue  $E_{t}$ . The state
 $|e,l,j\rangle$ is an eigenvector of  ${\hat p}h^{(seq)}{\hat p}$ with energy $e$ ($e>0$ ), the orbital angular momentum $l$ and the total angular momentum $j$. Discretization of the energy integral in (\ref{C6}) yields:
\begin{eqnarray}	
\sum_{c} \sum_{n=1}^{n=+\infty}|c,e_{n}\rangle \langle c,e_{n}|\Delta_{e}=1 
\label{completude}  
\end{eqnarray}
where the channel is defined by $c=(t,l,j;J)$. In the above expression, $\Delta_{e}$ is the discretization step and the states $|e_{n},l,j\rangle$ are normalized as follows:
\begin{eqnarray*}
\langle e_{n},l,j|e_{n'},l,j\rangle=\frac{\delta_{n,n'}}{\Delta_{e}}
\end{eqnarray*}
Inserting four times the completness relation (\ref{completude}) in (\ref{g1}), one finds:
\begin{eqnarray}
&&\delta^{(seq)}(E)=\sum_{c_{1},c_{2},c_{3},c_{4}}\sum_{n_{1},n_{2},n_{3},n_{4}=1}^{+\infty}
\langle {\tilde \Phi}^{A}_{i}|H_{QP}|c_{1},e_{n_1}\rangle \langle c_{1},e_{n_1}|\frac{1}{E^{+}-M}
|c_{2},e_{n_2}\rangle \nonumber \\ 
&&\times \langle c_{2},e_{n_2}|H_{Q'P'}\frac{1}{E^{+}-{\hat p}h^{(seq)}{\hat p}-H_{P'P'}}H_{P'Q'}
|c_{3},e_{n_3}\rangle \langle c_{3},e_{n_3}|\frac{1}{E^{+}-{\hat p}h^{(seq)}{\hat p}-H_{Q'Q'}}|c_{4},e_{n_4}\rangle 
\nonumber  \\ 
&&\times \langle c_{4},e_{n_4}|H_{PQ}|{\tilde \Phi}^{A}_{i} \rangle \Delta_{e}^{4} 
\label{mat_elem2}
\end{eqnarray}
Let us first consider the matrix element: $\langle c_{1},e_{n_1}|E^{+}-M|c_{2},e_{n_2} \rangle$, which can be written as:
\begin{eqnarray}	
&&\langle c_{1},e_{n_1}|E^{+}-M|c_{2},e_{n_2} \rangle \label{mat_el_seq2} 
\nonumber \\
&&=\langle t_{1},e_{n_1},l_{1},j_{1};J|E^{+}-{\hat p}h^{(seq)}{\hat p}-H_{Q'Q'}
-H_{Q'P'}\frac{1}{E^{+}-{\hat p}h^{(seq)}{\hat p}-H_{P'P'}}H_{P'Q'}|t_{2},e_{n_2},l_{2},j_{2};J\rangle 
\nonumber \\ 
&&=(E-E_{t_{1}}-e_{n_1})\delta_{c_{1},c_{2}}\delta_{e_{n_1},e_{n_2}}\frac{1}{\Delta_e} 
\nonumber \\
&&~~~~~~~~-\langle t_{1}|H_{Q'P'}\frac{1}{E^{+}-e_{n_1}-
H_{P'P'}}H_{P'Q'}  |t_{2}\rangle \hspace{0.3cm}\delta_{(e_{n_1},l_{1},j_{1}),(e_{n_2},l_{2},j_{2})}\frac{1}{\Delta_e}
\label{g3}
\end{eqnarray} 
Here, $|t_{1} \rangle$ is an eigenvector of  $H_{Q'Q'}$ with an eigenvalue $E_{t_{1}}$ and:
\begin{eqnarray}
\langle t_{1}|H_{Q'P'}\frac{1}{E^{+}-e_{n_1}-H_{P'P'}}H_{P'Q'}  |t_{2}\rangle 
=\sum_{t',l',j'}\int_{0}^{+\infty} dr w_{t_{1};t',l',j'}(r) \omega_{t_{2};t',l',j'}^{(+)(E-e_{n_1})}(r) \label{H_eff_Q'}
\end{eqnarray} 
$ t',l'$ and $j'$ in (\ref{H_eff_Q'})
denote the bound state of a daughter nucleus, the orbital angular momentum and the total angular  momentum of the second emitted proton, respectively.
$w_{t_{1};t',l',j'}(r)$  denotes the projection of the source term on the emission channel: 
$c'=(t',l',j';J_{t_1})$, where  $J_{t_1}$  is the total angular momentum of $|t_1\rangle$. $\omega_{t_{2};t',l',j'}^{(+)(E-e_{n_1})}$ is the projection on the same channel of the continuation of the state $|t_2\rangle$ in the continuum. The expression  (\ref{H_eff_Q'}) corresponds to the emission of a second proton  with energy $E-e_{n_1}$ from a nucleus  A-1. For
 $e_{n_1}>E$, the second emission is impossible and, consequently, the expression (\ref{H_eff_Q'}) becomes real. The source term and the continuation of the SM wave function in the continuum is calculated similarly as in the standard SMEC. Thus, the matrix element (\ref{mat_el_seq2}) takes a form:
\begin{eqnarray*}
&&\langle c_{1},e_{n_1}|E^{+}-M|c_{2},e_{n_2} \rangle \\ 
&=&\left[(E-E_{t_{1}}-e_{n_1})\delta_{c_1,c_2}\delta_{e_{n_1},e_{n_2}}
-\sum_{t',l',j'}\int_{0}^{+\infty} dr w_{t_{1};t',l',j'}(r) 
\omega_{t_{2};t',l',j'}^{(+)(E-e_{n_1})}(r)
\delta_{(e_{n_1},l_{1},j_{1}),(e_{n_2},l_{2},j_{2})} \right ]\frac{1}{\Delta_e}
\end{eqnarray*} 

The inverse operator $1/(E^{+}-M)$ in the matrix element $\delta^{(seq)}(E)$ (cf  eq. (\ref{mat_elem2})) is obtained by diagonalizing $E^{+}-M$ for each energy $e_{n_1}$ in the basis of ${\cal Q'}$.

The term $\langle c_{3},e_{n_3}|(E^{+}-{\hat p}h^{(seq)}{\hat p}-H_{Q'Q'})^{-1}|c_{4},e_{n_4}\rangle$  in (\ref{mat_elem2}) equals: 
\begin{eqnarray}
\langle c_{3},e_{n_3}|(E^{+}-{\hat p}h^{(seq)}{\hat p}-H_{Q'Q'})^{-1}|c_{4},e_{n_4}\rangle=[E-e_{n_3}-E_{t_3}]
^{-1}\delta_{(c_3,c_4)}\delta_{(e_{n_3},e_{n_4})}\frac{1}{\Delta_e}
\end{eqnarray}
and the term $\langle {\tilde \Phi}^{A}_{i}|H_{QP}|c_{1},e_{n_1}\rangle$ is:
\begin{eqnarray}
\langle {\tilde \Phi}^{A}_{i}|H_{QP}|c_{1},e_{n_1}\rangle =\int dr {~} w_{{\tilde \Phi}^{A}_{i};t_{1},l_{1},j_{1}}^{*}(r) u_{e_{n_1},l_{1},j_{1}}(r)  
\end{eqnarray}
Here, $w_{{\tilde \Phi}^{A}_{i};t_{1},l_{1},j_{1}}^{*}(r)$ is the source term projection for the emission of the first proton and $u_{e_{n_1},l_{1},j_{1}}(r)$ is the radial wave function of the state
$|e_{n_1},l_1,j_1\rangle$.
The calculation of 
$\langle c_{2},e_{n_2}|H_{Q'P'}(E^{+}-{\hat p}h^{(seq)}{\hat p}-H_{P'P'})^{-1}H_{P'Q'}|c_{3},e_{n_3}\rangle$ is identical to the calculation of the term in eq. (\ref{H_eff_Q'}).
Hence, the matrix element $\delta^{(seq)}(E)$  becomes:
\begin{eqnarray}
&&\delta^{(seq)}(E)=\sum_{t_1,t_{2},t_{3},l,j}\sum_{n=1}^{+\infty}
\int_{0}^{+\infty} dr {~} w_{{\tilde \Phi}^{A}_{i};t_{1},l,j}^{*}(r) u_{e_{n},l,j}(r) 
\nonumber \\
&\times&  \langle t_1|\frac{1}{E^{+}-e_n-H_{Q'Q'}
-H_{Q'P'}[E^{+}-e_n-H_{P'P'}]^{-1}H_{P'Q'}}|t_2\rangle
\nonumber \\
&\times& \sum_{t',l',j'}\int_{0}^{+\infty} dr' w_{t_{2};t',l',j'}^{*}(r') \omega_{t_{3};t',l',j'}^{(+)(E-e_{n})}(r')
\frac{1}{E^{+}-e_{n}-E_{t_3}} 
\nonumber \\ 
&\times& \int_{0}^{+\infty} dr w_{{\tilde \Phi}^{A}_{i};t_{3},l,j}(r) u_{e_{n},l,j}^{*}(r) ~~\Delta_{e} 
\label{y_1}
\end{eqnarray} 
and the partial width for the sequential 2p emission is given by eq. (\ref{szer}).

For $e_n>E$, the emission of a second proton is impossible. Hence, the contribution of different terms with $e_n>E$ is real. Since we are interested in calculating the emission width, therefore the energy summation in (\ref{y_1}) is restricted to an interval from  0 to $E$.

If the sequential decay occurs through a resonance in the A-1 intermediate nucleus, then one has to consider the operator $H_{QP}\tilde{G}_{P}^{(+)}(E)H_{PQ}$ (cf eq. (\ref{H_eff_seq2})).
Assuming (\ref{assum1})-(\ref{assum4}), one obtains:
\begin{eqnarray}
H_{QP}\frac{1}{E^{+}-{\hat p}h^{(seq)}{\hat p}-H_{Q'Q'}-H_{Q'P'}
\left[{E^{+}-{\hat p}h^{(seq)}{\hat p}-H_{P'P'}}\right]^{-1}H_{P'Q'}}H_{PQ}
\end{eqnarray}
Inserting two completness relations (cf eq. (\ref{completude})), one finds  the matrix element:
\begin{eqnarray}
&&\delta^{(seq)}(E)=\sum_{t_1,t_{2},l,j}\sum_{n=1}^{+\infty}
\int_{0}^{+\infty} dr {~} w_{{\tilde \Phi}^{A}_{i};t_{1},l,j}^{*}(r) u_{e_{n},l,j}(r) 
\nonumber \\
&\times&  \langle t_1|\frac{1}{E^{+}-e_n-H_{Q'Q'}
-H_{Q'P'}[E^{+}-e_n-H_{P'P'}]^{-1}H_{P'Q'}}|t_2\rangle
\nonumber \\
&\times& \int_{0}^{+\infty} dr w_{{\tilde \Phi}^{A}_{i};t_{3},l,j}(r) u_{e_{n},l,j}^{*}(r) ~~\Delta_{e} 
\label{y_x}
\end{eqnarray}  
from which the partial decay width can be calculated.


\section{The source term for the emission of (2p)  cluster} 
\label{source_annexe}

Projected source term for a direct emission of two protons as a cluster is:
\begin{eqnarray}
w_{i,c}(R)=R \langle t^{(int)},0s,(L_{rel},S),J_{2p};J,R|H_{TQ}|\phi_i^{(int)}\rangle 
\label{dfw}
\end{eqnarray}
where $c=(t^{(int)},0s,(L_{rel},S);J_{2p};J)$ is the decay channel in ${\cal T}$. $t^{(int)}$ is the intrinsic state of a daughter nucleus, $0s$ is the intrinsic state of a cluster, $L_{rel}$ is the relative angular momentum between the  cluster and the nucleus A-2, and  $S$ is the spin of the  cluster. $L_{rel}$ and $S$ are coupled to  $J_{2p}$, and $J=J_{t^{(int)}}+J_{2p}$ is the total angular momentum of total system  
$[A-2]\otimes[2]$. $R$ in (\ref{dfw}) denotes the relative coordinate between the daughter nucleus and the cluster.
The source term $w_{i,c}(R)$ is localized and can be developed in the harmonic oscillator basis:
\begin{eqnarray*}
w_{i,c}(R)=R \sum_{N_{rel}} {\cal R}_{N_{rel}L_{rel}}(R)\langle t^{(int)},0s,(L_{rel},S);J_{2p};J;N_{rel}|H_{TQ}|\phi^{int}\rangle
\end{eqnarray*}
where ${\cal R}_{N_{rel}L_{rel}}(R)$ is the harmonic oscillator wave function characterized by the radial quantum number  $N_{rel}$ and the angular momentum  $L_{rel}$. 

In the formalism of second quantization, the coupling operator between ${\cal Q}$ and ${\cal T}$ (cf eq. (\ref{hqt})) is:
\begin{eqnarray}
\label{nowe_01}
H_{TQ}= -\sum_{\tiny
\begin{array}{c} \alpha \leq \beta \\ \gamma \leq \delta \\ \Gamma  \end{array}}
\frac{1}{\sqrt{1+\delta_{\alpha,\beta}}} 
\frac{1}{\sqrt{1+\delta_{\gamma,\delta}}}
V_{\alpha,\beta,\gamma,\delta}^{\Gamma} [a^{\dagger}_{\alpha}a^{\dagger}_{\beta}]^{\Gamma}[\tilde{a}_{\gamma} \tilde{a}_{\delta}]^{\Gamma}
\end{eqnarray}
where $V_{\alpha,\beta,\gamma,\delta}^{\Gamma}$ is the antisymmetrized, reduced matrix element of the residual interaction. $\alpha,...\delta$ are the eigenstates of the one-body potential with an origin in the laboratory frame.
Since $H_{TQ}$ is expressed in the laboratory frame and $w_{i,c}(R)$ is calculated in the frame associated with the relative coordinate $R$, therefore one has to change the coordinate system in order to calculate the projection of the source.  
Let us consider the non-spurious SM states $|t_i\rangle$ and $|\Phi_i^{A}\rangle$  which correspond to intrinsic states  $|t_i^{(int)}\rangle$ and $|\phi_i^{(int)}\rangle$. By definition, these states 
can be written as:
\begin{eqnarray}	
|t_i\rangle=|t_i^{(int)}\rangle |\Phi_{00}^{A-2}\rangle   \hspace {1cm},\hspace {1cm}
|\Phi_i^{A}\rangle=|\phi_i^{(int)}\rangle |\Phi_{00}^{A}\rangle
\end{eqnarray}
where $|\Phi_{00}^{A-2}\rangle$ and $|\Phi_{00}^{A}\rangle$ are the ground states of the center of mass of nuclei A-2 and A, respectively.
Let us consider the following matrix element expressed in the laboratory frame:
\begin{eqnarray}
\langle t_i,0s,(L,S);J_{2p};J,N|H_{TQ}|\Phi_i^{A}\rangle \label{in_da_lab}
\end{eqnarray}
$N$  and $L$  in (\ref{in_da_lab}) are the oscillator quantum numbers characterizing the state of a cluster with respect to the origin fixed in the laboratory frame.
Using Moshinsky transformation, one can write (\ref{in_da_lab}) as follows:
\begin{eqnarray*}
\sum_{N_{G},L_{G},N_{rel},L_{rel}}\langle 0 0 N L;L| N_{G},L_{G},N_{rel},L_{rel};L\rangle
\langle N_{G}, L_{G}, t^{(int)},0s,L_{rel},S,J,N_{rel}|H_{TQ}|\Phi_i^{A}\rangle
\end{eqnarray*}
where $\langle 0 0 N L;L| N_{G},L_{G},N_{rel},L_{rel};L\rangle$ is a Moshinsky coefficient.
 $N_{G}$ and $L_{G}$ in the above equation are the quantum numbers corresponding to the motion of total system with respect to the laboratory frame. The residual interaction does not act on the center of mass coordinates of the system, what implies: $$N_{G}=0, L_{G}=0, N_{rel}=N, L_{rel}=L \ .$$ 
One obtains:

\begin{eqnarray}
\langle t_i,0s,(L,S);J_{2p};J,N|H_{TQ}|\Phi_i^{A}\rangle =\langle 0 0 N L;L| 0 0 N L;L\rangle 
\langle t^{(int)},0s,(L,S);J_{2p};J,N|H_{TQ}|\phi_i^{(int)}\rangle \nonumber \\
\end{eqnarray}
Hence, the matrix element in the laboratory frame has been transformed into the matrix element in the relative coordinates.
Using the analytical expressions for the Moshinsky coefficients, one can rewrite (\ref{dfw}) as:
\begin{eqnarray}
w_{i,c}(R)= \sum_{N} \left(\frac {A}{A-2}\right)^{(2N+L)/2} u_{NL}(R)
\langle t_i,0s,(L,S);J_{2p};J,N|
H_{TQ}|\Phi_i^{A}\rangle
\end{eqnarray}
where:  $u_{NL}(R)=R {\cal R}_{NL}(R)$.
Using the Wigner-Eckart theorem, one writes $w_{i,c}(R)$ as:
\begin{eqnarray}
\label{wc_r}  
&&w_{i,c}(R)=-\sum_{\tiny\begin{array}{c} (\alpha \leq \beta) \in cont \\ (\gamma \leq \delta) \in disc \\ \Gamma, N  \end{array}}
\frac{1}{\sqrt{1+\delta_{\gamma,\delta}}}\left(\frac {A}{A-2}\right)^{(2N+L)/2} u_{NL}(R) 
\nonumber 
\\
&&\times \hat{J} \left\{\begin{array}{ccc} J_{t_i}&J_{2p}&J\\J_{\Phi_i^{A}}&0&J\\ \Gamma&\Gamma&0 \end{array}\right\}
\langle t_i||[\tilde{a}_{\gamma} \tilde{a}_{\delta}]^{\Gamma}||\Phi_i^{A}\rangle 
\langle 0s,L,S,J_{2p},N||(\alpha \beta)^{\Gamma}\rangle \langle(\alpha,\beta)^{\gamma}||V||( \gamma \delta)^{\Gamma}\rangle
\end{eqnarray}
Let us consider a term:
\begin{eqnarray}
\sum_{(\alpha\leq\beta) \in cont}
\frac{1}{\sqrt{1+\delta_{\gamma,\delta}}}\langle 0s,L,S,J_{2p},N||(\alpha \beta)^{\Gamma}\rangle \langle(\alpha,\beta)^{\Gamma}||V||( \gamma \delta)^{\Gamma}\rangle   \nonumber 
\end{eqnarray}
in eq. (\ref{wc_r}).
Inserting twice the completness relation defined with the harmonic oscillator wave functions
and integrating over the energy of states  $\alpha$ and $\beta$, one obtains:
\begin{eqnarray}
&&\sum_{\tiny\begin{array}{c} (n_1,l_{1},j_{1})\leq (n_{2},l_{2},j_{2}) \\ (n_{3},l_{3},j_{3})\leq (n_{4},l_{4},j_{4}) \end{array}} \hat{\Gamma}^{2}
\langle 0s,L,S,J_{2p},N|(n_1,l_{1},j_{1},n_{2},l_{2},j_{2})^{\Gamma}\rangle \langle(n_{3},l_{3},j_{3},n_{4},l_{4},j_{4})^{\Gamma}|V|(\gamma \delta)^{\Gamma}\rangle \nonumber \\ %
&&\times \sum_{ \tiny\begin{array}{c} (l_{\alpha},j_{\alpha}) \leq (l_{\beta},j_{\beta}) \end{array} } \frac{1}{\sqrt{1+\delta_{j_{1},j_{2}}}\sqrt{1+\delta_{j_{3},j_{4}}}} 
\frac{1}{\sqrt{1+\delta_{\gamma,\delta}}}
\nonumber  \\
&&\times \left\{ 
\langle j_{1}|1-{\hat q}_{l_{\alpha},j_{\alpha}}|j_{3}\rangle \langle j_{2}|1-{\hat q}_{l_{\beta}j_{\beta}}|j_{4}\rangle 
-(-1)^{\Gamma-j_{\alpha}-j_{\beta}} \langle j_{1}|1-{\hat q}_{l_{\alpha},j_{\alpha}}|j_{4} \rangle \langle j_{2}|1-{\hat q}_{l_{\beta}j_{\beta}}|j_{3}\rangle \right\} 
\label{A_8}
\end{eqnarray}
where $j_{1}$ represents the state  $(n_{1},l_{1},j_{1})$, and similarly for $j_{2},j_{3},j_{4}$. 
${\hat q}_{l_{\alpha},j_{\alpha}}$ and ${\hat q}_{l_{\beta}j_{\beta}}$ are the projectors on proton (quasi-)bound states having quantum numbers $l_{\alpha},j_{\alpha}$  and  $l_{\beta},j_{\beta}$, respectively.
The term $\langle 0s,L,S,J_{2p},N|(n_1,l_{1},j_{1},n_{2},l_{2},j_{2})^{\Gamma}\rangle$  
in (\ref{A_8}) equals:
\begin{eqnarray}
\frac{1}{\sqrt{2(1+\delta_{j_{1},j_{2}})}}
 \hat{j_{1}} \hat{j_{2}} \hat{S} \hat{L} (1-(-1)^{S+1+l})
\left\{\begin{array}{ccc} l_{1}&s_1&j_{1}\\l_{2}&s_2&j_{2}\\L&S&\Gamma \end{array}\right\}
\langle n_1,l_{1},n_{2},l_{2},L|N,L,0,0,L\rangle      
\label{A_9}
\end{eqnarray}
The term $\langle(n_{3},l_{3},j_{3},n_{4},l_{4},j_{4})^{\Gamma}|V|(\gamma \delta)^{\Gamma}\rangle$ in (\ref{A_8}) equals:
\begin{eqnarray}
&&\sum_{i \leq j} \langle i j|(\gamma,\delta)^{\Gamma}\rangle \langle (j_{3},j_{4})^{\Gamma}|V|(i,j)^{\Gamma}\rangle \nonumber\\
&&=\sum_{i\leq j} \langle i j|(\gamma,\delta)^{\Gamma}\rangle 
\sum_{\tiny\begin{array}{c} L_{1} S_{1} N_{1}' l_{1}' n_{1}' l_{1}' \\  \\  L_{2} S_{2} N_{2}' l_{2}' n_{2}' l_{2}'\end{array} }
\frac {1}{ \sqrt{2(1+\delta_{i,j})}\sqrt{2(1+\delta_{j_{3},j_{4}})}}(1 - (-1)^{S_{1}+l_{1}'+1})(1 - (-1)^{S_{2}+l_{2}'+1}) \nonumber \\  \nonumber\\
&&\times\left\{\begin{array}{ccc} l_{j_{3}}&s_1&j_{j_{3}}\\l_{j_{4}}&s_2&j_{j_{4}}\\L_{1}&S_{1}&\Gamma \end{array}\right\}
\hat{j_{j_{3}}} \hat{j_{j_{4}}} \hat{L_{1}} \hat{S_{1}} 
\left\{\begin{array}{ccc} l_{i}&s_1&j_{i}\\l_{j}&s_2&j_{j}\\L_{2}&S_{2}&\Gamma \end{array}\right\}
\hat{j_{i}} \hat{j_{j}} \hat{L_{2}} \hat{S_{2}} \nonumber \\ 
&&\times \langle n_{j_{3}}l_{j_{3}},n_{j_{4}},l_{j_{4}},L_{1} |N_1',L_{1}',n_1',l_{1}',L_{1}\rangle 
\times \langle n_{i}l_{i},n_{j},l_{j},L_{2} |N_{2}',L_{2}',n_{2}',l_{2}',L_{2}\rangle  \nonumber 
\\
&&\times \langle N_1',L_{1}',n_1',l_{1}',L_{1},S_{1},\Gamma|V|N_{2}',L_{2}',n_{2}',l_{2}',L_{2},S_{2},\Gamma \rangle
\label{newC_10}
\end{eqnarray}
Since the residual two-body interaction  $V^{(res)}$ does not act on the center of mass of two particles, we have: $$N_1'=N_{2}' ~,~L_{1}'=L_{2}'  \ . $$
Moreover, in the case of Wigner-Bartlett interaction which is used in this work, the spin $(S_{1}=S_{2})$ and intrinsic angular momentum $(l_{1}'=l_{2}')$ are conserved, what implies: $L_{1}=L_{2}$.
Hence, (\ref{newC_10}) takes a  form:
\begin{eqnarray}
&&\sum_{i \leq j} \langle i j|(\gamma,\delta)^{\Gamma}\rangle \langle (j_{3},j_{4})^{\Gamma}|V|(i,j)^{\Gamma}\rangle \nonumber\\
&&=\sum_{i \leq j}  2 \langle ij|(\gamma,\delta)^{\Gamma}\rangle \sum_{L_{1},S_{1},N_{1}',L_{1}',n_{1}',l_{1}',n_{2}'}
\frac {1}{\sqrt{1+\delta_{i,j}}\sqrt{1+\delta_{j_{3},j_{4}}}} 
\left\{\begin{array}{ccc} l_{j_{3}}&s_1&j_{j_{3}}\\l_{j_{4}}&s_2&j_{j_{4}}\\L_{1}&S_{1}&\Gamma \end{array}\right\}
\hat{j_{j_{3}}} \hat{j_{j_{4}}} \hat{L_{1}} \hat{S_{1}} \nonumber \\ \nonumber \\
&&\times\left\{\begin{array}{ccc} l_{i}&s_1&j_{i}\\l_{j}&s_2&j_{j}\\L_{1}&S_{1}&\Gamma \end{array}\right\}
\hat{j_{i}} \hat{j_{j}} \hat{L_{1}} \hat{S_{1}} 
\langle n_{j_{3}}l_{j_{3}},n_{j_{4}},l_{j_{4}},L_{1} |N_1',L_{1}',n_1',l_{1}',L_{1}\rangle 
\nonumber \\
&&\times\langle n_{i}l_{i},n_{j},l_{j},L_{1} |N_1',L_{1}',n_{2}',l_{1}',L_{1}\rangle
\langle N_1',L_{1}',n_1',l_{1}',L_{1},S_{1}|V|N_1',L_{1}',n_{2}',l_{1}',L_{1},S_{1}\rangle \nonumber \label{A_11} \\
\end{eqnarray}
Inserting (\ref{A_8}), (\ref{A_9}), and (\ref{A_11}) in eq. (\ref{wc_r}), one obtains:
\begin{eqnarray}
&&w_{i,c}(R)= -\sum_{\Gamma,(\gamma \leq \delta) \in disc,N} u_{NL}(R)  
 \left(\frac{A}{A-2}\right)^{(2N+L)/2}
  \hat{J}(\hat{\Gamma})^{2} \left\{\begin{array}{ccc} J_{t_i}&J_{2p}&J\\J_{\Phi_i^{A}}&0&J\\ \Gamma&\Gamma&0 \end{array}\right\}   \nonumber \\ 
&&\times  \sum_{\tiny \begin{array}{c} \\ (l_{\alpha},j_{\alpha}) \leq (l_{\beta},j_{\beta} ) \\ (n_1,l_{1},j_{1}) \leq (n_{2},l_{2},j_{2}) \\ (n_{3},l_{3},j_{3})\leq (n_{4},l_{4},j_{4}) \end{array}   }
\frac{1}{1+\delta_{j_{1},j_{2}}} \frac{1}{1+\delta_{j_{3},j_{4}}}
\frac{1}{\sqrt{1+\delta_{\gamma,\delta}}}
 \langle t_i||[\tilde{a_{\gamma}}\tilde{a_{\delta}}]^{\Gamma}||\Phi_i^{A}\rangle \nonumber  \\ %
&& \times \left\{ \langle j_{1}|1-{\hat q}_{l_{\alpha},j_{\alpha}}|j_{3}\rangle \langle j_{2}|1-{\hat q}_{l_{\beta}j_{\beta}}|j_{4}\rangle - (-1)^{\Gamma-j_{\alpha}-j_{\beta}} \langle j_{1}|1-{\hat q}_{l_{\alpha},j_{\alpha}}|j_{4} \rangle \langle j_{2}|1-{\hat q}_{l_{\beta}j_{\beta}}|j_{3}\rangle \right\} 
\nonumber  \\ 
&&\times \hat{j_{1}} \hat{j_{2}} \hat{L} \hat{S} \sqrt{2}  \left\{\begin{array}{ccc} l_{1}&s_1&j_{1}\\l_{2}&s_2&j_{2}\\L&S&\Gamma \end{array}\right\} \langle n_1,l_{1},n_{2},l_{2},L|NL00,L\rangle \nonumber  \\ 
&&\times\sum_{i\leq j} \frac{1}{\sqrt{1+\delta_{i,j}}} 2 \langle ij|(\gamma,\delta)^{\Gamma}\rangle 
\left\{\begin{array}{ccc} l_{j_{3}}&s_1&j_{j_{3}}\\l_{j_{4}}&s_2&j_{j_{4}}\\L_{1}&S_{1}&\Gamma \end{array}\right\}
\hat{j_{j_{3}}} \hat{j_{j_{4}}} \hat{L_{1}} \hat{S_{1}} 
\left\{\begin{array}{ccc}l_{i}&s_1&j_{i}\\l_{j}&s_2&j_{j}\\L_{1}&S_{1}&\Gamma \end{array}\right\}
 \nonumber  \\  
&&\times\hat{j_{i}} \hat{j_{j}} \hat{L_{1}} \hat{S_{1}}
\langle n_{j_{3}}l_{j_{3}},n_{j_{4}},l_{j_{4}},L_{1} |N_1',L_{1}',n_1',l_{1}',L_{1}\rangle
 \nonumber \\  
&&\times\langle n_{i}l_{i},n_{j},l_{j},L_{1} |N_1',L_{1}',n_{2}',l_{1}',L_{1}\rangle \langle N_1',L_{1}',n_1',l_{1}',L_{1},S_{1}|V|N_1',L_{1}',n_{2}',l_{1}',L_{1},S_{1}\rangle
\end{eqnarray}
%

\section{The source term  for the direct 2p emission}
 \label{annexe_source_dir}
The projected source term for the direct 2p emission is:
\begin{eqnarray}
w_{i,c}(\rho)=\rho^{5/2} \langle t,K,(l_{x},l_{y}),L,S;J_{2p};J,\rho|H_{TQ}|\Phi_i^{A}\rangle
\end{eqnarray}
where the emission channel is defined as: $c=(t,K,(l_{x},l_{y}),L,S;J_{2p};J)$ (cf sect. \ref{para_direct}). 
Using the second quantization form (\ref{nowe_01})  for  $H_{TQ}$, one obtains:
\begin{eqnarray*}
w_{i,c}(\rho)=-\rho^{5/2} \langle t,K,(l_{x},l_{y}),L,S;J_{2p};J,\rho|
\sum_{\tiny\begin{array}{c} \alpha \leq \beta \\
\gamma \leq \delta   \\ \Gamma \end{array}}
V_{\alpha, \beta, \gamma, \delta}^{\Gamma} \frac{1}{\sqrt{1+\delta_{\alpha,\beta}}} 
\frac{1}{\sqrt{1+\delta_{\gamma,\delta}}}
[a^{+}_{\alpha}a^{+}_{\beta}]^{\Gamma}[\tilde{a_{\gamma}}\tilde{a_{\delta}}]^{\Gamma}
|\Phi_i^{A}\rangle \nonumber 
\end{eqnarray*}
Since the state represented in bra has two protons in the continuum and in the ket all nucleons are in (quasi-)bound states, the operators $a^{+}_{\alpha}a^{+}_{\beta}$  can only annihilate particles in the continuum states which are present in the bra vector. Hence:
\begin{eqnarray*}
w_{i,c}(\rho)=-\rho^{5/2} \langle t,K,(l_{x},l_{y}),L,S;J_{2p};J,\rho|
\sum_{\tiny\begin{array}{c} \alpha \leq \beta \in cont\\
\gamma \leq \delta  \in disc \\ \Gamma \end{array}}
\frac{1}{\sqrt{1+\delta_{\alpha,\beta}}} 
\frac{1}{\sqrt{1+\delta_{\gamma,\delta}}}
V_{\alpha, \beta, \gamma, \delta}^{\Gamma}[a^{+}_{\alpha}a^{+}_{\beta}]^{\Gamma}[\tilde{a_{\gamma}}\tilde{a_{\delta}}]^{\Gamma}|\Phi_i^{A}\rangle 
\nonumber 
\end{eqnarray*}
Separating the part of the operator acting on continuum states from the part acting on (quasi-)bound states and applying Wigner-Eckart theorem, one finds:
\begin{eqnarray}
&&w_{i,c}(\rho)=-\rho^{5/2}
\sum_{\tiny\begin{array}{c} \alpha \leq \beta \in cont\\
\gamma \leq \delta  \in disc \\ \Gamma \end{array}}
\frac{1}{\sqrt{1+\delta_{\gamma,\delta}}}
\hat{J}
\left\{\begin{array}{ccc} I_{t}&J_{2p}&J\\J&0&J\\ \Gamma & \Gamma & 0 \end{array}\right\}  
 \langle t||[\tilde{a_{\gamma}}\tilde{a_{\delta}}]^{\Gamma}||\Phi_i^{A}\rangle \nonumber\\ 
&& \times \langle K,(l_{x},l_{y}),L,S;J_{2p},\rho||(\alpha,\beta)^{\Gamma}\rangle
 \langle (\alpha, \beta)^{\Gamma}||V||(\gamma, \delta)^{\Gamma} \rangle \label{w_rho0}
\end{eqnarray}
Let us consider the sum over energies of proton states $\alpha$ and $\beta$ in the above expression:
\begin{eqnarray}
&&\sum_{\alpha \leq \beta \in cont} 
\langle K,(l_{x},l_{y}),L,S;J_{2p},\rho||(\alpha,\beta)^{\Gamma}\rangle
\langle (\alpha, \beta)^{\Gamma}||V||(\gamma, \delta)^{\Gamma} \rangle \nonumber \\
&&=\sum_{\bar{\alpha} \leq \bar{\beta}} \sum_{e_\alpha, e_\beta }
\langle K,(l_{x},l_{y}),L,S;J_{2p},\rho||
 (\alpha,\beta)^{\Gamma}\rangle
\langle (\alpha, \beta)^{\Gamma}||V||(\gamma, \delta)^{\Gamma} \rangle \nonumber \\ 
&&=\sum_{\bar{\alpha} \leq \bar{\beta}} 
\langle K,(l_{x},l_{y}),L,S;J_{2p},\rho||(1_{\bar{\alpha}}-{\hat q}_{\bar{\alpha}})
\otimes(1_{\bar{\beta}}-{\hat q}_{\bar{\beta}})||V||(\gamma, \delta)^{\Gamma} \rangle 
\end{eqnarray}
where $\bar{\alpha}$  denotes the quantum numbers $l_{\alpha}$, 
$j_{\alpha}$ and $\bar{\beta}$ the quantum numbers $l_{\beta}$, $j_{\beta}$. The operator
${\hat q}_{\bar{\alpha}}$ is the projector on proton (quasi-)bound s.p. states with the orbital angular momentum
$l_{\alpha}$ and total angular momentum $j_{\alpha}$. In a similar way, 
${\hat q}_{\bar{\beta}}$ projects on the subspace of proton(quasi-)bound states with the angular momentum
 $l_{\beta}$ and total angular momentum $j_{\beta}$.
 Hence, the projected source term $w_{i,c}(\rho)$ can be written as:
\begin{eqnarray}
&&w_{i,c}(\rho)=-\rho^{5/2}
\sum_{\tiny\begin{array}{c} \bar{\alpha} \leq \bar{\beta} \\
\gamma \leq \delta  \in disc \\ \Gamma \end{array}}
\frac{1}{\sqrt{1+\delta_{\gamma,\delta}}}
\hat{J}
\left\{\begin{array}{ccc} I_{t}&J_{2p}&J\\J&0&J\\ \Gamma & \Gamma & 0 \end{array}\right\}  
 \langle t||[\tilde{a_{\gamma}}\tilde{a_{\delta}}]^{\Gamma}||\Phi_i^{A}\rangle \nonumber\\ 
&& \times \langle K,(l_{x},l_{y}),L,S;J_{2p},\rho||(1_{\bar{\alpha}}-{\hat q}_{\bar{\alpha}})
\otimes(1_{\bar{\beta}}-{\hat q}_{\bar{\beta}})||V||(\gamma, \delta)^{\Gamma} \rangle 
\label{w_rho}
\end{eqnarray}
The term $(1_{\bar{\alpha}}-{\hat q}_{\bar{\alpha}})\otimes(1_{\bar{\beta}}-{\hat q}_{\bar{\beta}})$ in (\ref{w_rho}) can be written in a following form:
\begin{eqnarray}
(1_{\bar{\alpha}}-{\hat q}_{\bar{\alpha}})
\otimes(1_{\bar{\beta}}-{\hat q}_{\bar{\beta}})=1_{\bar{\alpha}}\otimes 1_{\bar{\beta}}
-1_{\bar{\alpha}}\otimes {\hat q}_{\bar{\beta}}-{\hat q}_{\bar{\alpha}}\otimes 1_{\bar{\beta}}+{\hat q}_{\bar{\alpha}}\otimes {\hat q}_{\bar{\beta}} 
\label{prod}
\end{eqnarray}
The sum over quantum numbers  $\bar{\alpha}$ and $\bar{\beta}$ yields for the first term in (\ref{prod}) the identity in the space of two-particle states:
\begin{eqnarray}
\sum_{\bar{\alpha} \leq \bar{\beta}} 1_{\bar{\alpha}}\otimes 1_{\bar{\beta}}=1\otimes 1
\end{eqnarray}
Hence, the contribution of this term in  $w_{i,c}(\rho)$ (eq. (\ref{w_rho})) can be written as:
\begin{eqnarray}
-\rho^{5/2}\sum_{\gamma \leq \delta  \in disc ; \Gamma }\frac{1}{\sqrt{1+\delta_{\gamma,\delta}}}
\hat{J}\left\{\begin{array}{ccc} I_{t}&J_{2p}&J\\J&0&J\\ \Gamma & \Gamma & 0 \end{array}\right\}  
 \langle t||[\tilde{a_{\gamma}}\tilde{a_{\delta}}]^{\Gamma}||\Phi_i^{A}\rangle 
\langle K,(l_{x},l_{y}),L,S;J_{2p},\rho||V||(\gamma, \delta)^{\Gamma} \rangle \nonumber \\
\label{q1}
\end{eqnarray}
States $\gamma$ and $\delta$ are (quasi-)bound and can be expanded in the harmonic oscillator basis:
\begin{eqnarray}
|(\gamma \delta)^{\Gamma}\rangle=\sum_{n_{\gamma} \leq n_{\delta}}
\langle n_{\gamma} l_{\gamma} j_{\gamma} n_{\delta} l_{\delta} j_{\delta};\Gamma|
(\gamma \delta )^{\Gamma}\rangle |n_{\gamma},l_{\gamma},j_{\gamma} n_{\delta},l_{\delta},j_{\delta};\Gamma\rangle 
\label{devel} 
\end{eqnarray}
where $n_{\gamma}$ and $n_{\delta}$ are the radial quantum numbers associated with the harmonic oscillator expansion. Passing from j-j to L-S coupling scheme and applying  Moshinsky transformation, one rewrites expression (\ref{devel}) in a following form:
\begin{eqnarray}
|(\gamma \delta)^{\Gamma}\rangle&=&\sum_{n_{\gamma} \leq n_{\delta}}
\langle n_{\gamma} l_{\gamma} j_{\gamma} n_{\delta} l_{\delta} j_{\delta};\Gamma|
(\gamma \delta)^{\Gamma}\rangle 
\sum_{N',L',n',l'}\sum_{\lambda',S'} \tilde{\eta}_{\lambda',S'}(n_{\gamma},n_{\delta}) \nonumber  \\
& \times& \langle n_{\gamma}l_{\gamma}n_{\delta}l_{\delta} |N'L'n'l';\lambda'\rangle
|N'L'n'l';\lambda' S'\rangle \label{dev2}
\end{eqnarray}
where the coefficient $\tilde{\eta}_{\lambda',S'}(n_{\gamma},n_{\delta})$ is given by:
\begin{eqnarray}
\tilde{\eta}_{\lambda',S'}(n_{\gamma},n_{\delta})=
\frac{1+(-1)^{S'+l'}}
{\sqrt{2(1+\delta_{n_{\gamma},n_{\delta}}
\delta_{l_{\gamma},l_{\delta}}\delta_{j_{\gamma},j_{\delta}})}}
\times \hat{j_{\gamma}}\hat{j_{\delta}}\hat{S'}\hat{\lambda'} 
\left\{\begin{array}{ccc} l_{\gamma} & s_{\gamma} & j_{\gamma} 
\\l_{\delta} &s_{\delta} & j_{\delta} \\ \lambda' &  S'& \Gamma\end{array}\right\} 
\end{eqnarray}
Introducing the expansion (\ref{dev2}) in the following term of (\ref{q1}), one obtains:  
\begin{eqnarray}
&&\langle K,(l_{x},l_{y}),L,S;J_{2p},\rho||V||(\gamma \delta)^{\Gamma}\rangle \nonumber \\
&=&\sum_{n_{\gamma} \leq n_{\delta}} \sum_{N',L',n',l',\lambda',S'}
\langle n_{\gamma},l_{\gamma},j_{\gamma} n_{\delta},l_{\delta},j_{\delta};\Gamma|
(\gamma \delta)^{\Gamma}\rangle 
 \tilde{\eta}_{\lambda',S'}(n_{\gamma},n_{\delta}) \nonumber  \\
&&\langle n_{\gamma}l_{\gamma}n_{\delta}l_{\delta} |N'L'n'l';\lambda\rangle
\times \langle K,(l_{x},l_{y}),L,S;J_{2p},\rho||V||N'L'n'l';\lambda';S'; \Gamma\rangle 
\label{term2}
\end{eqnarray}
where matrix elements of the residual interaction are:
\begin{eqnarray}
&&\langle K,(l_{x},l_{y}),L,S;J_{2p},\rho||V||N'L'n'l';\lambda';S'; \Gamma\rangle \nonumber \\
%
&&= \hat{\Gamma}\int d\alpha d \hat{\theta_x} d\Omega_S 
\cos^{2}(\alpha)\sin^{2}(\alpha) 
\Psi_{K}^{l_{x_k},l_{y_k}}(\alpha) Y^{*}_{l_x}(\hat{\theta_x})\chi^{*}_S(\Omega_S)
 V \left (\frac{\rho \cos(\alpha)}
{\sqrt{\mu_x}},\hat{\theta_x},\Omega_S \right) \nonumber \\
&&\times {\cal R}_{N'L'}\left (\frac{\sqrt{2}\rho \sin(\alpha)}{\sqrt{\mu_y}} \right )
{\cal R}_{n'l'}\left (\frac{\rho \cos(\alpha)}{\sqrt{2\mu_x}} \right )
 Y_{l'}(\hat{\theta_x})\chi_S'(\Omega_S) \delta_{l_{y},L'} 
 \label{V}
\end{eqnarray}
$\mu_x$ and $\mu_y$ in (\ref{V})
are the dimensionless reduced masses associated with different subsystems related to ${\bf x}$ and ${\bf y}$ directions in the
Jacobi coordinates system {\bf T} (cf sect. \ref{para_direct}).
 ${\cal R}_{N'L'}(\sqrt{2}\rho \sin(\alpha)/\sqrt{\mu_y})$ and ${\cal R}_{n'l'}(\rho \cos(\alpha)/\sqrt{2\mu_x})$ are the radial functions in the harmonic oscillator basis associated with the center of mass motion of two protons with respect to the nucleus A-2 and the relative motion of those two protons, respectively.

Contribution of the second term in  (\ref{prod}) to $w_{i,c}(\rho)$  is:
\begin{eqnarray}
&&\rho^{5/2} \sum_{\tiny\begin{array}{c} \bar{\alpha} \leq \bar{\beta} \\
\gamma \leq \delta  \in disc \\ \Gamma \end{array}}
\frac{1}{\sqrt{1+\delta_{\gamma,\delta}}}
\hat{J}\left\{\begin{array}{ccc} I_{t}&J_{2p}&J\\J&0&J\\ \Gamma & \Gamma & 0 \end{array}\right\}  
 \langle t||[\tilde{a_{\gamma}}\tilde{a_{\delta}}]^{\Gamma}||\Phi_i^{A}\rangle \nonumber\\ 
&&\times \langle K,(l_{x},l_{y}),L,S;J_{2p},\rho||1_{\bar{\alpha}}\otimes {\hat q}_{\bar{\beta}}
||V||(\gamma, \delta)^{\Gamma} \rangle 
\label{mars}
\end{eqnarray}
The last term in  (\ref{mars}) is calculated by inserting two completness relations in  the harmonic oscillator basis. One gets:
\begin{eqnarray}
&&\langle K,(l_{x},l_{y}),L,S;J_{2p},\rho||1_{\bar{\alpha}}\otimes {\hat q}_{\bar{\beta}}
||V||(\gamma, \delta)^{\Gamma} \rangle \nonumber \\ 
&=&\sum_{\tiny\begin{array}{c} n_1, l_1, j_1  \\ n_2, l_2 , j_2 \end{array}}
\sum_{\tiny\begin{array}{c} n'_1, l'_1, j'_1  \\ n'_2, l'_2 , j'_2 \end{array}}
\langle K,(l_{x},l_{y}),L,S;J_{2p},\rho||n_1,l_1,j_1 n_2,l_2,j_2\rangle \nonumber \\
&& \times\langle n_1,l_1,j_1 n_2,l_2,j_2||1_{\bar{\alpha}}\otimes {\hat q}_{\bar{\beta}}||
n'_1,l'_1,j'_1 n'_2,l'_2,j'_2\rangle  \times\langle n'_1,l'_1,j'_1 n'_2,l'_2,j'_2||V||(\gamma, \delta)^{\Gamma} \rangle \nonumber \\
\end{eqnarray}
To calculate the overlap $\langle K,(l_{x},l_{y}),L,S;J_{2p},\rho||n_1,l_1,j_1 n_2,l_2,j_2\rangle$, we make first a transformation from j-j to L-S coupling scheme and then use the Moshinsky transformation for a state $|n_1,l_1,j_1 n_2,l_2,j_2\rangle$ (cf (\ref{dev2})).

Contribution of the third term in (\ref{prod}) to $w_{i,c}(\rho)$ is:
\begin{eqnarray}
&&\rho^{5/2}\sum_{\tiny\begin{array}{c} \bar{\alpha} \leq \bar{\beta} \\
\gamma \leq \delta  \in disc \\ \Gamma \end{array}}
\frac{1}{\sqrt{1+\delta_{\gamma,\delta}}}
\hat{J}\left\{\begin{array}{ccc} I_{t}&J_{2p}&J\\J&0&J\\ \Gamma & \Gamma & 0 \end{array}\right\}  
 \langle t||[\tilde{a_{\gamma}}\tilde{a_{\delta}}]^{\Gamma}||\Phi_i^{A}\rangle \nonumber\\ 
&&\times \langle K,(l_{x},l_{y}),L,S;J_{2p},\rho||{\hat q}_{\bar{\alpha}}\otimes1_{\bar{\beta}}
||V||(\gamma, \delta)^{\Gamma} \rangle 
\end{eqnarray}
This term is calculated in the same way as the term  (\ref{mars}).

Finally, the contribution of the fourth term in (\ref{prod})  to $w_{i,c}(\rho)$ is:
\begin{eqnarray}
&&-\rho^{5/2}\sum_{\tiny\begin{array}{c} \bar{\alpha} \leq \bar{\beta} \\
\gamma \leq \delta  \in disc \\ \Gamma \end{array}}
\frac{1}{\sqrt{1+\delta_{\gamma,\delta}}}
\hat{J}\left\{\begin{array}{ccc} I_{t}&J_{2p}&J\\J&0&J\\ \Gamma & \Gamma & 0 \end{array}\right\}  
 \langle t||[\tilde{a_{\gamma}}\tilde{a_{\delta}}]^{\Gamma}||\Phi_i^{A}\rangle \nonumber\\ 
&&\times \langle K,(l_{x},l_{y}),L,S;J_{2p},\rho|| {\hat q}_{\bar{\alpha}}\otimes {\hat q}_{\bar{\beta}}
||V||(\gamma, \delta)^{\Gamma} \rangle \label{term4}
\end{eqnarray}
To calculate this term, we insert two completness relations in the harmonic oscillator basis.

\section{Calculation of the function  $\omega^{(+)}$ for the direct 2p emission}\label{Annexe_direct}

The state $|\omega_j^{(+)}\rangle$ which is the continuation of $|\Phi_j^{A}\rangle$ in
${\cal T}$ subspace, is the solution of the inhomogeneous  Schr\"{o}dinger equation (\ref{o_q}).
$|\omega^{(+)}_j\rangle$ can be expanded in the channel representation (in the Jacobi coordinates system {\bf T}) as:
\begin{eqnarray}
|\omega^{(+)}_j\rangle&=&\rho^{5}\sum_{c}\int d\rho  |t,K,(l_{x},l_{y}),L,S,J_{2p};J,\rho\rangle  \langle  t,K,(l_{x},l_{y}),L,S,J_{2p};J,\rho|
\omega^{(+)}_j\rangle  \nonumber \\
&=& \rho^{\frac{5}{2}}\sum_{c}\int d\rho |c,\rho\rangle \omega^{(+)}_{j,c}(\rho)
\label{e1new}
\end{eqnarray}
where $\omega^{(+)}_{j,c}(\rho)$ is defined as:
\begin{eqnarray}
\omega^{(+)}_{j,c}(\rho)=\rho^{5/2}\langle c,\rho|\omega^{(+)}_j\rangle
\end{eqnarray}
Inserting (\ref{e1new}) into the equation (\ref{o_q}), and then projecting on the channel
 $c$, one finds:
\begin{eqnarray}
\rho^{\frac{5}{2}}\langle c,\rho|E-H_{TT}|\sum_{c'} \int d\rho' \rho'^{\frac{5}{2}}|c' \rho'\rangle \omega^{(+)}_{j,c'}(\rho')=w_{j,c}(\rho)  \label{eq_cc}
\end{eqnarray}
where $w_{j,c}(\rho)$  is the projection of the source term  $w_j$ on the channel  $c$ 
(cf appendix \ref{annexe_source_dir}) and $H_{TT}$ is given in (\ref{extra_03}).
 By definition,  $|\omega^{(+)}_j\rangle$ belongs to ${\cal T}$. Hence,  one can rewrite  (\ref{eq_cc}) as:
\begin{eqnarray}
\rho^{\frac{5}{2}}\langle c,\rho|{\hat T}(E-H)|\sum_{c'} \int d\rho' \rho'^{\frac{5}{2}}|c' \rho'\rangle \omega^{(+)}_{j,c'}(\rho')=w_{j,c}(\rho) \label{e4}
\end{eqnarray}
The projection operator  ${\hat T}$ assures that the particles in the continuum do not occupy the (quasi-)bound states of the daughter nucleus. 

In the following discussion of this appendix, we shall leave out ${\hat T}$ and replace it by the identity operator $I_d$. The effect of the projection operator  ${\hat T}$ will be taken into account subsequently by the method described in the appendix \ref{annexe_proj}. Let us  define by $|\omega^{(+),0}_{j}\rangle$ the solution of equation (\ref{e4}) without the projection operator ${\hat T}$ . One obtains:
\begin{eqnarray}
\rho^{\frac{5}{2}}\langle c,\rho|(E-H)|\sum_{c'} \int d\rho' \rho'^{\frac{5}{2}}|c' \rho'\rangle \omega^{(+),0}_{j,c'}(\rho')=w_{j,c}(\rho) \label{e5}
\end{eqnarray}
where $\omega^{(+),0}_{j,c}(\rho)$ is:
\begin{eqnarray}
\omega^{(+),0}_{j,c}(\rho)=\rho^{5/2}\langle c,\rho|\omega^{(+),0}_j\rangle
\end{eqnarray}
%
%
%
\subsection{Calculation of the term $E-{\tilde H}^{(A-2)} -\hat{{\cal K}}$}\label{app_najdl}
Let us calculate the contribution of the term $E-{\tilde H}^{(A-2)} -\hat{{\cal K}}$ in eq. (\ref{e5}). The corresponding matrix element is:
\begin{eqnarray}
&&\langle c \rho|E-{\tilde H}^{(A-2)} -\hat{\cal K}|c' \rho'\rangle \nonumber  \\
&&=\left[(E-E_{t})
+\frac{\hbar^{2}}{2m}\left\{\frac{\partial^{2}}{\partial \rho^{2}}-\frac{(K+3/2)(K+5/2)}{\rho^{2}}\right\}\right] \delta_{c,c'}
\frac{\delta(\rho-\rho')}{\rho'^{5}} \label{nrf}
\end{eqnarray}
where $E_{t}$  is the intrinsic energy of the state  $t$. 
Hence one obtains (cf  eq. (\ref{e5})):
\begin{eqnarray}
&&\rho^{5/2} \langle c \rho|E-{\tilde H}^{(A-2)} -\hat{\cal K}|
\sum_{c'} \int d\rho' \rho'^{\frac{5}{2}}|c' \rho'\rangle \omega^{(+),0}_{j,c'}(\rho') \nonumber  \\
&&=\left[(E-E_{t})
+\frac{\hbar^{2}}{2m}\left\{\frac{\partial^{2}}{\partial \rho^{2}}-\frac{(K+3/2)(K+5/2)}{\rho^{2}}\right\}\right]\omega^{(+),0}_{j,c}(\rho) \nonumber 
 \\ \label{j100}
\end{eqnarray}
%
%
\subsection{Contribution due to the interaction between two protons in the continuum}

To illustrate the calculation of the interaction terms, we shall consider the Gaussian interaction:
$V^{(res)}(i,j)={\bar V}_0 \exp[{ -\beta^{2}({\bf r}_{i}-{\bf r}_{j})^{2}}]$.
In the coordinate system  {\bf T} (cf sect. \ref{para_direct}), the residual interaction between the two protons in the continuum is:
\begin{eqnarray}
V^{(res)}(A-1,A)&=&{\bar V}_0 e^{ -\beta^{2}(x/\sqrt{\mu_x})^{2}}={\bar V}_0 e^{ -\beta^{2}(\rho \cos (\alpha)/\sqrt{\mu_x})^{2}}
\end{eqnarray}
where $\mu_x$ is the dimensionless reduced mass of two protons. Hence, one obtains:
\begin{eqnarray}
\langle c \rho| V^{(res)}(A-1,A)|c' \rho'\rangle 
&=&\langle K,\rho|V^{(res)}(A-1,A)|K',\rho'\rangle 
\delta_{t,t'} \delta_{l_{x},l_{x}'} \delta_{l_{y},l_{y}'} \delta_{L,L'} \delta_{S,S'}
\delta_{J_{2p},J_{2p}'} \nonumber \\ 
&=&\int d\alpha \cos^{2}(\alpha) \sin^{2}(\alpha) \Psi_{K}^{l_{x},l_{y}}(\alpha)
\Psi_{K'}^{l'_{x},l'_{y}}(\alpha)
\left[{\bar V}_0 e^{ -\beta^{2}(\rho \cos (\alpha)/\sqrt{\mu_x})^{2}}\right] 
 \nonumber\\
&\times& \frac{\delta(\rho-\rho')}{\rho^{5}} \delta_{t,t'}\delta_{l_{x},l_{x}'} \delta_{l_{y},l_{y}'} \delta_{L,L'} \delta_{S,S'}
\delta_{J_{2p},J_{2p}'}\equiv
V^{(res)}_{cc'}(\rho)\frac{\delta(\rho-\rho')}{\rho^{5}} \nonumber 
\\
\label{h1}
\end{eqnarray}
The Coulomb interaction between two protons in the continuum  $V^{(C)}_{cc'}(\rho)$ is calculated similarly. Hence,  the final expression is (cf eqs. (\ref{e5}), (\ref{h1})):
\begin{eqnarray}
&& \rho^{5/2} \langle c \rho| V^{(res)}(A-1,A)+V^{(C)}(A-1,A)|\sum_{c'} \int d\rho' \rho'^{\frac{5}{2}}|c' \rho'\rangle \omega^{(+),0}_{j,c'}(\rho')   
\nonumber \\
&=&\sum_{c'}\left[ V^{(res)}_{cc'}(\rho)+ V^{(C)}_{cc'}(\rho)\right]\omega^{(+),0}_{j,c'}(\rho)
 \label{j101}
\end{eqnarray}

\subsection {Contributions from  the one-body potential}

Let us calculate the matrix element of the one-body potential $v_{0}$ between channels $c$ and $c'$ in eq. (\ref{e5}):
\begin{eqnarray}
\langle c \rho|v_{0}(A-1)+v_{0}(A)|c' \rho'\rangle  
\label{d1}
\end{eqnarray}
$v_{0}$ is the finite depth potential of a Woods-Saxon type with the spin-orbit term and the Coulomb potential of the daughter nucleus. Channels are defined in the coordinate system {\bf T}. This system  is not convenient for the calculation of the terms (\ref{d1}) because $v_0$ is a function of variables ${\bf x}_1$ and ${\bf x}_2$ which are associated with the coordinate systems  {\bf Y} (cf sect. \ref{para_direct} and Fig. \ref{jaz_fig}).  The change from the coordinate system {\bf T} to a coordinate system {\bf Y} is done using the Raynal-Revai transformation \cite{raynal}. 

Let us consider the transformation from the coordinate system (3) to the coordinate system (2) (cf Fig. \ref{jaz_fig}). The hyperspherical functions are transformed as follows:
\begin{eqnarray}
\label{extra_04}
{\cal Y}_{KL}^{l_{x_3},l_{y_3}}(\Omega^{3}_5)
=\sum_{l_{x_{2}},l_{y_{2}}} \langle l_{x_{2}} l_{y_{2}}| l_{x_{3}} l_{y_{3}}\rangle_{KL}  
{\cal Y}_{KL}^{l_{x_2},l_{y_2}}(\Omega^{2}_5)
\end{eqnarray}
where $\langle l_{x_{2}} l_{y_{2}}| l_{x_{3}} l_{y_{3}}\rangle_{KL}$  are the Raynal-Revai coefficients. The hypermoment $K$ and the angular momentum  $L$ are conserved by this transformation which
 corresponds to a rotation in the space of Jacobi coordinates. 
The summation over angular momenta $l_{x_2}$  and $l_{y_2}$ associated with directions $x_2$ and $y_2$ in the system {\bf Y} defined by the coordinates  (2) (cf Fig. \ref{jaz_fig}), is constrained as follows:
\begin{eqnarray}
&&|l_{x_2}-l_{y_2}|\leq L \leq l_{x_2}+l_{y_2}  \\    
&& K=2n+l_{x_2}+l_{y_2} \nonumber
\end{eqnarray}
where $n\geq 0$ is an integer number. Analytical expression for the Raynal-Revai coefficients can be found in \cite{raynal}. Hence, one obtains:
\begin{eqnarray}
&&\langle c \rho|v_{0}(A-1)|c' \rho'\rangle \nonumber \\ 
&&=\langle t,K,(l_{x},l_{y})L,S,J_{2p};J \rho|v_0(A-1)
|t',K',(l_{x}',l_{y}')L',S',J_{2p}';J\rho'\rangle \nonumber \\
&&=\sum_{l_{x_{2}},l_{y_{2}},l_{x_{2}}',l_{y_{2}}'}
\langle l_{x_{2}},l_{y_{2}} |l_{x},l_{y}\rangle_{KL}
\langle l_{x_{2}}',l_{y_{2}}'|l_{x}',l_{y}'\rangle_{K'L'}  \delta_{t,t'}\delta_{J_{2p},J_{2p}'}
 \nonumber \\
&&\times  \langle K,(l_{x_{2}},l_{y_{2}})L,S,J_{2p};J\rho|v_0 (A-1)
|K',(l_{x_{2}}',l_{y_{2}}')L',S',J_{2p}';J\rho' \rangle  
\label{coupl_WS2}
\end{eqnarray}
Performing the transformation from  $L-S$ to $j-j$ coupling scheme, one obtains for the last term in eq. (\ref{coupl_WS2}):
\begin{eqnarray}
&&\langle K,(l_{x_{2}},l_{y_{2}})L,S,J_{2p};J\rho|v_0(A-1)
|K',(l_{x_{2}}',l_{y_{2}}')L',S',J_{2p}';J\rho'\rangle \nonumber \\
&&=\sum_{j_{x_{2}},j_{y_{2}},j_{x_{2}'},j_{y_{2}}'}\hat{L}\hat{S}\hat{j_{x_{2}}}\hat{j_{y_{2}}}
\hat{L'}\hat{S'}\hat{j_{x_{2}}'}\hat{j_{y_{2}}'}
\left\{\begin{array}{ccc} l_{x_{2}} & l_{y_{2}}&L\\s_{1}&s_{2}&S\\ j_{x_{2}} &j_{y_{2}}&J_{2p}  \end{array}\right\} \nonumber \\
&&\times \left\{\begin{array}{ccc} l_{x_{2}}' & l_{y_{2}}'&L'\\s_{1}&s_{2}&S'\\ j_{x_{2}}' &j_{y_{2}}'&J_{2p}'  \end{array}\right\}
\langle K,(j_{x_{2}},j_{y_{2}}),J_{2p};\rho|v_0(A-1)|K',(j_{x_{2}}',j_{y_{2}}'),J_{2p}';\rho'\rangle ~~\label{last}
\end{eqnarray}
where: 
\begin{eqnarray}
&&\langle K,(j_{x_{2}},j_{y_{2}}),J_{2p};\rho|v_0(A-1)|K',(j_{x_{2}}',j_{y_{2}}'),J_{2p}';\rho'\rangle \nonumber  \\
&&=\int d\alpha \cos^{2}(\alpha) \sin^{2}(\alpha) \Psi_{K}^{l_{x_2},l_{y_2}}(\alpha)
\Psi_{K'}^{l_{x_2}',l_{y_2}'}(\alpha)v^{l_{x_2},j_{x_2}}_0(\rho \cos (\alpha)/\sqrt{\mu_x}) 
\nonumber \\ 
&&\times \frac{\delta(\rho-\rho')}{\rho^{5}} \delta_{l_{x_2},l_{x_2}'}
\delta_{j_{x_2},j_{x_2}'} \delta_{l_{y_2},l_{y_2}'}
\delta_{j_{y_2},j_{y_2}'} \equiv V_{cc'}^{v_0(A-1)}(\rho)\frac{\delta(\rho-\rho')}{\rho^{5}}
 \label{coupl_WS3} 
 \end{eqnarray}
In this expression $\mu_x$ is the dimensionless reduced mass of a system [A-2]$\otimes$p and
$v_{0}^{l_{x_2},j_{x_2}}(\rho \cos (\alpha)/\sqrt{\mu_x})$ is the radial part of the one-body potential $v_0$ for angular momenta $l_{x_2}$, $j_{x_2}$.
 Similarly, one proceeds to calculate the term for a second proton, and eq. (\ref{d1}) takes a form:
\begin{eqnarray}
\langle c \rho|v_{0}(A-1)+v_{0}(A)|c' \rho'\rangle \equiv \left\{V_{cc'}^{v_0}(\rho)\right\}
\frac{\delta(\rho-\rho')}{\rho^{5}} 
\label{f3}
\end{eqnarray}
Hence, the total contribution from the one-body potentials becomes (cf eqs. (\ref{e5}), (\ref{f3})):
\begin{eqnarray}
&&\rho^{\frac{5}{2}}\langle c,\rho|\left\{ v_{0}(A-1) + v_{0}(A)\right\}|
\sum_{c'} \int d\rho' \rho'^{\frac{5}{2}}|c' \rho'\rangle \omega^{(+),0}_{j,c'}(\rho')    \nonumber \\ 
&&=\sum_{c'}\left\{V_{cc'}^{v_0}(\rho)  \right\} \omega^{(+),0}_{j,c'}(\rho)
 \nonumber \\
\label{kin4}
\end{eqnarray}
\subsection {Contributions due to the interaction between protons in the continuum and nucleons in the daughter nucleus}

Let us calculate now the contribution from the residual two-body interaction between protons in the 
continuum and nucleons in the nucleus A-2 (cf eq. (\ref{e5})):
\begin{eqnarray}
\rho^{\frac{5}{2}}\langle c,\rho|
 \sum_{i\leq j}^{j\geq A-1}V^{(res)}(i,j)|\sum_{c'} \int d\rho' \rho'^{\frac{5}{2}}|c' 
\rho'\rangle \omega^{(+),0}_{j,c'}(\rho')  \label{cont_res}
\end{eqnarray}
In the second quantization form, this term is:
\begin{eqnarray}
&&\rho^{\frac{5}{2}}\langle c,\rho| 
 \sum_{i\leq j}^{j\geq A-1}V^{(res)}(i,j)|\sum_{c'} \int d\rho' \rho'^{\frac{5}{2}}|c' 
\rho'\rangle \omega^{(+),0}_{j,c'}(\rho') \nonumber \\
\label{term1}  
 \\ 
&=& \rho^{\frac{5}{2}}\langle c,\rho|-\sum_{\tiny\begin{array}{c} (\alpha \in disc \leq \beta \in cont) \\ (\gamma\in disc \leq \delta \in cont) \\ \Gamma  \end{array}}
 V_{\alpha,\beta,\gamma,\delta}^{\Gamma} \left\{[a^{\dagger}_{\alpha}a^{\dagger}_{\beta}]^{\Gamma}[\tilde{a}_{\gamma} \tilde{a}_{\delta}]^{\Gamma}\right\}^{0}|\sum_{c'} \int d\rho' \rho'^{\frac{5}{2}}|c' 
\rho'\rangle \omega^{(+),0}_{j,c'}(\rho')  \nonumber \\ 
\label{x_0}
\end{eqnarray}  
where $V_{\alpha,\beta,\gamma,\delta}^{\Gamma}$ is the reduced matrix element coupled to $\Gamma$ of the residual interaction. 
Here we take into account the projection operator ${\hat T}$ by  constraining summation over s.p. states  $\alpha$, $\beta$, $\gamma$, $\delta$:
\begin{eqnarray}
&&\alpha \in disc \leq \beta \in cont \nonumber \\
&&\gamma\in disc \leq \delta \in cont \nonumber
\end{eqnarray}   
The creation and annihilation operators are coupled in such a way that the operators acting on (quasi-)bound states are separated from those acting on the continuum states. One obtains:
\begin{eqnarray}
\left\{[a^{\dagger}_{\alpha}a^{\dagger}_{\beta}]^{\Gamma}[\tilde{a}_{\gamma} \tilde{a}_{\delta}]^{\Gamma}\right\}^{0}
=-\sum_{J'}\hat{J'}^{2}\hat{\Gamma}^{2}\left\{\begin{array}{ccc} j_{\alpha} & j_{\beta}& \Gamma \\j_{\gamma}&j_{\delta}& \Gamma\\  J'&J'&0  \end{array}\right\}
\left\{[a^{\dagger}_{\alpha} \tilde{a_{\gamma}}]^{J'}
[a^{\dagger}_{\beta} \tilde{a_{\delta}}]^{J'}\right\}^{0} 
\end{eqnarray} 
Inserting this expression in eq. (\ref{x_0}) and applying the Wigner-Eckart theorem, one gets:
\begin{eqnarray}
&&\hat{J} \rho^{5/2} \sum_{\tiny\begin{array}{c} (\alpha \in disc \leq \beta \in cont) \\ (\gamma\in disc \leq \delta \in cont) \\ \Gamma ,J' \end{array}} V_{\alpha,\beta,\gamma,\delta}^{\Gamma} 
\hat{\Gamma}^{2}\hat{J'}^{2}
\left\{\begin{array}{ccc} j_{\alpha} & j_{\beta}& \Gamma \\j_{\gamma}&j_{\delta}& \Gamma\\  J'&J'&0  \end{array}\right\} 
\left\{\begin{array}{ccc} I_{t} & J_{2p}& J \\I_{t'} & J_{2p}' & J \\  
J'&J'&0  \end{array}\right\}  
\langle t||(a^{\dagger}_{\alpha} \tilde{a_{\gamma}})^{J'}||t'\rangle  \nonumber \\ 
&&\times
\langle K,(l_{x},l_{y})L,S,J_{2p}, \rho||(a^{\dagger}_{\beta} \tilde{a_{\delta}})^{J'}||\sum_{c'} \int d\rho' \rho'^{\frac{5}{2}}|K',(l_{x}',l_{y}')L',S',J_{2p}', \rho'\rangle 
\omega^{(+),0}_{j,c'}(\rho') \nonumber \\
\label{DCDC}
\end{eqnarray}   
Let us consider the operator:
\begin{eqnarray}
\sum_{\tiny\begin{array}{c} (\alpha \in disc, \beta \in cont)
 \\ (\gamma\in disc, \delta \in cont) \\ \Gamma \end{array}}
 V_{\alpha,\beta,\gamma,\delta}^{\Gamma}(a^{\dagger}_{\beta} \tilde{a_{\delta}})^{J'}_{M'}
\label{z_0}
\end{eqnarray}
in (\ref{DCDC}). Inserting the completness relation twice, one obtains:
\begin{eqnarray}
&&\sum_{\tiny\begin{array}{c} (\alpha \in disc, \beta \in cont)
 \\ (\gamma\in disc, \delta \in cont) \\ \Gamma \end{array}}
 V_{\alpha,\beta,\gamma,\delta}^{\Gamma}(a^{\dagger}_{\beta} \tilde{a_{\delta}})^{J'}_{M'}=
 \nonumber \\ 
&&\sum_{\tiny\begin{array}{c} (\alpha \in disc, e_\beta >0)
 \\ (\gamma\in disc, e_\delta>0) \\ \Gamma \end{array}}
 V_{\alpha,\beta,\gamma,\delta}^{\Gamma} 
\sum_{j_{\beta},j_2,j_{\delta},j'_2,J_1,J_2,M_1,M_2}
 \int dr_1 dr_2 dr_1' dr_2' r_1^{2}{r'_1}^{2} r_2^{2}{r'_2}^{2}  |j_{\beta},r_1,j_2,r_2,J_1,M_1\rangle \nonumber \\
&&\times \langle j_{\beta},r_1,j_2,r_2,J_1,M_1|
(a^{\dagger}_{\beta} \tilde{a_{\delta}})^{J'}_{M'}
|j_{\delta},r'_1,j'_2,r'_2,J_2,M_2\rangle  \langle j_{\delta},r'_1,j'_2,r'_2,J_2,M_2| 
\label{z_5}
\end{eqnarray}
where an index $j$ stands for the quantum numbers $l$ and $j$ of a given proton , {\it e.g.} $j_2$ denotes quantum numbers $l_2$, $j_2$. $e_{\beta}$ and $e_{\delta}$ are the single particle energies associated with the states $\beta$ and $\delta$.
 To simplify the demonstration, let us consider the case of two particles in the continuum which do not correspond to the same ensemble of quantum numbers $j$, {\it i.e.} $j_{\beta}$ differs from $j_2$ and, similarly, $j_{\delta}$ differs from $j'_2$. Applying the Wigner-Eckart theorem in (\ref{z_5}), one finds:
\begin{eqnarray}
&&\sum_{\tiny\begin{array}{c} (\alpha \in disc, e_\beta >0)
 \\ (\gamma\in disc, e_\delta>0 ) \\ \Gamma \end{array}}
 V_{\alpha,\beta,\gamma,\delta}^{\Gamma}
\sum_{j_{\beta},j_2,j_{\delta},j'_2,J_1,J_2,M_1,M_2}
 \int dr_1 dr_2 dr_1' dr_2' r_1^{2}{r'_1}^{2} r_2^{2}{r'_2}^{2}
|j_{\beta},r_1,j_2,r_2,J_1,M_1\rangle \nonumber \\
&&\times \langle J_2 M_2 J'M' |J_1 M_1\rangle (\hat{J'})^{2} \hat{J_2}\hat{j_2}
\left\{\begin{array}{ccc} j_{\beta} & j_2& J_1 \\j_{\gamma}&j'_2& J_2\\  J'&0 &J'
  \end{array}\right\} 
\frac{u_{\beta}(r_1)}{r_1} \frac{u_{\delta}(r'_1)}{r'_1}  
\delta_{j_2,j'_2} \frac{\delta(r_2-r'_2)}{r_2^{2}} 
\langle j_{\delta},r'_1,j'_2,r'_2,J_2,M_2| \nonumber \\ 
\label{f4}
\end{eqnarray}
where  $u_{\beta}(r_1)$ and $u_{\delta}(r'_1)$ are the radial wave functions of states $\beta$ and $\delta$, respectively. 

We shall sum now over energies of the states $\beta$ and $\delta$ in (\ref{f4}). The term which  depends on these energies can be written as:
\begin{eqnarray}
&&\sum_{e_\beta >0, e_\delta >0} \int dr_a dr_b dr_1 dr_1' r_1 r'_1
 u_{\alpha}(r_a)u_{\beta}(r_b)u_{\beta}(r_1) u_{\delta}(r'_1) 
\nonumber \\
&&\times \left[v^{\Gamma}_{\bar{\alpha},\bar{\beta},\bar{\gamma},\bar{\delta}}(r_a,r_b)u_{\gamma}(r_a)
u_{\delta}(r_b) -(-1)^{\phi}
v^{\Gamma}_{\bar{\alpha},\bar{\beta},\bar{\delta},\bar{\gamma}}(r_a,r_b)u_{\gamma}(r_b)u_{\delta}(r_a)\right] \label{f5}
\end{eqnarray}
where $v^{\Gamma}_{\bar{\alpha},\bar{\beta},\bar{\gamma},\bar{\delta}}(r_a,r_b)$ are the angle integrated, unsymmetrized  matrix elements of the residual interaction, and the phase $\phi$ equals:
$\phi=j_{\gamma}+j_{\delta}-\Gamma$.  
Using the completness relation for s.p. states:
\begin{eqnarray}
\sum_{n} u_{e_n,l,j,\tau_{z}}(r) u_{e_n,l,j,\tau_{z}}(r')
+ \int_{0}^{+\infty} de ~u_{e,l,j,\tau_{z}}(r) u_{e,l,j,\tau_{z}}(r')=\delta(r-r') 
\label{zcompl}
\end{eqnarray}
one finds for the first term of (\ref{f5}):
\begin{eqnarray}
&& \int dr_a dr_b dr_1 dr'_1 r_1 r'_1
 u_{\alpha}(r_a) v^{\Gamma}_{\bar{\alpha},\bar{\beta},\bar{\gamma},\bar{\delta}}(r_a,r_b)u_{\gamma}(r_a)
[ \delta(r_b-r_1)
 -\sum_{n_{\beta}}u_{n_{\beta}}(r_b)  u_{n_{\beta}}(r_1)]\nonumber \\
&&\times[ \delta(r_b-r'_1) -\sum_{n_{\delta}} u_{n_{\delta}}(r_b) 
u_{n_{\delta}}(r'_1)] \label{f6}
\end{eqnarray}
where $u_{n_{\beta}}(r_b)$  and $u_{n_{\delta}}(r_b)$ are the radial functions of proton (quasi-)bound states with quantum numbers $j_{\beta}\equiv[l_{\beta},j_{\beta}]$ and  $j_{\delta}\equiv[l_{\delta},j_{\delta}]$, respectively.
Similarly, the second term in (\ref{f5}) becomes:
\begin{eqnarray}
-(-1)^{\phi}&&\int dr_a dr_b dr_1 dr'_1 r_1 r'_1
 u_{\alpha}(r_a) v^{\Gamma}_{\bar{\alpha},\bar{\beta},\bar{\delta},\bar{\gamma}}(r_a,r_b)u_{\gamma}(r_b)
[ \delta(r_b-r_1)
 -\sum_{n_{\beta}}u_{n_{\beta}}(r_b)  u_{n_{\beta}}(r_1)]\nonumber \\
&&\times[ \delta(r_a-r'_1) -\sum_{n_{\delta}} u_{n_{\delta}}(r_a) 
u_{n_{\delta}}(r'_1)] \label{f7}
\end{eqnarray}
Using expressions (\ref{f4}), (\ref{f5}), (\ref{f6}), and (\ref{f7}), the operator (\ref{z_0}) can be written as: 
\begin{eqnarray}
&&\sum_{\tiny\begin{array}{c} (\alpha \in disc, \beta \in cont)
 \\ (\gamma\in disc, \delta \in cont) \\ \Gamma \end{array}}
 V_{\alpha,\beta,\gamma,\delta}^{\Gamma}(a^{\dagger}_{\beta} \tilde{a_{\delta}})^{J'}_{M'}=
\nonumber \\
&&
=\sum_{\tiny\begin{array}{c} (\alpha \in disc )
 \\ (\gamma\in disc) \\ \Gamma \end{array}} 
\sum_{j_{\beta},j_2,j_{\delta},J_1,J_2,M_1,M_2}
\hat{\Gamma}(\hat{J'})^{2}\hat{J_2}\hat{j_2}\langle J_2 M_2 J'M' |J_1 M_1\rangle 
\left\{\begin{array}{ccc} j_{\beta} & j_2& J_1 \\j_{\gamma}&j_2& J_2\\  J'&0 &J'
\end{array}\right\}
 \nonumber \\
&& \int dr_a dr_b dr_1 dr'_1 r_1 r'_1 dr_2 r_2^{2}
\times |j_{\beta},r_1,j_2,r_2,J_1,M_1\rangle  \langle j_{\delta},r'_1,j_2,r_2,J_2,M_2| 
u_{\alpha}(r_a)
 \nonumber \\   
&&\times \left\{u_{\gamma}(r_a) v^{\Gamma}_{\bar{\alpha},\bar{\beta},\bar{\gamma},\bar{\delta}}(r_a,r_b)
[ \delta(r_b-r_1)
 -\sum_{n_{\beta}}u_{n_{\beta}}(r_b)  u_{n_{\beta}}(r_1)]
[ \delta(r_b-r'_1) -\sum_{n_{\delta}} u_{n_{\delta}}(r_b) 
u_{n_{\delta}}(r'_1)] \right.\nonumber \\ 
&& \left.  -(-1)^{\phi} u_{\gamma}(r_b) v^{\Gamma}_{\bar{\alpha},\bar{\beta},\bar{\delta},\bar{\gamma}}(r_a,r_b)
[ \delta(r_b-r_1)
 -\sum_{n_{\beta}}u_{n_{\beta}}(r_b)  u_{n_{\beta}}(r_1)]
[ \delta(r_a-r'_1) -\sum_{n_{\delta}} u_{n_{\delta}}(r_a) 
u_{n_{\delta}}(r'_1)]\right\} \nonumber \\
 \label{A_10}
\end{eqnarray}
or as:
\begin{eqnarray}
&&\sum_{\tiny\begin{array}{c} (\alpha \in disc)
 \\ (\gamma\in disc) \\ \Gamma \end{array}} 
\sum_{j_\beta,j_2,j_{\delta},J_1,J_2,M_1,M_2}
\hat{\Gamma}(\hat{J'})^{2}\hat{J_2}\hat{j_2}\langle J_2 M_2 J'M' |J_1 M_1\rangle 
\left\{\begin{array}{ccc} j_{\beta} & j_2& J_1 \\j_{\gamma}&j_2& J_2\\  J'&0 &J'
\end{array}\right\} \int dr_2 r_2^{2} dr_1 dr'_1 r_1 r'_1  \nonumber \\
&&|j_{\beta},r_1,j_2,r_2,J_1,M_1\rangle  f(r_1,r_1')
\langle j_{\delta},r_1',j_2,r_2,J_2,M_2| 
\label{f26ww}
\end{eqnarray}
The form of the operator $f(r_1,r_1')$  in the above expression 
can be found easily from  (\ref{A_10}). 

Let us now calculate the operator  (\ref{A_10}) for antisymmetrized two-particle states $|a b\rangle$ and 
$|c d\rangle $:
\begin{eqnarray}
 \label{zarx}
&&\langle a b| \sum_{\tiny\begin{array}{c} (\alpha \in disc)
 \\ (\gamma\in disc) \\ \Gamma \end{array}} 
\sum_{j_{\beta}j_2,j_{\delta}J_1,J_2,M_1,M_2}
\hat{\Gamma}(\hat{J'})^{2}\hat{J_2}\hat{j_2}\langle J_2 M_2 J'M' |J_1 M_1\rangle 
\left\{\begin{array}{ccc} j_{\beta} & j_2& J_1 \\j_{\gamma}&j_2& J_2\\  J'&0 &J'
\end{array}\right\} \int dr_2 r_2^{2}dr_1 dr'_1 r_1 r'_1 \nonumber \\
&& \times |j_{\beta},r_1,j_2,r_2,J_1,M_1\rangle  f(r_1,r_1')
\langle j_{\delta},r_1',j_2,r_2,J_2,M_2|cd\rangle
 \nonumber \\ 
&&= \sum_{\tiny\begin{array}{c} (\alpha \in disc)
 \\ (\gamma\in disc) \\ \Gamma \end{array}} 
\sum_{j_{\beta},j_2,j_{\delta},J_1,J_2,M_1,M_2}
\hat{\Gamma}(\hat{J'})^{2}\hat{J_2}\hat{j_2}\langle J_2 M_2 J'M' |J_1 M_1\rangle 
\left\{\begin{array}{ccc} j_{\beta} & j_2& J_1 \\j_{\gamma}&j_2& J_2\\  J'&0 &J'
\end{array} \right\}
\int dr_2 r_2^{2} dr_1 dr'_1 r_1 r'_1\nonumber \\
&& \times \langle a b|j_{\beta},r_1,j_2,r_2,J_1,M_1\rangle_{(nas)}~2f(r_1,r_1')~
 _{(nas)}\langle j_{\delta},r_1',j_2,r_2,J_2,M_2| cd\rangle
\end{eqnarray}
where $'nas'$ means that the considered state is {\em not} antisymmetrized. Formally, above expression suggests that the operator (\ref{z_0}) acts only on the first proton  in different unsymmetrized states referenced by  $r_1$ and $r_1'$. Indeed, the particle in a state $|j_2,r_2\rangle$ enters in the calculation only through the angular momentum coupling. \\
\\
To calculate the last term in (\ref{DCDC}), one goes from L-S to j-j coupling scheme and then applies the Raynal-Revai transformation to obtain:
\begin{eqnarray}
&&\langle K,(l_{x},l_{y})L,S,J_{2p}, \rho||(a^{\dagger}_{\beta} \tilde{a_{\delta}})^{J'}||K',(l_{x}',l_{y}')L',S',J_{2p}',\rho'\rangle  \nonumber \\ 
%
&&=\sum_{l_{x_{2}},l_{y_{2}},l_{x_{2}}',l_{y_{2}}'}
\langle l_{x_{2}},l_{y_{2}}|l_{x},l_{y}\rangle_{KL}  
\langle l_{x_{2}}',l_{y_{2}}'|l_{x}',l_{y}'\rangle_{K'L'} 
\sum_{j_{x_{2}},j_{y_{2}},j_{x_{2}}',j_{y_{2}}'}\hat{L}\hat{S}\hat{j_{x_{2}}}\hat{j_{y_{2}}}
\hat{L'}\hat{S'}\hat{j_{x_{2}}'}\hat{j_{y_{2}}'}  \nonumber \\ 
&&\times \left\{\begin{array}{ccc} l_{x_{2}} & l_{y_{2}} & L\\ s_{1} &s_{2} & S \\ j_{x_{2}} &j_{y_{2}}&J_{2p}  \end{array}\right\} 
\left\{\begin{array}{ccc} l_{x_{2}}' & l_{y_{2}}'&L'\\s_{1}&s_{2}&S'\\ j_{x_{2}}' &j_{y_{2}}'&J_{2p}'  \end{array}\right\}  
\langle K,(j_{x_{2}},j_{y_{2}}),J_{2p},\rho||(a^{\dagger}_{\beta} \tilde{a_{\delta}})^{J'}||K',(j_{x_{2}}',j_{y_{2}}'),J_{2p}',\rho'\rangle 
\nonumber \\
\label{dd3}
\end{eqnarray}   
As discussed above (cf expression (\ref{zarx})), one may consider that only a proton labeled by 
$j_{x_2}$ on the l.h.s., and a proton labeled by  $j_{x_2}'$ on the r.h.s. enter in this calculation. Hence, the reduced matrix element becomes:
\begin{eqnarray}
&& \langle K,(j_{x_{2}},j_{y_{2}}),J_{2p},\rho||(a^{\dagger}_{\beta} \tilde{a_{\delta}})^{J'}||K',(j_{x_{2}}',j_{y_{2}}'),J_{2p}',\rho'\rangle  \nonumber \\ 
&&=\hat{J}_{2p}\hat{J'}_{2p}(\hat{J'})^{2}\hat{j_{y_2}}
\left\{\begin{array}{ccc} j_{x_{2}} & j_{y_{2}}&J_{2p}\\ j_{x_2}' & j_{y_{2}}'&J_{2p}'\\ J' & 0 &J'  \end{array}\right\}
\int d\alpha 
d\alpha' \cos^{2}(\alpha)\sin^{2}(\alpha)\cos^{2}(\alpha')\sin^{2}(\alpha') 
\nonumber \\ 
&&\times \Psi_{K}^{l_{x_{2}},l_{y_{2}}}(\alpha) \Psi_{K'}^{l_{x_{2}}',l_{y_{2}}'}(\alpha')  \ \frac{u_{\beta}(\rho \cos\alpha/\sqrt{\mu_{x}} )}{\rho\cos\alpha/\sqrt{\mu_{x}}}  
  \frac{u_{\delta}(\rho' \cos \alpha'/\sqrt{\mu_{x}} )}{\rho' \cos\alpha'/\sqrt{\mu_{x}}}
\frac{\delta(\rho \sin \alpha-\rho' \sin \alpha')}{( \rho \sin \alpha)^{2} } 
\nonumber \\
&&\times \delta_{l_{x_{2}},l_{\beta}} \delta_{j_{x_{2}},j_{\beta}} 
\delta_{l_{x_{2}}',l_{\delta}} \delta_{j_{x_{2}}',j_{\delta}} 
\delta_{l_{y_{2}},l_{y_{2}}'} \delta_{j_{y_{2}},j_{y_{2}}'} \nonumber \\
\end{eqnarray}  
Using this last expression and (\ref{dd3}) one can write (\ref{DCDC}) as:
\begin{eqnarray}
&&-\rho^{5/2} \sum_{\tiny \begin{array}{c} c',(\alpha,\gamma)\in disc ,\Gamma,J' \\
  l_{y_{2}},j_{y_2},l_{\beta},j_{\beta},l_{\delta}, j_{\delta}\end{array}} 
\int_{0}^{+\infty} d_{e_{\beta}} d_{e_{\delta}} 
 \Big\{ A(\Gamma,\alpha,\beta,\gamma,\delta,l_{y_{2}},\rho)+B(\Gamma,\alpha,\beta,\gamma,\delta,l_{y_{2}},\rho)\Big\}
\nonumber \\ 
&&\times\hat{\Gamma}^{2}\hat{J'}^{2}(-1)^{\phi'}
\left\{\begin{array}{ccc} j_{\alpha} & j_{\gamma}& J' \\j_{\delta}&j_{\beta}& \Gamma \end{array}\right\} 
\left\{\begin{array}{ccc} I_{t} & I_{t}' & J' \\J_{2p}' & J_{2p} & J \end{array}\right\}  
\langle t||(a^{\dagger}_{\alpha} \tilde{a}_\gamma)^{J'}||t'\rangle  \nonumber \\   \nonumber \\ 
&&\times \langle l_{x_{2}},l_{y_{2}}|l_{x},l_{y}\rangle_{KL} 
\langle l_{x_{2}}',l_{y_{2}}'|l_{x}',l_{y}'\rangle_{K'L'} 
\hat{L}\hat{S}\hat{j_{\beta}}\hat{j_{\delta}}
\hat{L'}\hat{S'}\hat{j_{y_{2}}}^{2} \hat{J_{2p}} \hat{J_{2p}'} \nonumber \\  \nonumber \\ 
&&\times \left\{\begin{array}{ccc} l_{\beta} & l_{y_{2}}&L\\s_{1}&s_{2}&S\\ j_{\beta} &j_{y_{2}}&J_{2p}  \end{array}\right\} 
\left\{\begin{array}{ccc} l_{\delta} & l_{y_{2}}&L'\\s_{1}&s_{2}&S'\\ j_{\delta} &j_{y_{2}}&J_{2p}'  \end{array}\right\}
\left\{\begin{array}{ccc} J_{2p} & J_{2p} ' & J' \\j_{\delta}&j_{\beta}&j_{y_2} \end{array}\right\} 
\nonumber \\ 
\label{bb0}
\end{eqnarray}
where $\phi'=1+j_{\gamma}+j_{y_2}+\Gamma+J_{2p}+J_{2p}'+J+J'+I_{t}'$. \\
\\
$A(\Gamma,\alpha,\beta,\gamma,\delta,l_{y_{2}},\rho)$ in (\ref{bb0}) is equal to: 
\begin{eqnarray}
&&A(\Gamma,\alpha,\beta,\gamma,\delta,l_{y_{2}},\rho)  
=\int dr_{a}dr_{b}u_{\alpha}(r_{a})u_{\beta}(r_{b})v^{\Gamma}_{\bar{\alpha}\bar{\beta}
\bar{\gamma}\bar{\delta}}(r_{a},r_{b})u_{\gamma}(r_{a})u_{\delta}(r_{b}) \nonumber \\ 
&&\times  \int d\rho' \rho'^{\frac{5}{2}} d\alpha d\alpha' \cos^{2}\alpha\sin^{2}\alpha\cos^{2}\alpha'\sin^{2}\alpha' 
\Psi_{K}^{l_{\beta},l_{y_{2}}}(\alpha) \Psi_{K'}^{l_{\delta},l_{y_{2}}}(\alpha')  \nonumber \\ %
&&\times \frac{u_{\beta}(\rho \cos\alpha/\sqrt{\mu_{x}} )}{\rho \cos\alpha/\sqrt{\mu_{x}}}
\frac{u_{\delta}(\rho' \cos\alpha'/\sqrt{\mu_{x}} )}{\rho' \cos\alpha'/\sqrt{\mu_{x}}}
\frac{\delta(\rho \sin \alpha-\rho' \sin \alpha')}{( \rho \sin \alpha)^{2} }
\omega^{(+),0}_{j,c'}(\rho')
\end{eqnarray}
\\
and $B(\Gamma,\alpha,\beta,\gamma,\delta,\rho)$ is:
\begin{eqnarray}
&&B(\Gamma,\alpha,\beta,\gamma,\delta,l_{y_{2}},\rho)=
-(-1)^{\Gamma-j_{\gamma}-j_{\delta}}
\int dr_{a}dr_{b}u_{\alpha}(r_{a})u_{\beta}(r_{b})
v^{\Gamma}_{\bar{\alpha}\bar{\beta}
\bar{\delta}\bar{\gamma}}(r_a,r_b)
u_{\gamma}(r_{b})u_{\delta}(r_{a}) \nonumber \\
&&\times  \int d\rho' \rho'^{\frac{5}{2}} d\alpha d\alpha' \cos^{2}\alpha\sin^{2}\alpha\cos^{2}\alpha'\sin^{2}\alpha' 
\Psi_{K}^{l_{\beta},l_{y_{2}}}(\alpha) \Psi_{K'}^{l_{\delta},l_{y_{2}}}(\alpha')  \nonumber \\ %
&&\times \frac{u_{\beta}(\rho \cos\alpha/\sqrt{\mu_{x}} )}{\rho \cos\alpha/\sqrt{\mu_{x}}}
\frac{u_{\delta}(\rho' \cos\alpha'/\sqrt{\mu_{x}} )}{\rho' \cos\alpha'/\sqrt{\mu_{x}}}
\frac{\delta(\rho \sin \alpha-\rho' \sin \alpha')}{( \rho \sin \alpha)^{2}}
\omega^{(+),0}_{j,c'}(\rho') 
\end{eqnarray}
\\
Using the completness relation for s.p. states and integrating over energies $e_{\beta}$ and  $e_{\delta}$, the first term in (\ref{bb0})  becomes: 
\begin{eqnarray}
&&\rho^{5/2}\int_{0}^{+\infty} d_{e_{\beta}} d_{e_{\delta}} 
 A(\Gamma,\alpha,\beta,\gamma,\delta,l_{y_{2}},\rho)  \nonumber \\
&&=\rho^{5/2} \int d\rho' \rho'^{\frac{5}{2}} dr_{a}dr_{b} d\alpha d\alpha' \cos^{2}\alpha\sin^{2}\alpha\cos^{2}\alpha'\sin^{2}\alpha'  \nonumber \\ 
&&\times u_{\alpha}(r_{a})v_{\bar{\alpha}\bar{\beta}
\bar{\gamma}\bar{\delta}}^{\Gamma}(r_{a},r_{b})u_{\gamma}(r_{a})
\Psi_{K}^{l_{\beta},l_{y_{2}}}(\alpha) \Psi_{K'}^{l_{\delta},l_{y_{2}}}(\alpha') 
\nonumber  \\ 
&&\times\frac{\delta(\rho \sin \alpha-\rho' \sin \alpha')}{( \rho \sin \alpha)^{2} }
 \left[ \frac{\delta(r_{b}-\rho \cos \alpha/\sqrt{\mu_{x}})}{r_{b}}- \frac{u_{n_{\beta}}(r_{b})u_{n_{\beta}}(\rho\cos \alpha/\sqrt{\mu_{x}})}{(\rho\cos \alpha/\sqrt{\mu_{x}})}\right]  \nonumber \\ 
&&\times\left[ \frac{\delta(r_{b}-\rho' \cos \alpha'/\sqrt{\mu_{x}})}{r_{b}}- \frac{u_{n_{\delta}}(r_{b})u_{n_{\delta}}(\rho'\cos \alpha'/\sqrt{\mu_{x}})}{(\rho'\cos \alpha'/\sqrt{\mu_{x}})}\right] \omega^{(+),0}_{j,c'}(\rho') 
\label{bb1}
\end{eqnarray}
\\
The role of terms:
$$u_{n_{\beta}}(r_{b})u_{n_{\beta}}(\rho\cos \alpha/\sqrt{\mu_{x}})/(\rho\cos \alpha/\sqrt{\mu_{x}})$$ 
and
$$u_{n_{\delta}}(r_{b})u_{n_{\delta}}(\rho'\cos \alpha'/\sqrt{\mu_{x}})/(\rho'\cos \alpha'/\sqrt{\mu_{x}})$$ 
in (\ref{bb1}) is to project on the (quasi-)bound s.p. states. Their outcome 
will be taken into account by the method described in appendix \ref{annexe_proj} and, 
at this point, we remove them. Integrating over $r_b$, $\rho'$ and $\alpha_2$, one can rewrite (\ref{bb1}) as follows:
\begin{eqnarray}
\int d\alpha \cos^{2}\alpha\sin^{2}\alpha
\Psi_{K}^{l_{\beta},l_{y_{2}}}(\alpha) 
\int dr_a u_{\alpha}(r_{a})u_{\gamma}(r_{a})
v_{\bar{\alpha}\bar{\beta}\bar{\gamma}\bar{\delta}}^{\Gamma}(r_{a},\rho \cos \alpha/\sqrt{\mu_{x}}))
\Psi_{K'}^{l_{\delta},l_{y_{2}}}(\alpha) \omega^{(+),0}_{j,c'}(\rho) \nonumber \\
\label{j_999}
\end{eqnarray}

Similarly, using the completness relation for s.p. states and integrating over energies  $e_{\beta}$ and  $e_{\delta}$, the second term in  (\ref{bb0}) takes the form: 
\begin{eqnarray}
&&\rho^{5/2}\int_{0}^{+\infty} d_{e_{\beta}} d_{e_{\delta}} 
B(\Gamma,\alpha,\beta,\gamma,\delta,l_{y_{2}},\rho)  \nonumber \\ 
&&=-(-1)^{\Gamma-j_{\gamma}-j_{\delta}}\rho^{5/2}\int d\rho' \rho'^{\frac{5}{2}} dr_{a}dr_{b} d\alpha d\alpha' \cos^{2}\alpha\sin^{2}\alpha\cos^{2}\alpha'\sin^{2}\alpha' \nonumber  \\ 
&&\times u_{\alpha}(r_{a})v_{\bar{\alpha}\bar{\beta}\bar{\delta}\bar{\gamma}}^{\Gamma}(r_{a},r_{b})u_{\gamma}(r_{b})
\Psi_{K}^{l_{\beta},l_{y_{2}}}(\alpha) \Psi_{K'}^{l_{\delta},l_{y_{2}}}(\alpha') \nonumber  \\ 
&&\times\frac{\delta(\rho \sin \alpha-\rho' \sin \alpha')}{( \rho \sin \alpha)^{2} }
\times \left[ \frac{\delta(r_{b}-\rho \cos \alpha/\sqrt{\mu_{x}})}{r_{b}}- \frac{u_{n_{\beta}}(r_{b})u_{n_{\beta}}(\rho \cos \alpha/\sqrt{\mu_{x}})}{(\rho \cos \alpha/\sqrt{\mu_{x}})}\right]  \nonumber \\ 
&&\times\left[ \frac{\delta(r_{a}-\rho' \cos \alpha'/\sqrt{\mu_{x}})}{r_{a}}- \frac{u_{n_{\delta}}(r_{a})u_{n_{\delta}}(\rho'\cos \alpha'/\sqrt{\mu_{x}})}{(\rho'\cos \alpha'/\sqrt{\mu_{x}})}\right] \omega^{(+),0}_{j,c'}(\rho') 
\end{eqnarray}
Integrating over  $r_a$, $r_b$ and $\alpha$, and neglecting the projection operators on the (quasi-)bound s.p. states, one obtains:
\begin{eqnarray}
&& 
-(-1)^{\Gamma-j_{\gamma}-j_{\delta}}\int d\rho' \rho'^{\frac{3}{2}} \rho^{-\frac{3}{2}} d\alpha' 
\cos \alpha'\sin^{2}\alpha'
\nonumber  \\ 
&&\times 
u_{\alpha}(\rho' \cos \alpha'/\sqrt{\mu_{x}})
v_{\bar{\alpha}\bar{\beta}\bar{\delta}\bar{\gamma}}^{\Gamma}(\rho' \cos \alpha'/\sqrt{\mu_{x}},\rho \cos \alpha^{0}/\sqrt{\mu_{x}})
u_{\gamma}(\rho \cos \alpha^{0}/\sqrt{\mu_{x}})\nonumber \\
&& 
\times \Psi_{K}^{l_{x_{2}},l_{y_{2}}}(\alpha^{0}) 
\Psi_{K'}^{l_{x_{2}}',l_{y_{2}}}(\alpha')  \omega^{(+),0}_{j,c'}(\rho') 
\label{J_52}
\end{eqnarray}
where: 
\begin{eqnarray*}
\alpha^{0}=\arcsin\left(\frac{\rho' \sin \alpha_2}{\rho}\right) ~ \ .
\end{eqnarray*}

Using expressions  (\ref{j_999}) and (\ref{J_52}), one can calculate the term  (\ref{bb0}). Let us write this term in the form:
\begin{eqnarray}
\sum_{c'}\left[ V^{(loc)}_{cc'}(\rho)+ V^{(nl)}_{cc'}(\rho) \right ]\omega^{(+),0}_{j,c'}(\rho)
 \label{j102}
\end{eqnarray}
where $V^{(loc)}_{cc'}(\rho)$  and $V^{(nl)}_{cc'}(\rho)$  are, respectively, local and non-local potentials. %
Using expressions  (\ref{j100}), (\ref{j101}), (\ref{kin4}), (\ref{bb0})  and (\ref{j102}), one can write eq. (\ref{o_q}) as:
\begin{eqnarray}
&& 
\left[E-E_{t}+\frac{\hbar^{2}}{2m}\left\{\frac{\partial^{2}}{\partial \rho^{2}}
-\frac{(K+3/2)(K+5/2)}{\rho^{2}}\right\}\right]\omega_{j,c}^{(+),0}(\rho) \nonumber \\ 
\nonumber \\ 
&&+\sum_{c'}\left( V^{(res)}_{cc'}(\rho)+ V^{(C)}_{cc'}(\rho)
+ V_{cc'}^{(v_0)}(\rho)
+V^{(loc)}_{cc'}(\rho)+ V^{(nl)}_{cc'}(\rho) \right)\omega^{(+),0}_{j,c'}(\rho)=
w_{j,c}(\rho) 
\nonumber \\
\label{j103}
\end{eqnarray}
or:
\begin{eqnarray}
\sum_{c'} H_{cc'}(\rho)\omega^{(+),0}_{j,c'}(\rho)=w_{j,c}(\rho)
\end{eqnarray} 
where the channel-channel coupling potentials are:
\begin{eqnarray}
&&H_{cc'}(\rho)=\left[E-E_{t}+\frac{\hbar^{2}}{2m}\left\{\frac{\partial^{2}}{\partial \rho^{2}}
-\frac{(K+3/2)(K+5/2)}{\rho^{2}}\right\}\right]\delta_{cc'} \nonumber \\ 
&&+V^{(res)}_{cc'}(\rho)+ V^{(C)}_{cc'}(\rho)
+ V_{cc'}^{(v_0)}(\rho)+V^{(loc)}_{cc'}(\rho)+ V^{(nl)}_{cc'}(\rho) 
\end{eqnarray}


\section{Projection operator in ${\cal T}$-subspace  } \label{annexe_proj}
In appendix \ref{Annexe_direct}, we have discussed the calculation of the function $|\omega^{(+),0}_j\rangle$ 
neglecting the projection operator ${\hat T}$:
\begin{eqnarray}
(E-H)|\omega^{(+),0}_j\rangle=|w_j\rangle \label{oak0}
\end{eqnarray}
Below, we shall show how this  operator can be included effectively, {\it i.e} how one can solve equation:
$(E-H_{TT})|\omega^{(+)}_j\rangle=|w_j\rangle$.
By definition, $|\omega^{(+)}_j\rangle$ belongs to ${\cal T}$. Let us  now define 'forbidden states' for the state $|\omega^{(+)}_j\rangle$.  In the coordinate system  {\bf Y} 
(cf sect. \ref{para_direct}), they can be expressed as:
\begin{eqnarray} 
\Phi_{t,n,J_{t,n},s,J}=t(\xi) u_{n}(x_1)f_{s}(y_1)|\{(I_{t},j_{n}),J_{t,n}\},j_{s};J\rangle
\label{sp_row1}
\end{eqnarray}
Here, $t(\xi)$ is a state of the nucleus  A-2 and $u_{n}(x_1)$ is a (quasi-)bound state of the one-body potential  $h_0$. $I_{t}$ is the total angular momentum of a state  $t(\xi)$ and $j_n$ is the total angular momentum of the state $u_{n}(x_1)$. $I_t$ and $j_n$ are coupled to $J_{t,n}$.
$\{f_{s}(y_{i})\}$ in (\ref{sp_row1}) is a complete set of spline functions \cite{Nun} and $j_s$ is the angular momentum associated with the spline function $f_s$. $J_{t,n}$ and $j_s$ are coupled to the total angular momentum $J$. To simplify notation, we shall write $|\Phi_{t,n,J_{t,n},s,J}\rangle$ as 
$|\Phi_{e,J}\rangle$, where $e \equiv (t,n,J_{t,n},s)$. States $|\Phi_{e,J}\rangle$ satisfy:
\begin{eqnarray}
\hat{T}|\Phi_{e,J}\rangle=0
\end{eqnarray}
 and are called ``forbidden'' in the sense that as $|\omega^{(+)}_j\rangle$ belongs to ${\cal T}$ then:
\begin{eqnarray}
\langle \omega_j^{(+)}|\Phi_{e,J}\rangle=0 
\end{eqnarray}
for all $e$. The set of states
 $|\Phi_{e,J}\rangle$  is constructed from all considered states $|t\rangle$ of a nucleus  A-2 and all (quasi)bound states of a potential $h_0$, using the complete set of spline functions $\{f_{s}(y_{i})\}$.

Solution of eq. (\ref{o_q}) is written in the form:
\begin{eqnarray} 
| \omega^{(+)}_j\rangle=|\omega^{(+),0}_j\rangle+\sum_{e} \lambda_e  |\omega^{(+)}_e\rangle \label{wwq}
\end{eqnarray} 
{\it i.e.} we want to determine coefficients  $\lambda_e$. $|\omega^{(+)}_e\rangle$ in eq. (\ref{wwq}) is a solution of the equation:
\begin{eqnarray} 
(E-H)| \omega^{(+)}_e\rangle=|\Phi_{e,J}\rangle
\label{sp_row2}
\end{eqnarray} 
The calculation of $|\omega^{(+)}_e\rangle$ is analogous to the calculation of $|\omega^{(+),0}_j\rangle$ (cf appendix \ref{Annexe_direct}). 
From (\ref{oak0}) and (\ref{sp_row2})  one can see that $|\omega^{(+)}_e\rangle$ fulfills the equation:
\begin{eqnarray}
(E-H)|\omega^{(+)}_j\rangle=|w_j\rangle +\sum_e \lambda_e |\Phi_{e,J}\rangle \label{oak1}
\end{eqnarray}
Coefficients  $\lambda_e$ in (\ref{wwq}) are chosen to assure that $|\omega^{(+)}_j\rangle$ belongs to ${\cal T}$, {\it i.e.} they satisfy:
\begin{eqnarray}
\langle \omega_j^{(+)}|\Phi_{e,J}\rangle=0 ~~~~~~~~~~~~~~~~~~~~ \forall ~~e  
\end{eqnarray}
or equivalently:
\begin{eqnarray}
\langle \omega_j^{(+),0}|\Phi_{e,J}\rangle
+\sum_{e'} \lambda_{e'} \langle \omega_{e'}^{(+)}| \Phi_{e,J}\rangle=0  ~~~~~~~~~~~~~~~\forall~~e
\end{eqnarray}
Hence, $|\omega^{(+)}_j \rangle$  belongs to ${\cal T}$ and is the solution of equation (cf eq. \ref{oak1}):
\begin{eqnarray}
(E-H{\hat T})|\omega_j^{(+)}\rangle=|w_j\rangle +\sum_e \lambda_e |\Phi_{e}\rangle
\label{rown_dod}
\end{eqnarray}
Multiplying eq. (\ref{rown_dod}) from the l.h.s. by the operator ${\hat T}$, one obtains eq. (\ref{o_q}), what completes the argument.

\end{document}